\documentclass[prd,aps,nofootinbib,showpacs,floats,letterpaper,floatfix,groupedaddress,eqsecnum]{revtex4}

\usepackage{dcolumn,epsfig}
\usepackage{amssymb,amsmath}

\def\nn{\nonumber}

\def\be{\begin{equation}}
\def\ee{\end{equation}}
\def\beq{\begin{eqnarray}}
\def\eeq{\end{eqnarray}}

\def\IL{\relax{\rm I\kern-.18em L}}

\def\nn{\nonumber}
\def\f{\frac}

\begin{document}

\title{Inspiral, merger and ringdown of unequal mass black hole binaries:\\
a multipolar analysis}

\author{Emanuele Berti} \email{berti@wugrav.wustl.edu}
\affiliation{McDonnell Center for the Space Sciences, Department of Physics, Washington University, Saint Louis,
Missouri 63130, USA}

\author{Vitor Cardoso} \email{vcardoso@phy.olemiss.edu}
\affiliation{Department of Physics and Astronomy, The University of Mississippi, University, MS 38677-1848, USA
\footnote{Also at Centro de F\'{\i}sica Computacional, Universidade de Coimbra, P-3004-516 Coimbra, Portugal}}

\author{Jos\'e A. Gonzalez} \email{jose.gonzalez@uni-jena.de}
\affiliation{Theoretical Physics Institute, University of Jena, Max-Wien-Platz 1, 07743, Jena, Germany
\footnote{Also at Instituto de F\a'{\i}sica y Matem\a'aticas,
Universidad Michoacana de San Nicol\a'as de Hidalgo, Edificio C-3,
Cd. Universitaria. C. P. 58040 Morelia, Michoac\a'an, M\a'exico}}

\author{Ulrich Sperhake} \email{ulrich.sperhake@uni-jena.de}
\affiliation{Theoretical Physics Institute, University of Jena, Max-Wien-Platz 1, 07743, Jena, Germany}

\author{Mark Hannam} \email{mark.hannam@uni-jena.de}
\affiliation{Theoretical Physics Institute, University of Jena, Max-Wien-Platz 1, 07743, Jena, Germany}

\author{Sascha Husa} \email{sascha.husa@uni-jena.de}
\affiliation{Theoretical Physics Institute, University of Jena, Max-Wien-Platz 1, 07743, Jena, Germany}

\author{Bernd Br\"ugmann} \email{bernd.bruegmann@uni-jena.de}
\affiliation{Theoretical Physics Institute, University of Jena, Max-Wien-Platz 1, 07743, Jena, Germany}

\date{\today}

\begin{abstract}
  We study the inspiral, merger and ringdown of unequal mass black hole
  binaries by analyzing a catalogue of numerical simulations for seven
  different values of the mass ratio (from $q=M_2/M_1=1$ to $q=4$).  We
  compare numerical and Post-Newtonian results by projecting the waveforms
  onto spin-weighted spherical harmonics, characterized by angular indices
  $(l\,,m)$. We find that the Post-Newtonian equations predict remarkably well
  the relation between the wave amplitude and the orbital frequency for each
  $(l\,,m)$, and that the convergence of the Post-Newtonian series to the
  numerical results is non-monotonic.  To leading order the total energy
  emitted in the merger phase scales like $\eta^2$ and the spin of the final
  black hole scales like $\eta$, where $\eta=q/(1+q)^2$ is the symmetric mass
  ratio. We study the multipolar distribution of the radiation, finding that
  odd-$l$ multipoles are suppressed in the equal mass limit. Higher multipoles
  carry a larger fraction of the total energy as $q$ increases.  We introduce
  and compare three different definitions for the ringdown starting time.
  Applying linear estimation methods (the so-called Prony methods) to the
  ringdown phase, we find resolution-dependent time variations in the fitted
  parameters of the final black hole. By cross-correlating information from
  different multipoles we show that ringdown fits can be used to obtain
  precise estimates of the mass and spin of the final black hole, which are in
  remarkable agreement with energy and angular momentum balance calculations.

\end{abstract}

\pacs{04.25.Dm, 04.25.Nx, 04.30.Db, 04.70.Bw}

\maketitle

\clearpage

\tableofcontents

\clearpage

\section{Introduction}

More than thirty years after the first numerical simulations of binary black
hole dynamics, the numerical relativity community is finally ready to compare
binary black hole simulations with experimental data. Thanks to a series of
recent breakthroughs, long term evolutions of inspiralling binary black holes
that last for more than one orbit have been obtained with several independent
codes, and accurate gravitational wave signals have been computed
\cite{Buonanno:2006ui,Bruegmann:2006at,Gonzalez:2006md,Baker:2006yw,Campanelli:2006gf,Herrmann:2006ks,Sperhake:2006cy,Szilagyi:2006qy,Pfeiffer:2007yz,Pretorius:2005gq,Baker:2006ha}.

The use of numerical waveforms as templates for gravitational wave detection
requires large-scale parameter studies, and correspondingly large
computational resources. The main current technical problems in the field are
the efficiency of the numerical simulations and the development of a ``data
analysis pipeline'', connecting numerical simulations with analytical
calculations of the early inspiral and late ringdown phases, and (eventually)
with gravitational wave searches in actual detector data.
To build a common language between the numerical relativity and data analysis
communities we must develop a deeper understanding of the physical content of
the simulations using analytical techniques, such as Post-Newtonian (PN)
theory and black hole perturbation theory. A better analytical understanding
of the simulations is important for many reasons:

\begin{itemize}

\item[(1)] To determine which regions of the parameter space (mass, spin
  magnitude and inclination, orbital separation, eccentricity...)  {\it must}
  be explored by numerical simulations, and which regions can be covered by
  (say) analytically-inspired interpolations of the numerical waveforms. This
  would obviously save a significant amount of computing time.

\item[(2)] To develop optimal strategies for the construction of detection
  templates, using a combination of numerical and analytical techniques.

\item[(3)] To understand details of the non-linear physics encoded in the
  strong-field merger gravitational waveforms, and extract as much science as
  possible from a detection.

\end{itemize}

In this paper we focus on point (3), and we try to develop a general framework
to quantitatively compare analytical calculations of the inspiral and ringdown
waveforms with the ``full'' waveforms produced by numerical simulations,
extending from late inspiral through merger and ringdown.

Our work can be considered an extension of the recent analysis by Buonanno,
Cook and Pretorius (\cite{Buonanno:2006ui}, henceforth BCP).  BCP studied
simulations of non-spinning, equal mass black hole binaries starting out at
three different initial separations.  In this work we examine a larger set of
simulations performed using the {\sc Bam} code
\cite{Bruegmann:2006at,Gonzalez:2006md} and the moving puncture method. We
consider seven different mass ratios ($q\equiv M_2/M_1\simeq 1$ to $q\simeq 4$
in steps of $\simeq 0.5$) with initial coordinate separation $D\simeq 7\,M$,
roughly corresponding to $\sim 2$ orbits before merger. For each mass ratio,
the simulations were carried out at three different resolutions. To explore
the effect of initial separation on the physical parameters of the remnant,
we also consider two runs at separation $D\simeq 8\,M$ (for $q=2$ and $q=3$),
and one run at separation $D\simeq 10\,M$ (for $q=1$).  We typically use an
extraction radius $r_{\rm ext}=30M$, with the exception of the $q=1$ run with
$D\simeq 10\,M$, in which case we extract gravitational waves at $r_{\rm
  ext}=30M,~40M$ and $50M$.

Section \ref{setup} contains details of our numerical setup.
In Section \ref{memory} we study in some detail a well-known issue with the
extraction of gravitational waveforms from numerical simulations: the problem
of fixing integration constants when we integrate the Weyl scalar $\Psi_4$
twice in time to obtain the gravitational wave amplitude $h$. Fixing the
integration constants to zero produces a systematic drift in $h$ and in its
first time derivative. This drift is sometimes referred to in the literature
as a ``memory effect'', but this is somewhat misleading. The so-called
``memory effect'' is really due to numerical errors, wrong initial conditions
and limitations of wave extraction techniques, and it should {\it not} be
confused with the Christodoulou memory, which is a true (if typically small)
physical effect due to the non-linearity of general relativity
\cite{memory-refs}. We find that the extraction radius is critical to reduce
the amplitude drift, and that resolution only seems to affect the drift for
low-amplitude components of the wave.

In Section \ref{sec:transition} we study the inspiral-merger transition.  We
start by projecting the 2.5PN gravitational wave amplitude for quasi-circular,
non-spinning binaries \cite{Blanchet:1996pi,Arun:2004ff,Kidder:2007gz} onto
spin-weighted spherical harmonics. In this way we obtain the spin-weighted
spherical harmonic components of the Weyl scalar as PN series in the binary's
orbital frequency: $\psi_{l\,,m}=\psi_{l\,,m}(M\Omega)$. We refer to this
analytical expression of the gravitational wave amplitudes as the {\it
  Post-Newtonian Quasi-Circular} (PNQC) approximation (see Section
\ref{sec:pnqc} for details, and Appendix \ref{app:multipoles} for a complete
list of all the multipolar components).

The PNQC approximation can be used in two ways. First, given the orbital
frequency evolution $\Omega(t)$, we can compute (an approximation to) the
multipolar components $\psi_{l\,,m}$.  Conversely, given the modulus of the
wave amplitude $|\psi_{l\,,m}|(t)$, we can numerically invert the PN
expansions to obtain a PNQC estimate of the orbital frequency: $\Omega\simeq
\omega_{\rm PNQC}$.  In Sections \ref{freq-ests} we compare $\omega_{\rm
  PNQC}$ with two alternative estimates of the orbital frequency, first
introduced in BCP: $\omega_{Dm}$ (an estimate obtained from the gravitational
wave frequency) and $\omega_c$ (computed from the punctures' coordinate
motion). Using these three different estimates of the orbital frequency, we
study the convergence of the PNQC approximation. We find that, as in the point
particle case \cite{Poisson:1995vs}, the convergence of the PN series is not
monotonic.  We also study the effect of resolution and wave extraction on the
agreement between PNQC results and numerical results.  We find that low
resolution increases numerical noise in the frequencies and amplitudes at late
times. A small extraction radius produces systematic errors at large
separations, where gravitational wavelengths are longer, but it does not
sensibly affect the ringdown phase.

In Section \ref{energy-j} we study in detail the total radiated energy $E_{\rm
  tot}$ and the final angular momentum $j_{\rm fin}$ as functions of the mass
ratio, providing fitting formulas for each of these quantities. We also
compare the energy and angular momentum fluxes with their PNQC estimates, and
we study (both analytically and numerically) the multipolar distribution of
the radiation. To leading order, we find that $E_{\rm tot}\sim \eta^2$ and
$j_{\rm fin}\sim \eta$, where $\eta\equiv q/(1+q)^2$ is the so-called
symmetric mass ratio, and we provide fitting formulas for these quantities. As
predicted by the PNQC approximation, odd-$l$ multipoles of the radiation are
suppressed in the equal mass limit.  As the mass ratio increases, higher
multipoles (with $l>2$) carry a larger fraction of the total energy: for
$q\gtrsim 2$, $l=3$ typically carries $\sim 10\%$ of the total energy (see
Table \ref{tab:summary} below).

In Section \ref{fitmerger} we turn our attention to the merger-ringdown
transition. During ringdown the waveform can be described as a superposition
of complex exponentials, the quasinormal modes (QNMs). In \cite{Berti:2007dg}
we argued that Prony methods (which are well-known in signal processing) are
in many ways ``optimal'' methods to extract QNM frequencies from a numerical
signal. After explaining our choice of the fitting window, in Section
\ref{wrjm} we use Prony methods and standard, non-linear least-squares fits to
look at the time dependence of the final black hole's parameters. We find
resolution-dependent deviations in these parameters from the values predicted
by linear black hole perturbation theory. These effects may be due to
non-linearities and/or to rotational mode coupling, but at present we cannot
exclude the possibility that they are, more trivially, an artifact of finite
numerical resolution. In Section \ref{sec:crosscorrelate} we show that, by
cross-correlating information from different multipolar components of the
ringdown waves, we can find an empirical ``best guess'' for the optimal time
to estimate the final black hole's mass and angular momentum. We argue that,
because of the no-hair theorem, this best guess corresponds to the last time
when the angular momenta (or masses) obtained by fitting the dominant
multipoles agree with each other. In support of this argument, we also show
that estimates of the mass and spin of the final black hole based on QNM fits
are in remarkable agreement with wave extraction methods.

Black hole QNMs do not form a complete set, and for this reason it is not
possible to define unambiguously the beginning of the ringdown phase. In
Section \ref{sec:rdstart} we consider three different definitions of the
ringdown starting time, two of which have already appeared in the literature
(but not in the context of binary black hole simulations).  The first
definition is based on looking for the time at which a QNM expansion provides
the best fit to the actual numerical waveform, in the sense of a suitably
defined norm \cite{Dorband:2006gg}.
Unfortunately, when applied to our numerical waveforms, this method is not
particularly useful. The reason is that the norm is quite flat (and even
worse, has some oscillations) over a wide range of starting times around the
minimum. A second, more useful definition looks for the time maximizing the
{\it energy content} of the QNM component of the waveform.  For this reason,
following Nollert \cite{Nollert}, we call it the Energy Maximized Orthogonal
Projection, or EMOP.
We find that the ``EMOP time'' $t_{\rm EMOP}$ and the maximum fraction of
energy carried by ringdown ($\simeq 42\%$) are remarkably independent of the
mass ratio $q$.  This is an indication that the ringdown waveform is in some
sense ``universal'': it does not depend too much on the details of the
pre-merger phase.  To our knowledge, the third definition of the ringdown
starting time has not been introduced before. It uses a detection-based
criterion, maximizing the ``effective energy'' deposited in a matched filter.

In the conclusions we present a list of open problems and directions for
future research.

To improve readability, some lengthy equations and technical material are
presented in the Appendices.

Appendix \ref{app:multipoles} lists the spin-weighted spherical harmonic
components of the Weyl scalar, up to and including 2.5PN terms in a PN
expansion of the waveforms.

Appendix \ref{app:postplunge} provides fits for the energy, angular momentum
and linear momentum radiated after the estimated time of formation of a common
apparent horizon (CAH). Since the total energy radiated in a simulation
depends on the initial separation of the binary, in this Appendix we also try
to provide estimates for the energy, angular momentum and linear momentum
radiated ``after plunge''. A problem here is that the Innermost Stable
Circular Orbit (ISCO) is a controversial concept for comparable-mass binaries,
and there is no unique way to define the beginning of the plunge phase. Given
these intrinsic ambiguities, we estimate the starting time of the plunge,
$t_{\rm ISCO}$, as the time when the orbital frequency $\Omega$ becomes
larger than the ISCO frequency computed in PN theory (at 2PN or 3PN order, to
bracket uncertainties). We also present a comparison of our results with PN
estimates of the post-plunge radiation by Blanchet {\it et al.}
\cite{Blanchet:2005rj}.

Computational resources and resolution limitations reduce the accuracy of
numerical simulations for large mass ratio. Unfortunately, many astrophysical
black hole binaries could have $q=10$ or larger (see eg. \cite{Berti:2006ew}
and references therein). It is important to determine the maximum value of $q$
that should be simulated in numerical relativity, or equivalently, the
smallest value of $q$ for which black hole perturbation theory can be
considered adequate for detection and/or parameter estimation. Appendix
\ref{app:pointparticles} collects some results from perturbation theory that
may be useful in this context. We point out that, for large mass ratio, our
numerical simulations seem to be in reasonable agreement with perturbative
calculations of particles plunging with large angular momentum into a
Schwarzschild black hole.

Finally, in Appendix \ref{app:polarization} we introduce quantitative measures
of the polarization state of the waveform. We show that the polarization of
the wave (as viewed from the normal to the orbital plane) is circular for both
inspiral and ringdown, with the exception of the unphysical portions of the
wave: the initial data burst and the final, noise-dominated part of the
ringdown waveform.

In all of this paper we adopt geometrical units ($c=G=1$). Unless otherwise
indicated, physical quantities are usually normalized to the total
Arnowitt-Deser-Misner (ADM) mass of the system $M$. The ADM masses of all
configurations presented in this study have been calculated using Ansorg's
\cite{Ansorg:2004ds} spectral solver for binary black hole puncture data. Due
to the spectral accuracy and the compactification of the coordinates, which
facilitates evaluation of the ADM mass at infinite radius, the uncertainties in
this quantity are negligible relative to those arising out of the numerical
time evolution.

\begin{table}[ht]
  \centering
  \caption{\label{tab:summary} Summary of the main results of this paper (see
    text). To convert from radiated momenta to kick velocities in km~s$^{-1}$,
    the numbers in this Table must be multiplied by $c/10^4\simeq 30$
    km~s$^{-1}$. Further details (including estimates of the uncertainties)
    are given in the bulk of the paper.}
\begin{tabular}{c|cc|ccc|cccc|cccc|c}
\hline \hline
$q$
& $j_{\rm fin}$ &$j_{\rm QNM}$
&$\frac{E_{\rm tot}}{M}$
&$(\%l=2,3)$
&$\frac{J_{\rm tot}}{M^2}$
&$\frac{E_{\rm EMOP}}{M}$
&$(\%l=2,3)$
&$\frac{J_{\rm EMOP}}{M^2}$
&$\frac{10^4 P_{\rm EMOP}}{M}$
&$\frac{E_{\rm ISCO}}{M}$
&$(\%l=2,3)$
&$\frac{J_{\rm ISCO}}{M^2}$
&$\frac{10^4 P_{\rm ISCO}}{M}$
&$\frac{E_{\rm filter}}{M}$
%
\\
\hline
$1.0$ & 0.689&0.684 &0.0372 &(96.3,0.4)  &0.246 &0.0185 &(97.3,0.7)  &0.0700 &0    &0.0266 &(99.1,0.5) &0.1231 &   0 &0.028   \\
$1.5$ & 0.665&0.664 &0.0340 &(94.8,2.0)  &0.229 &0.0174 &(91.9,2.1)  &0.0676 &2.32 &0.0245 &(97.1,2.4)  &0.1165 &2.05 &0.026   \\
$2.0$ & 0.626&0.626 &0.0286 &(91.8,4.6)  &0.196 &0.0142 &(91.5,5.1)  &0.0565 &8.77 &0.0208 &(93.5,5.4)  &0.0985 &3.17 &0.021   \\
$2.5$ & 0.584&0.581 &0.0238 &(89.2,6.8)  &0.167 &0.0119 &(92.4,7.6)  &0.0480 &9.42 &0.0172 &(91.1,8.0)  &0.0850 &3.68 &0.018   \\
$3.0$ & 0.543&0.544 &0.0200 &(86.8,8.7)  &0.143 &0.0103 &(85.4,9.3)  &0.0438 &8.99 &0.0146 &(88.1,10.1) &0.0737 &3.92 &0.015   \\
$3.5$ & 0.506&0.509 &0.0170 &(84.6,10.1) &0.124 &0.0089 &(84.2,9.6)  &0.0387 &8.29 &0.0123 &(86.2,11.8) &0.0645 &3.88 &0.012   \\
$4.0$ & 0.474&0.478 &0.0145 &(83.2,11.3) &0.108 &0.0078 &(82.1,10.4) &0.0345 &7.49 &0.0106 &(83.4,13.0) &0.0568 &3.82 &0.011   \\
\hline \hline
\end{tabular}
\end{table}

For reference, we find it useful to summarize some of our main results in
Table \ref{tab:summary}. There we list, for each mass ratio:

\begin{itemize}

\item[(1)]
the dimensionless angular momentum of the final black hole $J_{\rm fin}/M_{\rm
  fin}^2$ as estimated from wave extraction methods ($j_{\rm fin}$) and from
QNM fits ($j_{\rm QNM}$);

\item[(2)]
the total energy and angular momentum radiated in each simulation
($E_{\rm tot}/M$, $J_{\rm tot}/M^2$);

\item[(3)]
the energy, angular momentum and linear momentum radiated in ringdown, where
the ringdown starting time is chosen according to the EMOP criterion
($E_{\rm EMOP}/M$, $J_{\rm EMOP}/M^2$, $P_{\rm EMOP}/M$);

\item[(4)]
the energy, angular momentum and linear momentum radiated after plunge, where
the plunge is defined by the location of the 3PN ISCO
($E_{\rm ISCO}/M$, $J_{\rm ISCO}/M^2$, $P_{\rm ISCO}/M$);

\item[(5)]
the effective fraction of energy detected by a ringdown filter
($E_{\rm filter}/M$).

\end{itemize}

The Table also shows the fraction of energy being radiated in the two dominant
multipoles ($l=2,~3$).

\clearpage

\section{Numerical setup}
\label{setup}

The sequence of numerical simulations of unequal mass black hole binaries
studied in this work has been obtained with the {\sc Bam} code
\cite{Bruegmann:2006at} using the moving puncture method
\cite{Campanelli:2006gf, Baker:2006yw}. Specifically, we study here a subset
of the sequence used in Ref.~\cite{Gonzalez:2006md} to determine the maximum
recoil resulting from the inspiral of non-spinning black hole binaries. The
{\sc Bam} code has been described extensively in Ref.~\cite{Bruegmann:2006at}
and further details of the numerical simulations of the unequal mass binaries
are given in Ref.~\cite{Gonzalez:2006md}. Here we summarize the model
parameters relevant for our present study.

A sequence of quasi-circular initial data of non-spinning black hole binaries
is determined by the initial coordinate separation $D$, the mass ratio $q$ of
the black holes, and the initial momenta $P_i$ of each black hole. Approximate
values of $P_i$ appropriate for quasi-circular orbits were calculated using
the 3PN-accurate expression given in Section VII of
Ref.~\cite{Bruegmann:2006at}.  For most of the models we consider in this
work, the initial coordinate separation is $D\simeq 7\,M$ (denoted by ``D7'').
The mass ratio is varied
from $q\simeq 1.0$ to $q\simeq 4.0$ in steps of approximately $0.5$.  In order
to assess the impact of larger initial separations on our results, we also
construct models with larger initial separation: $D\simeq 10\,M$ for $q\simeq
1.0$, and $D\simeq 8\,M$ for $q\simeq 2.0$ and $q\simeq 3.0$. We will denote
these models by D10 and D8, respectively. The complete set of models is
summarized in Table \ref{tab:pars}.

\begin{table}[htb]
  \caption{Summary of the main physical parameters for the series of
    simulations studied in this work. $q$ denotes the mass ratio, $D$ is the
    initial coordinate separation, and $J$ the total angular momentum. We also
    list the simulation time at which the orbital frequency equals the orbital
    frequency at the 3PN Innermost Stable Circular Orbit or ISCO, $t_{\rm
      ISCO}^{\rm 3PN}$; an estimate of the time at which a CAH forms, $t_{\rm
      CAH}$; the time at which the energy flux has a maximum, $t_{\rm flux}$;
    and the time at which the modulus of the $l=m=2$ mode has a peak, $t_{\rm
      peak}$. All quantities are normalized to the ADM mass $M$. The final
    column lists the number $N$ of orbits until the estimated time of
    formation of the CAH.}
 \begin{tabular}{ccccccccc}
\hline \hline
 $q$   & $D/M$ & $J/M^2$
&$t_{\rm ISCO}^{\rm 3PN}/M$ & $t_{\rm CAH}/M$
& $t_{\rm flux}/M$ & $t_{\rm peak}/M$ &$N$\\
 \hline
 1.00  & 7.046 & 0.8845 &211.9 & 215.0 & 231.8 & 234.0 &1.94 \\ 
 1.49  & 7.044 & 0.8494 &213.5 & 218.2 & 234.3 & 236.4 &1.96 \\ 
 1.99  & 7.040 & 0.7870 &211.6 & 217.5 & 233.6 & 235.4 &1.93 \\ 
 2.48  & 7.036 & 0.7232 &213.2 & 221.0 & 236.2 & 238.1 &1.96 \\ 
 2.97  & 7.034 & 0.6649 &213.2 & 223.2 & 237.8 & 239.8 &1.98 \\ 
 3.46  & 7.030 & 0.6132 &215.0 & 226.8 & 240.6 & 242.5 &2.02 \\ 
 3.95  & 7.028 & 0.5679 &216.7 & 230.5 & 243.0 & 244.8 &2.06 \\ 
 \hline
 1.00  & 10.104 & 0.9826&972.7 & 973.7 & 992.3 & 994.5 &5.93 \\ 
 1.99  & 8.086 & 0.8214 &438.0 & 443.9 & 459.2 & 461.6 &3.44 \\ 
 2.97  & 8.038 & 0.6865 &392.2 & 402.9 & 414.5 & 418.1 &3.23 \\ 
\hline \hline
 \end{tabular}
 \label{tab:pars}
\end{table}

All models have been evolved in time using a resolution of $M_1/22.4$ near the
punctures, where $M_1$ is the puncture mass of the smaller hole.  The models
starting from an initial separation $D=7\,M$ have also been evolved using
resolutions of $M_1/25.6$ and $M_1/28.8$. In the remainder of this work we will
refer to these resolutions as low (LR), medium (MR) and high resolution (HR).
Gravitational waves have been extracted in the form of the Newman-Penrose
scalar $\Psi_4$.  Unless specified otherwise, we use an extraction radius
$r_{\rm ext}=30\,M$.  We decompose the resulting $\Psi_4$ into modes by
projection onto spherical harmonics of spin-weight $s=-2$ (see
Ref.~\cite{Bruegmann:2006at} for conventions) according to
\be \label{psi4dec}
Mr\Psi_4=Mr\,\sum_{l=2}^\infty \sum_{m=-l}^l
 \,{_{-2}}Y_{lm}(\theta\,,\phi)\, \psi_{l\,,m}\,.
\ee

\begin{figure*}[ht]
\begin{center}
\begin{tabular}{cc}
\epsfig{file=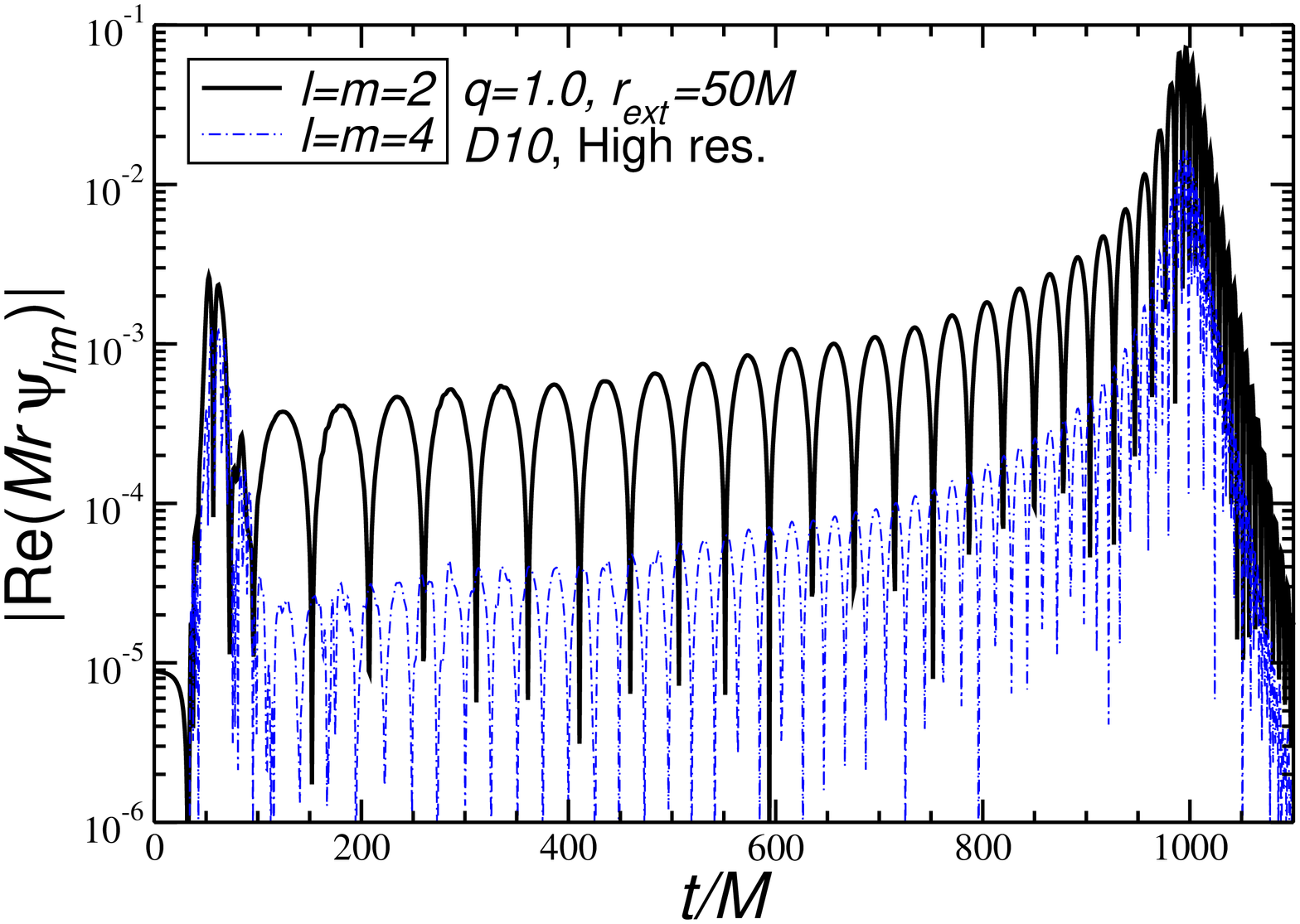,width=7cm,angle=-90} &
\epsfig{file=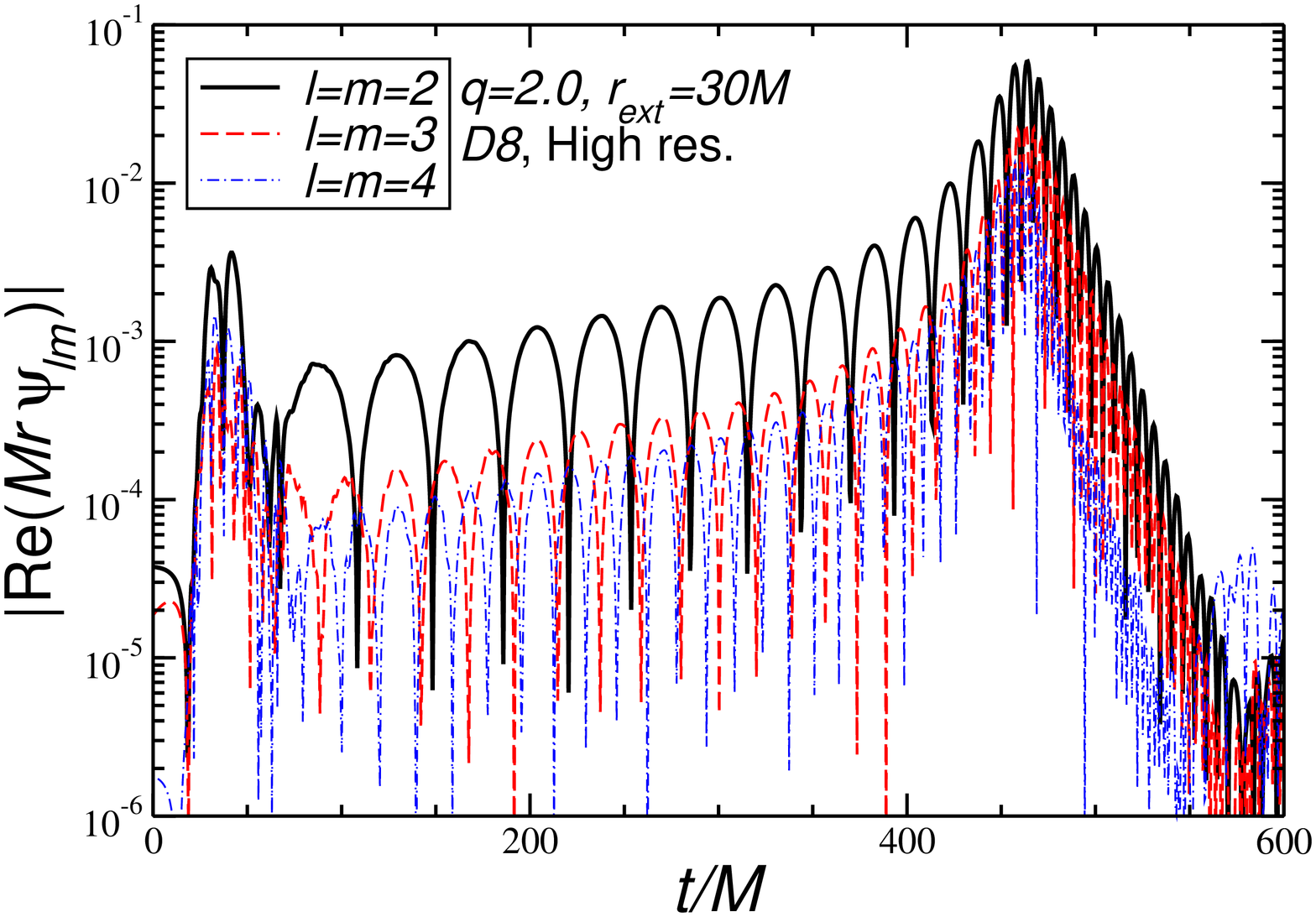,width=7cm,angle=-90} \\
\end{tabular}
\caption{$|{\rm Re}(Mr\,\psi_{l\,,m})|$ for $q=1.0$ (left) and $q=2.0$
  (right). For the equal mass ($q=1.0$) binary the $l=m=3$ component is
  strongly suppressed, and we do not show it.
  \label{wf1}}
\end{center}
\end{figure*}

In Fig.~\ref{wf1} we show examples of the resulting modes by plotting $|{\rm
  Re}(Mr\,\psi_{l\,,m})|$, the modulus of the real part of the waveforms.
Except for the spurious initial wave burst (visible up to about $t=50\,M$) and
for the final, noisy signal following the ringdown phase, the imaginary part
of $\psi_{l\,,m}$ is related to the real part by a phase shift of $\pi/2$ (see
Appendix \ref{app:polarization} for a more detailed discussion of the
polarization of the waveforms).  The figures demonstrate that the $l=2$,
$|m|=2$ modes dominate the gravitational wave emission in all simulations.
Contributions due to higher order modes become increasingly significant,
though, as the mass ratio is increased. In all of our models we find the
strongest contributions of higher-$l$ modes to result from $m=\pm l$. An
exception to this rule is the equal mass limit ($q=1$), where odd-$m$ modes
(including the $l=m=3$ component) are suppressed. For this reason, in the left
panel of Fig.~\ref{wf1} we only show modes with $l=2,\,\,4$.

\begin{figure*}[ht]
\begin{center}
\begin{tabular}{cc}
\epsfig{file=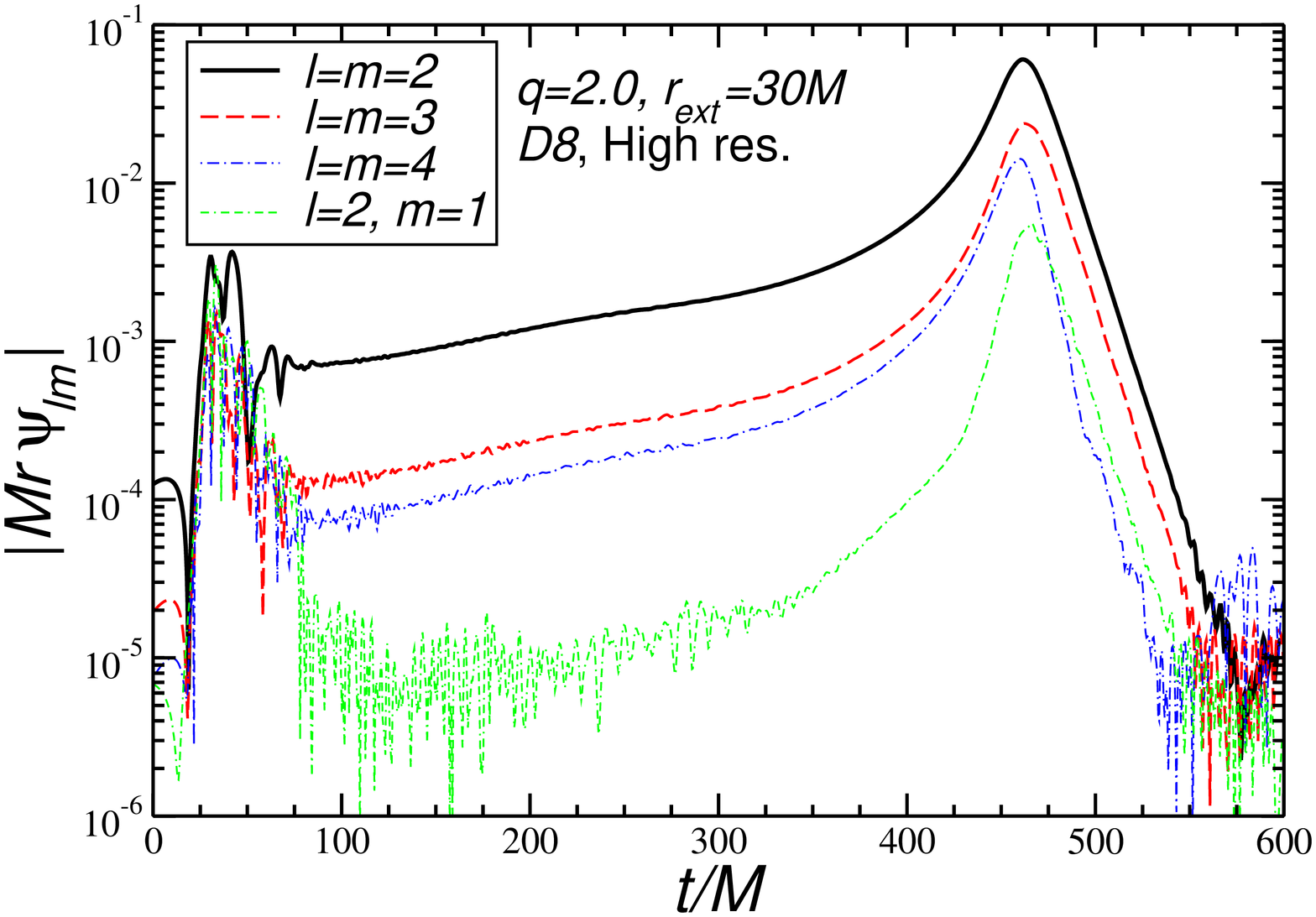,width=7cm,angle=-90} &
\epsfig{file=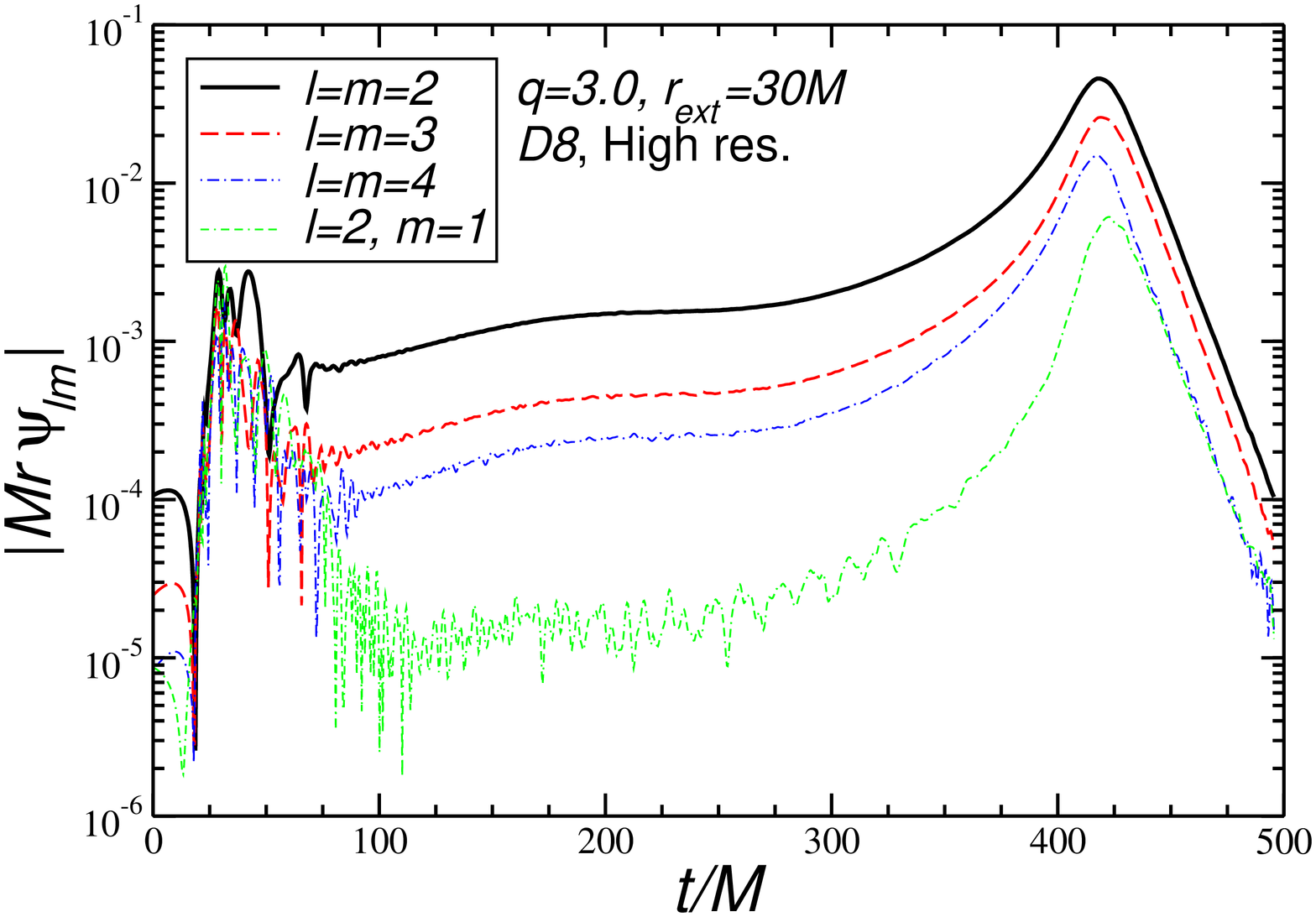,width=7cm,angle=-90} \\
\end{tabular}
\caption{$|Mr\,\psi_{l\,,m}|$ for different mass ratios. Each plot shows only
  some of the dominant components: $l=m=2,~3,~4$ and $(l=2,~m=1)$. The initial
  burst of radiation is induced by the initial data, and the wiggles at late
  times are due to numerical noise.
  \label{wf2}}
\end{center}
\end{figure*}

In Fig.~\ref{wf2} we plot the modulus $|Mr\,\psi_{l\,,m}|$ of the amplitude of
some of the dominant multipoles. Two features of this plot are worth
stressing: (i) the $l=m=4$ mode does not have a single, clear damping time in
the ringdown phase (this is particularly evident for $q=2.0$); (ii) the
amplitude modulation visible in the inspiral phase is induced by some
eccentricity in the initial data. This eccentricity seems to decrease during
the evolution, but estimates of the eccentricity damping are beyond the scope
of this paper.

The late-time, exponentially decaying portion of the waveforms is the ringdown
phase. As the wave amplitude decreases, numerical noise gradually starts
dominating the signal. In order to exclude this noisy part from the fitting of
damped sinusoids in the modelling of the ringdown part, discussed in Section
\ref{fitmerger} below, we introduce a cutoff time beyond which we no longer
use the waveforms.  The practical criterion to choose this late-time cutoff
will be discussed in more detail in Section \ref{fitmerger}.

\begin{figure*}[ht]
\begin{center}
\begin{tabular}{cc}
\epsfig{file=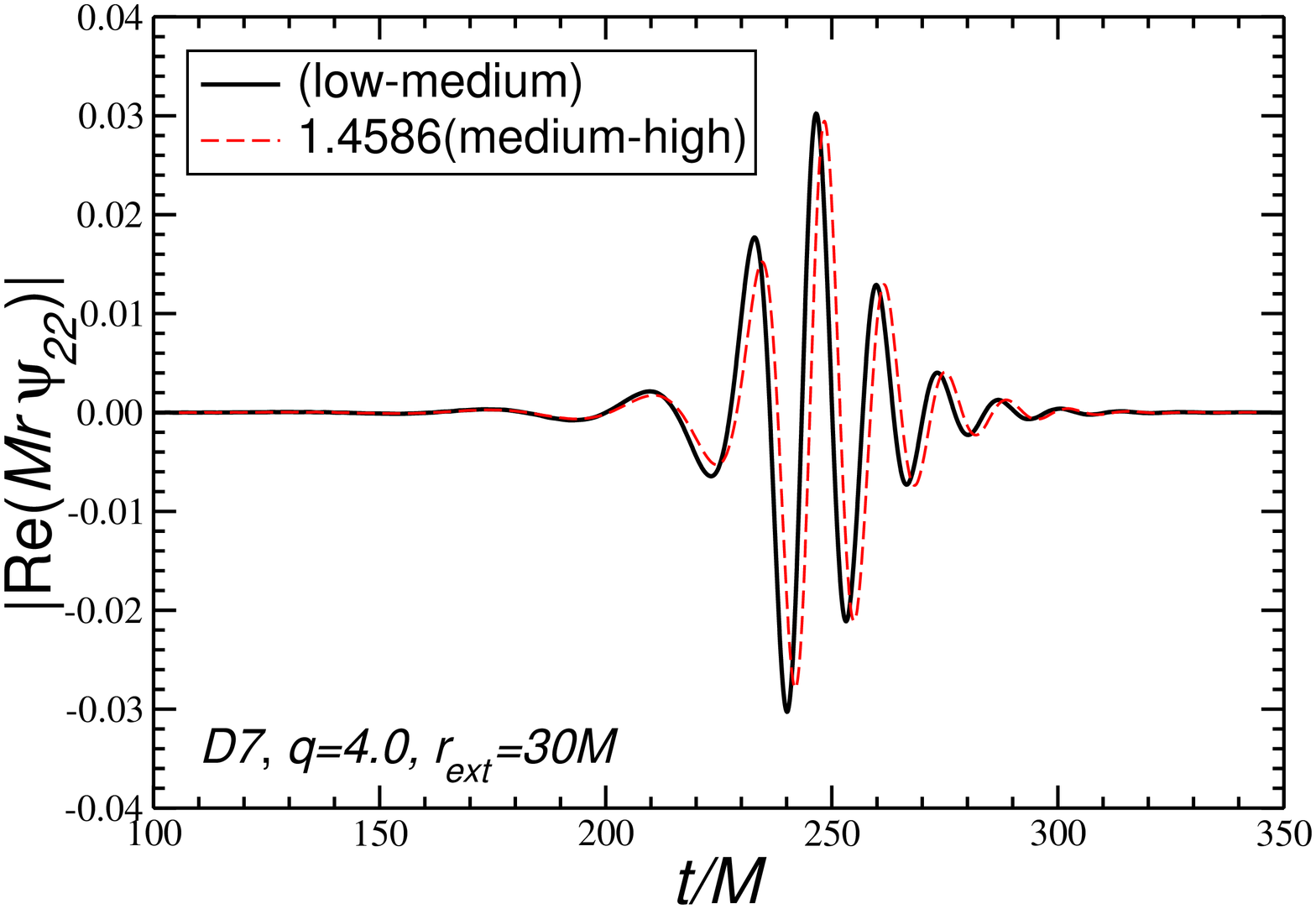,width=7cm,angle=-90} &
\epsfig{file=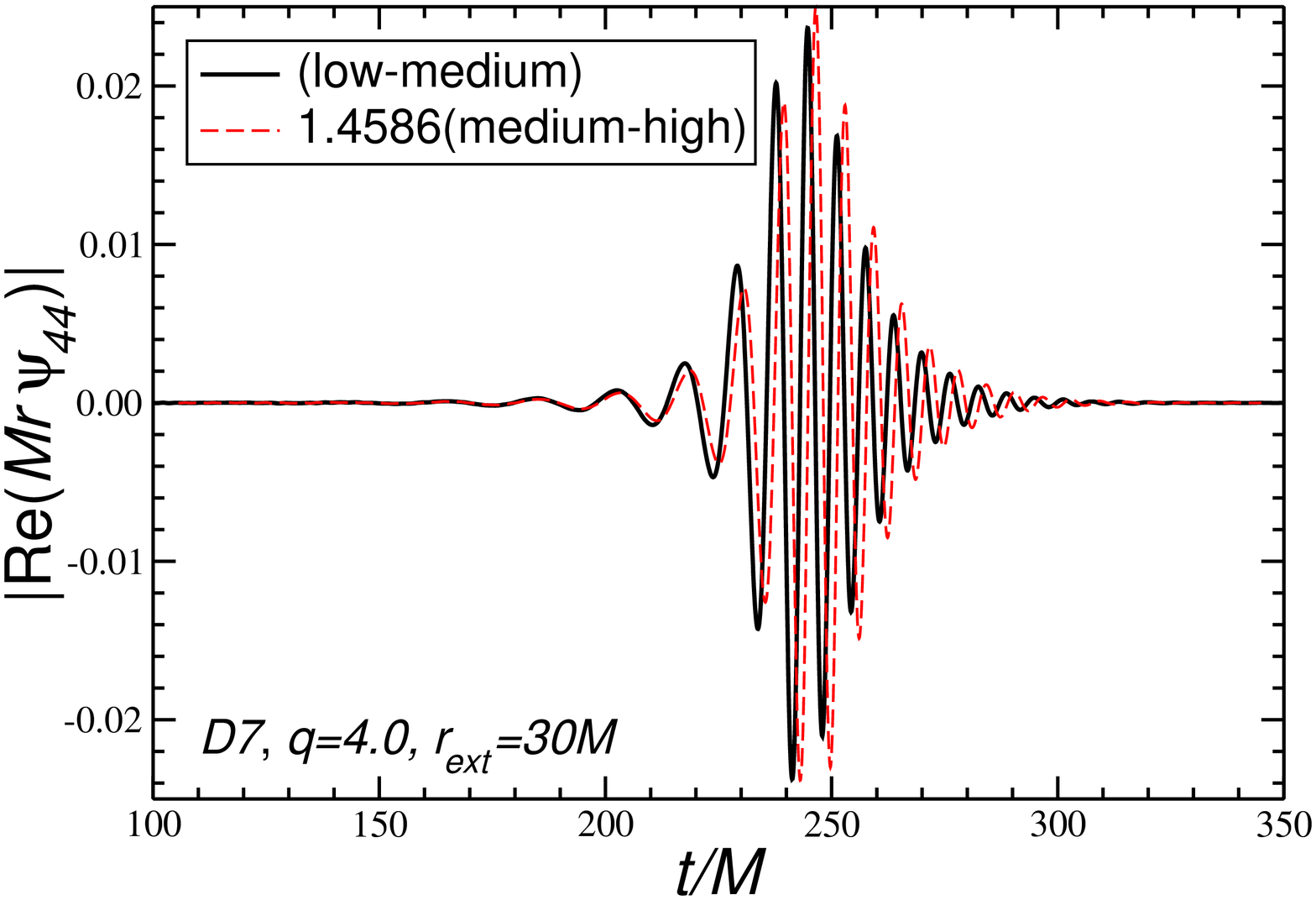,width=7cm,angle=-90} \\
\end{tabular}
\caption{Convergence plots for $Mr\psi_{22}$ (left) and $Mr\psi_{44}$
  (right). These plots show the differences between runs at different
  resolutions (as indicated in the inset), scaled to be consistent with
  second-order accuracy. They refer to run D7, mass ratio $q=4$ and $r_{\rm
    ext}=30M$.
  \label{fig: convergence}}
\end{center}
\end{figure*}

In order to assess the uncertainties arising from the discretization of the
Einstein equations, we performed a convergence analysis for mass ratios
$q=1,~2,~3$ and $4$. We find our results to converge at second order, with the
exception of the equal mass ($q=1$) case, where we have fourth order
convergence. Second order convergence is demonstrated in Fig.~\ref{fig:
  convergence}, where we show the real part of the $l=m=2$ and $l=m=4$
components of the scaled Newman-Penrose scalar $Mr\Psi_4$. Here we choose the
worst case (mass ratio $q=4$) but the convergence is still quite good,
especially considering that we do not apply any time shift to the waveforms.
We similarly find second order convergence for the radiated energy and momenta
(see \cite{Gonzalez:2006md} for further convergence plots). We are therefore
able to apply Richardson extrapolation and use the difference between the
values thus obtained and the high resolution numerical results as estimates
for the uncertainties associated with the finite differencing of the
equations.


For part of our analysis, we find it helpful to have estimates of the merger
time of the black hole binary. The most reliable estimate would be the
formation of a CAH.  In order to reduce the computational cost, however, all
simulations have been performed without using an apparent horizon finder, so
that we need to rely on alternative estimates. In the case of equal masses, we
follow Ref.~\cite{Baker:2006yw} and use the lapse function $\alpha$ to
estimate the black hole merger time as the time when the $\alpha=0.3$ regions
around each hole merge. Unfortunately, this criterion does not generalize
straightforwardly to unequal mass binaries. For these cases, we instead locate
the time when the ratio of the radial and tangential speeds of the punctures
is equal to 0.3, which corresponds roughly to the time when the black holes
reach the ``light ring'' in the effective-one-body model
\cite{Buonanno:2000ef}. This is discussed further in Section
\ref{sec:transition} below.

\subsection{Memory effects in subdominant multipoles}
\label{memory}

The effect of a gravitational wave on a detector in the far field of the
source is best described in terms of the transverse-tracefree part of the
metric.
The two polarization states, $h_+$ and $h_\times$, of the gravitational wave
are related to the curvature, expressed in terms of the complex Newman-Penrose
scalar $\Psi_4$, by
\begin{equation}\label{hpsi4}
  \Psi_4 = \ddot{h}_+ - i \ddot{h}_{\times}\,.
\end{equation}
Here $h_+(t-z) = h_{xx}^{TT} = - h_{yy}^{TT}$, $h_\times(t-z) = h_{xy}^{TT} =
h_{yx}^{TT}$ for a wave propagating in the $z$-direction.
Note that different conventions (typically for the Newman-Penrose scalar) are
used in the literature, correspondingly leading to different relations with
$h_+$ and $h_\times$.  Ref.~\cite{Baker:2002qf}, for example, has a factor $2$
in their Eq.~(5.3).

Given the Newman-Penrose scalar $\Psi_4$ for a particular mode, we thus have
to integrate twice in time to obtain $h_+$ and $h_\times$ and, in consequence,
fix two constants of integration, which correspond to the values and time
derivatives of $h_+$ and $h_\times$ at the initial time as functions on the
(celestial) sphere.  Integrations in time over $\Psi_4$ are also required to
compute the radiated energy, linear and angular momentum from the radiation
content:
\begin{eqnarray}
  \frac{dE}{dt} &=& \lim_{r\rightarrow \infty} \left[ \frac{r^2}{16\pi}
      \int_{\Omega} \left| \int_{-\infty}^{t} \Psi_4 d\tilde{t}
      \right|^2 d{\Omega}\right], \label{enflux}\\
  \frac{dP_i}{dt} &=& -\lim_{r \rightarrow \infty} \left[ \frac{r^2}{16\pi}
      \int_{\Omega} \ell_i \left| \int_{-\infty}^{t} \Psi_4 d\tilde{t}
      \right|^2 d{\Omega}\right], \\
  \frac{dJ_z}{dt} &=& -\lim_{r \rightarrow \infty} \left\{ \frac{r^2}{16\pi}
      \mathrm{Re}\left[ \int_{\Omega} \left( \partial_{\phi} \int_{-\infty}^t
      \Psi_4 d\tilde{t}
      \right)\left(
      \int_{-\infty}^t \int_{-\infty}^{\hat{t}}
      \overline{\Psi_4} d\tilde{t} d\hat{t}
      \right)d{\Omega} \right] \right\}, \label{FJnum}
\end{eqnarray}
where
\begin{equation}
  \ell_i = \left(-\sin \theta \cos \phi,\,\,\,-\sin \theta \sin \phi,\,\,\,
           -\cos \theta \right).
\end{equation}
The definitions above are based on time integrals which start in the infinite
past (at time $t=-\infty$), and thus capture the complete
gravitational wave signal.  Starting the time integrations at $t=-\infty$
corresponds to the limit of infinite extraction radius on the initial time
slice --- the slice would then extend all the way to spatial infinity, no part
of the waveform would be lost, and it would take an infinite time for the
waves to reach the extraction sphere.  With our current setup of the numerical
codes this situation cannot be handled, and we work with finite extraction
radii. The constants of integration would then correspond to the signal that
has been lost.  In order to accurately compute from the Newman-Penrose scalar
the radiated energy and momenta and the gravitational wave strain required by
data analysts, it is thus necessary to understand the influence of these
constants of integration, and ideally, how to choose them correctly\footnote{
  Koppitz {\it et al.} recently argued that the choice of integration
  constants may be important for an accurate calculation of the recoil
  velocity of the final black hole \cite{Koppitz:2007ev}.}.

Naively setting the constants to zero typically leads to a non-zero value and
slope of $h_+$ and $h_\times$ after the passage of the wave. This effect will
in general have contributions from the signal that has been lost due to a
finite extraction radius, from numerical error, and from the inherent
ambiguities of the extraction procedure at finite radius.  Furthermore, a time
independent gravitational wave {\em memory} effect is also possible and has
been described in the literature \cite{memory-refs,Arun:2004ff} (see also
\cite{Gleiser:2003md}). Apart from the effect due to an improper setting of
the constants of integration, the other effects will accumulate over time,
which may allow for some discrimination.

While the time independent phenomenon is physically expected (although it
should be small), the time-dependent drift phenomenon appears to be
counter-intuitive and we expect $h_+$ and $h_\times$ to settle down into a
stationary state at late times.  All unphysical effects in the non-zero value
and slope of $h_+$ and $h_\times$ after the passage of the wave should
converge away with resolution and increasing extraction radius. A rigorous
convergence test could attempt to identify a remaining physical gravitational
wave memory effect (indepent of time after the passage of the wave).
Consistency with the physical situation requires that the slope of $h_+$ and
$h_\times$ after the passage of the wave converge away with resolution and
increasing extraction radius.

\begin{table}[htb]
  \centering \caption{\label{tab:memory_slopes} {Average slope of the $l=2$,
      $m=2$ component of $h_+$ for different runs and mass ratios, obtained
      from linear regression.}
 }
\begin{tabular}{cccccccc}
\hline \hline
$q$ &run   & Resolution & $r_{\rm ex}$ & $10^5a^+_1$ & $r_{\rm ex}^4 a^+_1$ & $10^5a^{\times}_1$ & $r_{\rm ex}^2 a^{\times}_1$  \\
\hline
$1.0$& D10 & HR & 30 & 7.11 & 57.6 & 3.67 & 0.033 \\
$1.0$& D10 & HR & 40 & 2.24 & 57.3 & 2.37 & 0.038 \\
$1.0$& D10 & HR & 50 & 0.91 & 56.9 & 1.42 & 0.036 \\
\hline
$1.0$& D10 & MR  & 30 & 7.12 & 57.7 & 3.67 & 0.033 \\
$1.0$& D10 & MR  & 40 & 2.27 & 58.1 & 2.38 & 0.038 \\
$1.0$& D10 & MR  & 50 & 0.94 & 58.8 & 1.42 & 0.036 \\
\hline
$1.0$& D10 & LR  & 30 & 7.09 & 57.4 & 3.68 & 0.033 \\
$1.0$& D10 & LR  & 40 & 2.26 & 57.9 & 2.38 & 0.038 \\
$1.0$& D10 & LR  & 50 & 0.94 & 58.8 & 1.42 & 0.036 \\
\hline
\hline
$2.0$& D8 & LR  & 20 & 1.35 & 2.16 & 5.57 & 0.022 \\
$2.0$& D8 & LR  & 25 & 0.56 & 2.19 & 4.57 & 0.029 \\
$2.0$& D8 & LR  & 30 & 0.27 & 2.19 & 3.20 & 0.029 \\
\hline
\hline
$3.0$& D8 & LR  & 20 & 0.86 & 1.38 & 3.33 & 0.013 \\
$3.0$& D8 & LR  & 25 & 0.36 & 1.41 & 2.83 & 0.018 \\
$3.0$& D8 & LR  & 30 & 0.17 & 1.38 & 2.00 & 0.018 \\
\hline \hline
\end{tabular}
\end{table}

\begin{figure*}[ht]
\begin{center}
\begin{tabular}{cc}
\epsfig{file=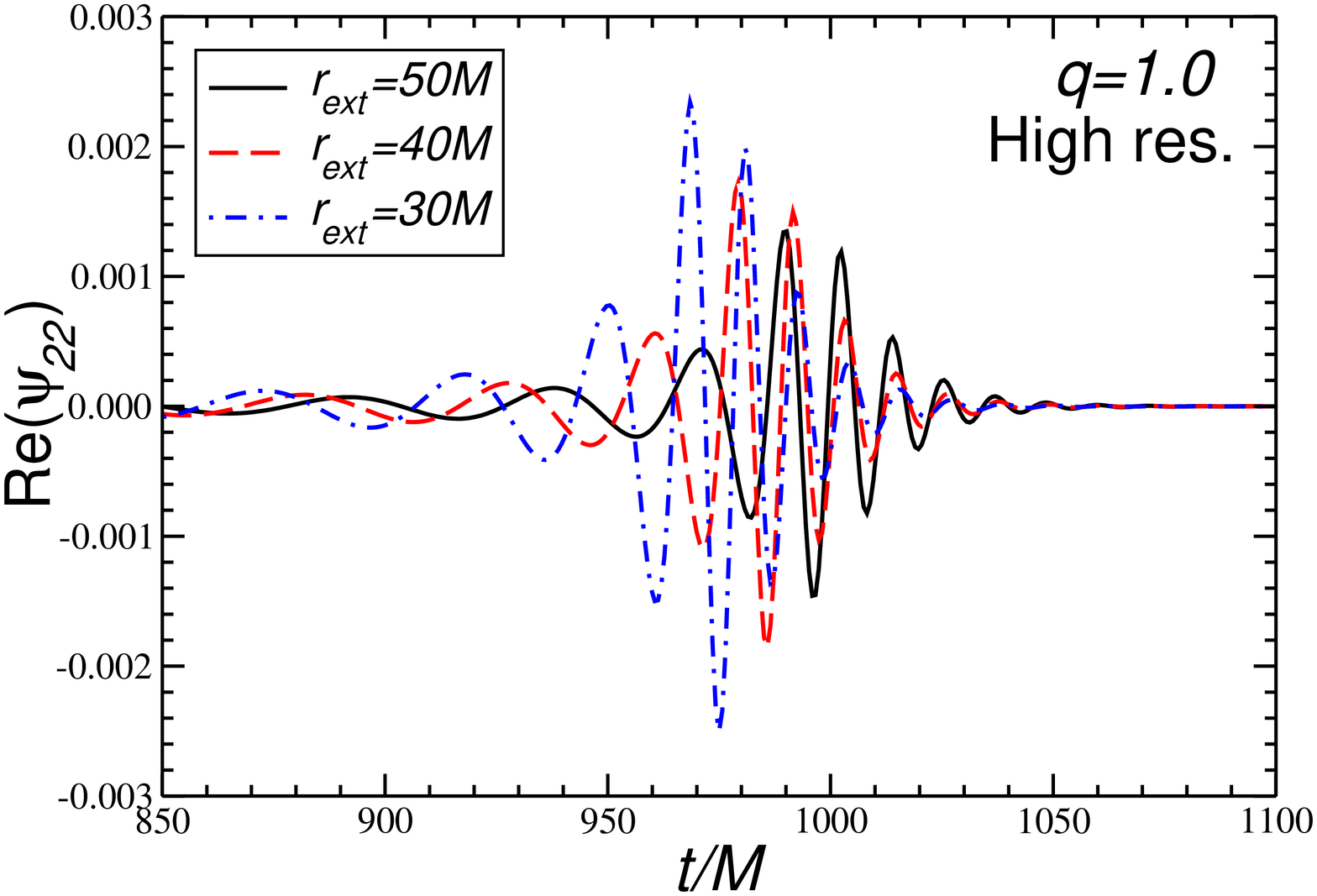,width=7cm,angle=-90}&
\epsfig{file=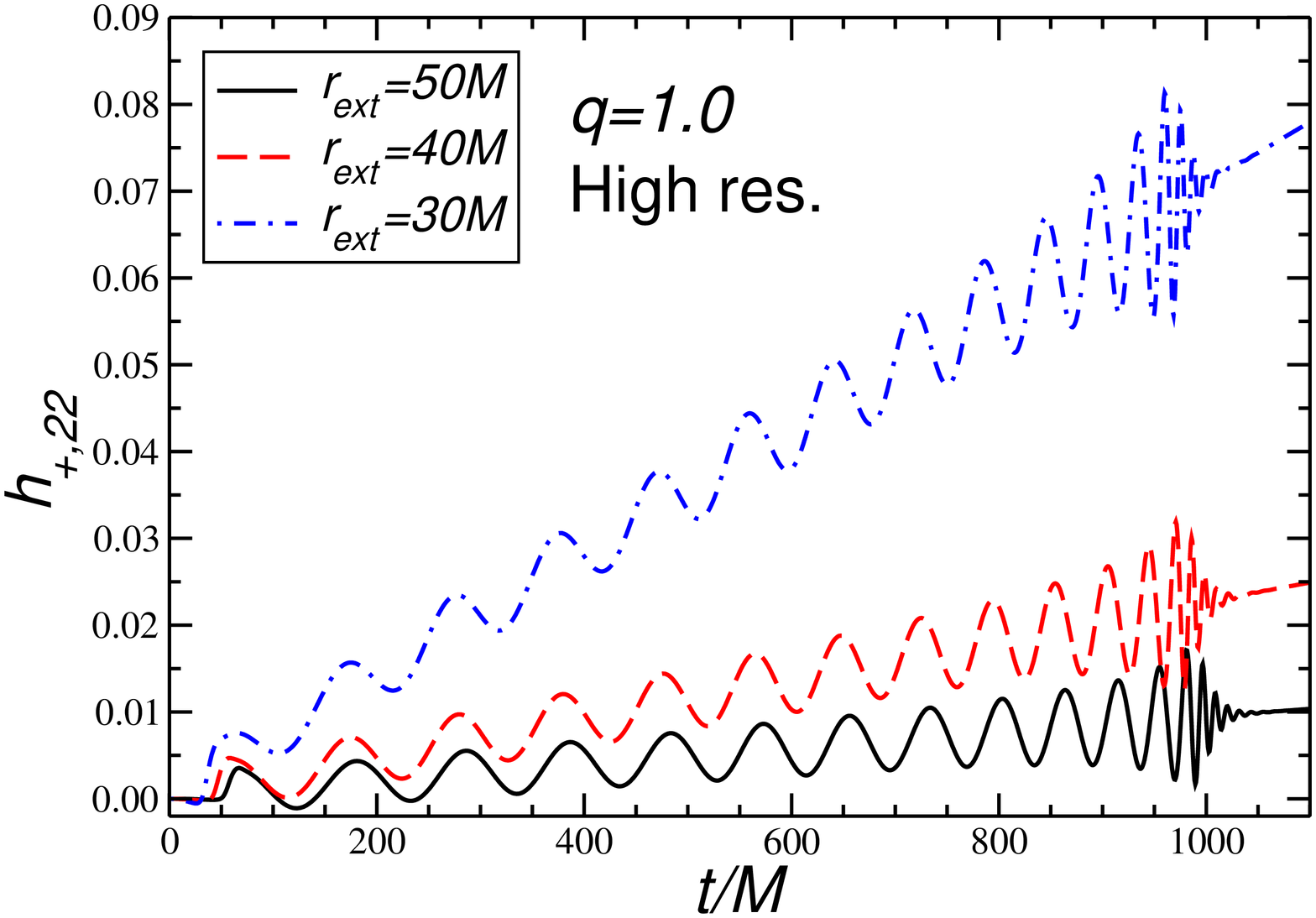,width=7cm,angle=-90}\\
\end{tabular}
\caption{Effect of changing the extraction radius on the ``memory effect''
  when we do not apply corrections to the integration constants. These plots
  refer to the dominant multipole ($l=m=2$) of run D10 with $q=1.0$.
  \label{PsiVsH}}
\end{center}
\end{figure*}

We study this effect in more detail by considering the D10 run with $q=1.0$.
In Fig.~\ref{PsiVsH} we plot the resulting $l=2$, $m=2$ contribution for $h_+$
obtained at different extraction radii.  The figure demonstrates two important
features of this memory effect. First, the linear growth starts right at the
beginning of the simulation, indicating that the memory effect is indeed
essentially due to a non-vanishing constant of integration, or that possibly
it is accumulated already in the early stages of the wave pulse (including the
artificial burst of radiation). Second, the slope decreases significantly if
we use larger extraction radii.

\begin{figure*}[ht]
\begin{center}
\begin{tabular}{cc}
\epsfig{file=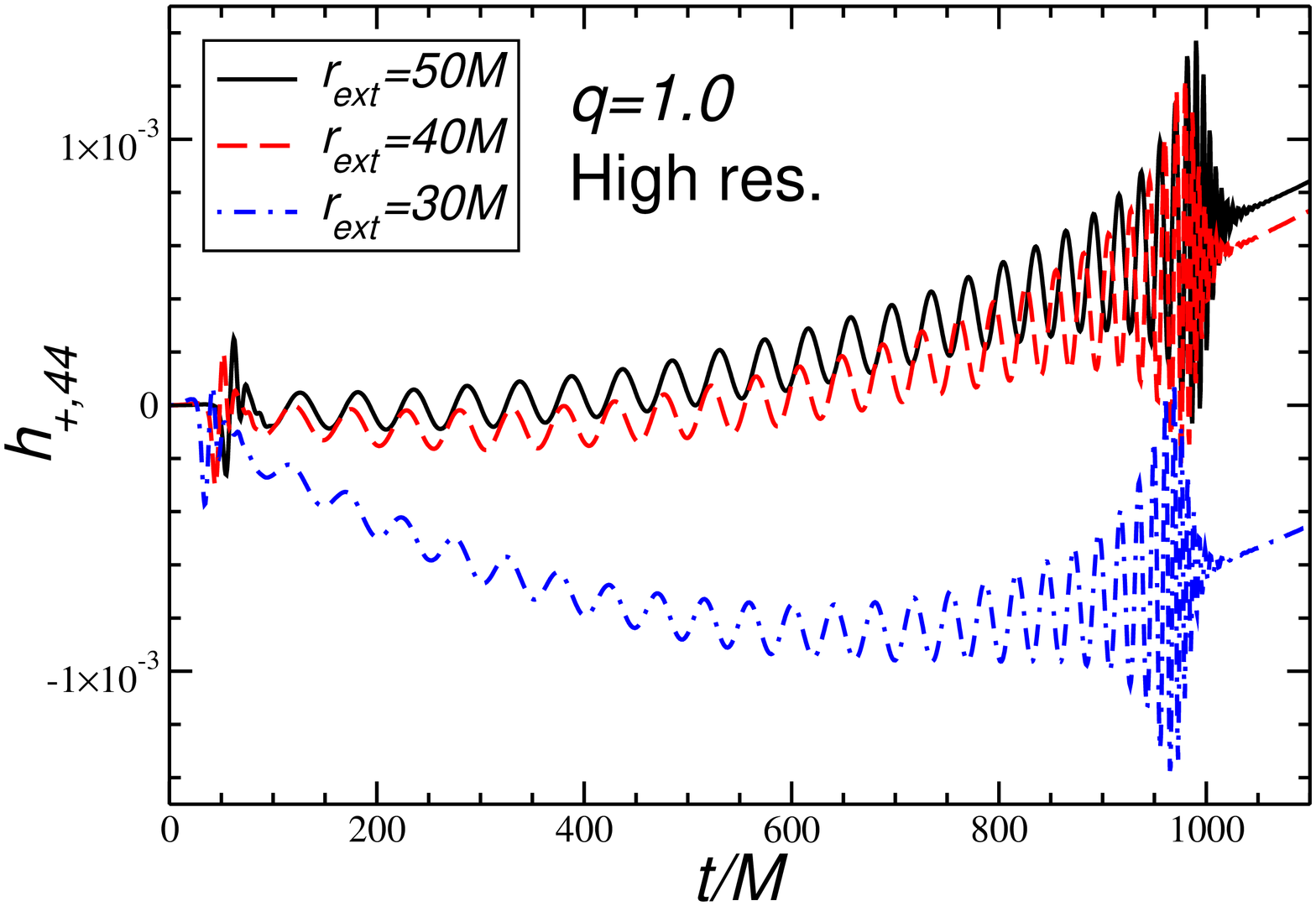,width=7cm,angle=-90}&
\epsfig{file=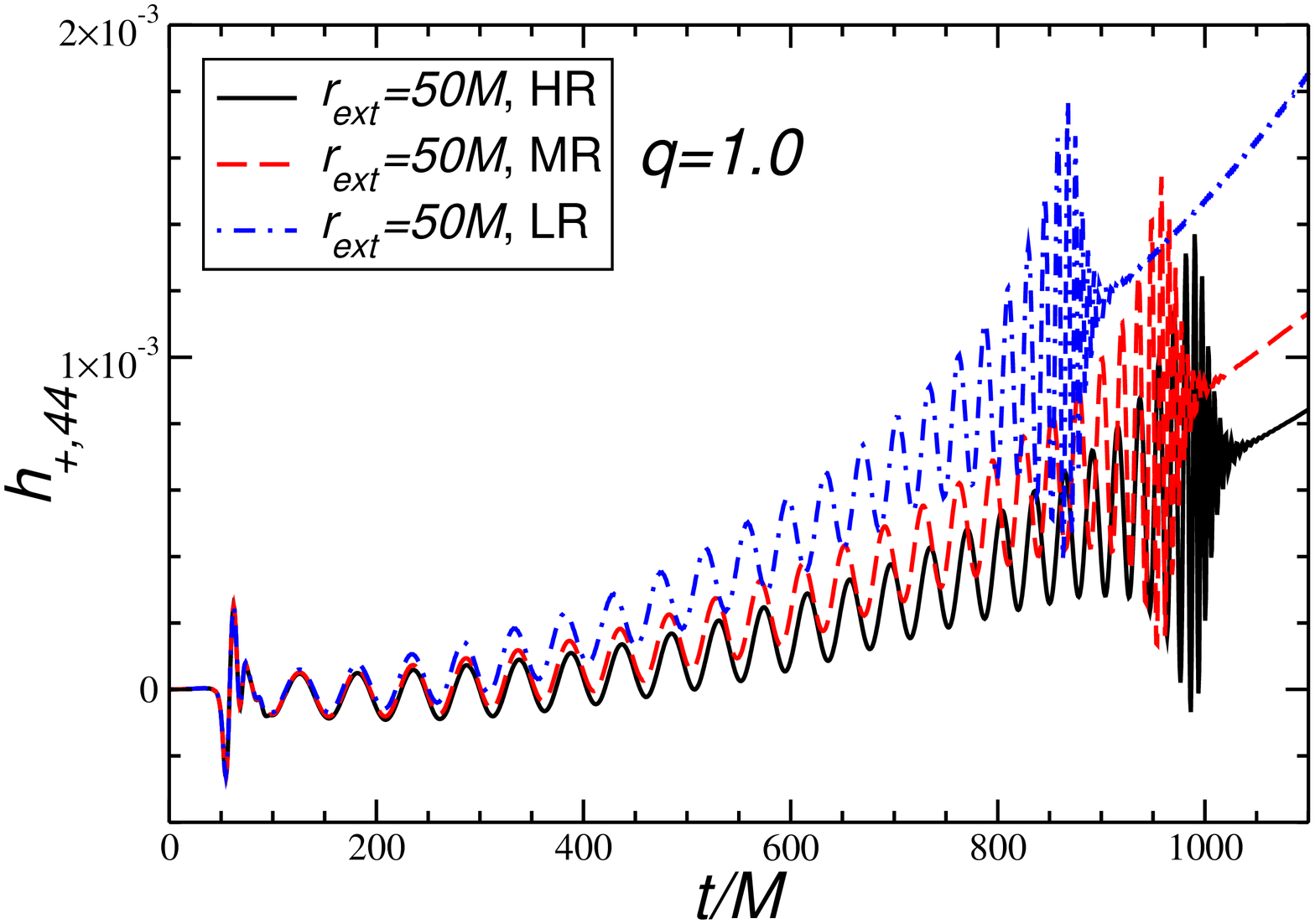,width=7cm,angle=-90}\\
\end{tabular}
\caption{Effect of changing the extraction radius on the ``memory effect''
  when we do not apply corrections to the integration constants. These plots
  refer to a small amplitude mode ($l=m=4$) of run D10 with $q=1.0$. The left
  panel shows the effect of changing the extraction radius at fixed
  resolution. In the right panel, we change the resolution at fixed extraction
  radius.
  \label{PsiVsH2}}
\end{center}
\end{figure*}

We next apply a least-squares fit of a linear function $f(t) = a_0 + a_1 t$ to
$h_+$ and $h_\times$ resulting from the simulations of models $q=1.0$, D10;
$q=2.0$, D8; and $q=3.0$, D8. The resulting slopes are labelled as $a_1^+$ and
$a_1^\times$ respectively in Table \ref{tab:memory_slopes}.  The table
demonstrates in the case of the model with $q=1.0$, D10 that the coefficients
are essentially independent of the grid resolution. Columns 5 to 7 of the
table indicate, however, that their dependence on the extraction radius can be
rather well approximated by power laws: $a_1^+\sim r_{\rm ex}^{-4}$, and
$a_1^\times \sim r_{\rm ex}^{-2}$.  This discrepancy between the $+$ and
$\times$ polarization modes is rather surprising, because the circular
polarization of the waves implies that they differ merely by a phase shift of
$\pi/2$. The key observation in this context is that this simple relation
between $h_+$ and $h_\times$ applies to the inspiral waveform but {\em not} to
the spurious initial wave burst. This is explicitly shown in Appendix
\ref{app:polarization}. We thus conclude that the slope is a consequence of
the omitted early wave signal (the constant of integration) or the initial
wave burst. We emphasize, however, that the decreasing impact of the initial
data pulse at larger radii is not due to it being dissipated away by numerical
viscosity, which would have manifested itself in a resolution dependence of
the slope.

The picture is somewhat more complicated in the case of the $l=4$, $m=4$ mode
plotted in Fig.~\ref{PsiVsH2}, where we do not only see a significant
dependence of the memory effect on resolution, but also a non-linear trend of
$h_+$, suggesting contributions from the ambiguities of wave extraction or
numerical error. Consistent with the issue being due to the ambiguities of the
wave extraction algorithm, we observe a significant decrease in the memory
effect at larger extraction radius. In contrast to the $l=2$, $m=2$ case, we
did not find a simple systematic dependence of the coefficients on extraction
radius and resolution.

We conclude this Section with a discussion of alternative choices for the
integration constants. For each polarization and each mode in the spectral
decompositions we work with (to take care of the angular dependence of these
constants of integration) we can fix the integration constants by demanding
that the time derivative of $h_{+/\times}$ vanishes at late times.  This can
be achieved by matching to a ringdown signal, or more heuristically by
subtracting the time average or time dependent polynomials, such as those
obtained by the fitting processes mentioned above.  The second constant of
integration, which may have a contribution from a {\it physical} memory
effect, is typically very small, and can be set to zero for many practical
purposes if the wave extraction radius is not too small.  From our
observations, we conclude that using a sufficiently large extraction radius is
certainly a highly recommended way of reducing spurious memory effects.

\clearpage

\section{The inspiral-merger transition}
\label{sec:transition}

\begin{figure*}[ht]
\begin{center}
\epsfig{file=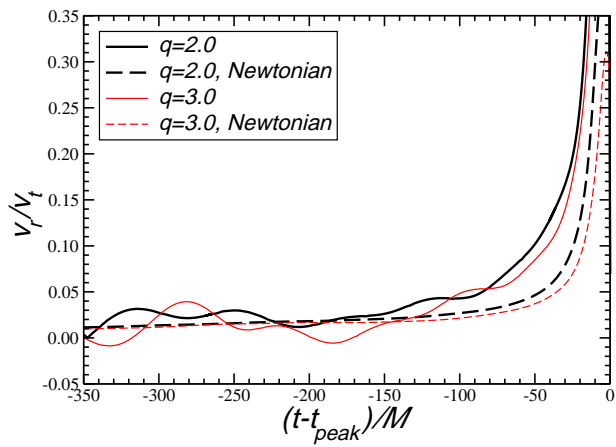,width=7cm,angle=-90}
\caption{Ratio of radial and tangential velocities for D8 runs and for two
  values of the mass ratio ($q=2.0$ and $q=3.0$). \label{velratio}}
\end{center}
\end{figure*}

The parameters chosen in Section \ref{setup} for the initial data do not give
perfect quasi-circular (non-eccentric) orbits.
This problem has been discussed in various
papers \cite{Buonanno:2006ui,Berti:2006bj,Miller:2003pd,Pfeiffer:2007yz}. A
simple way to visualize the residual eccentricity of the binary's orbit is to
compare the punctures' motion with predictions for circular, Newtonian orbits.
At leading order, the quadrupole formula predicts that the orbital radius
should evolve according to \cite{peters}:
\be v_r=\dot{r}=-\frac{64}{5}\frac{\eta\,M^3}{r^3}\,, \ee
where $\eta=M_1 M_2/M^2=q/(1+q)^2$ is the symmetric mass ratio. From the
relation between the orbital radius $r$ and the (Keplerian) orbital frequency
$\Omega$, $M=\Omega^2r^3$, we get the ratio of radial and tangential
velocities in unequal mass, circular orbit binaries:
\be\label{newtquad}
\f{v_r}{v_t}=
\left(\f{\dot r}{\Omega r}\right)=
-\f{64}{5}\eta (M\Omega)^{5/3}\,.
\ee
This formula is of course a rough approximation, being based on the quadrupole
formula and assuming a Keplerian orbit \cite{Buonanno:2006ui}.  In
Fig.~\ref{velratio} we show the ratio of radial and tangential velocities
$v_r/v_t$ obtained from the punctures' motion and from the Newtonian
quadrupole prediction. Curves labeled ``Newtonian'' are obtained by replacing
the punctures' orbital frequency $\Omega=\omega_c$ (see Section
\ref{PNQC-conv} below for details of the definition) in Eq.~(\ref{newtquad}).
Curves obtained from the actual puncture orbital motion clearly oscillate
around the Newtonian circular value, mainly because of the non-zero orbital
eccentricity. A similar effect was observed in Fig.~6 of BCP.

At early times in the evolution, say $(t-t_{\rm peak})\lesssim -100M$, the
ratio $|v_r/v_t|\lesssim 0.05$, and the orbit is (to a reasonably good
approximation) quasi-circular. At later times $v_r/v_t$ grows, as the motion
turns from inspiral to plunge. Given the computational cost of implementing an
apparent horizon finder during the evolution, we use the following rough
criterion to locate the formation of the apparent horizon. In the
effective-one-body model \cite{Buonanno:2000ef}, the ratio between radial and
tangential velocities is $v_r/v_t\sim 0.3$ (so that the motion is strongly
``plunging'') at the light ring $r=3M$.  Since the light ring should be close
to the location where a CAH forms, we simply {\it define} the time of
formation of a CAH $t_{\rm CAH}$ as the point where the ratio $v_r/v_t$, as
computed from the punctures' orbital motion, becomes larger than $0.3$ (see
also the related discussion around Table \ref{tab:pars}). Fig.~\ref{velratio}
shows that $v_r/v_t$ rises very steeply in this region, so we expect the error
introduced by our rough approximation to be at most of order a few $M$.

\subsection{\label{sec:pnqc}The Post-Newtonian quasi-circular approximation for the inspiral phase}

In this work we will perform extensive comparisons of numerical waveforms with
the PN approximation. For this purpose it is useful to decompose the Weyl
scalar $\Psi_4$ in spin-weighted spherical harmonic components according to
Eq.~(\ref{psi4dec}). The $\psi_{l\,,m}$'s can be obtained by taking two time
derivatives of the PN gravitational waveforms $h_{+\,,\times}$ according to
Eq.~(\ref{hpsi4}), and then computing
\be Mr\, \psi_{l\,,m}=Mr\, \int \sin\theta d\theta d\phi\, 
{_{-2}}Y_{l\,,m}^*(\theta\,,\phi) \Psi_4 \equiv Mr\, 
\int \sin\theta d\theta d\phi \,
{_{-2}}Y_{l\,,m}^*(\theta\,,\phi) \left (\ddot{h}_{+}-i\ddot{h}_{\times} \right )\,. \ee
The azimuthal dependence of the PN waveforms has the functional form $\left
  (\int \Omega dt-2M\Omega\ln{\Omega/\Omega_0}\right )-\phi$
\cite{Blanchet:1996pi}. Thus, the expansion of the {\it waveform} $h\equiv
h_+-ih_{\times}$ in spin-weighted spherical harmonics has a time dependence of
the form $\exp[-im( \int \Omega dt-2M\Omega\ln{\Omega/\Omega_0})]$.  For
consistency, we use the same convention\footnote{This definition does not
  include the Condon-Shortley phase, an extra factor of $(-1)^m$ which is
  needed for agreement with the usual definition of scalar ($s=0$) spherical
  harmonics.} on spin-weighted spherical harmonics as in
Refs.~\cite{Buonanno:2006ui,Bruegmann:2006at}. For the dominant, $l=m=2$
component of the waveform we get:
\beq
Mr\, (h_+-ih_{\times})_{2\,,2}&=& 8 \sqrt{\frac{\pi}{5}}\eta (M\Omega)^{2/3}
\left[1+\f{55\eta-107}{42}(M\Omega)^{2/3}+2\pi
(M\Omega)-\frac{2173+7483\eta-2047\eta^2}{1512}(M\Omega)^{4/3}\right.\nn\\
&+&\left. \left(\f{-107+34\eta}{21}\pi+\varpi i\eta
\right)(M\Omega)^{5/3} \right]e^{-im( \int \Omega
dt-2M\Omega\ln{\Omega/\Omega_0})}\,. \label{h22m} 
\eeq
Here $\Omega_0$ is an arbitrary constant \cite{Blanchet:1996pi} and the
orbital angular velocity $\Omega$ is a time dependent quantity, the 3.5PN
expansion of which can be found (for example) in \cite{Blanchet:2004ek}.

Kidder {\it et al.} \cite{Kidder:2007gz} recently corrected an inconsistency
in the derivation of radiation reaction terms of Ref.~\cite{Blanchet:2004ek}.
They also argued that radiation reaction terms are in fact negligible in the
2.5PN waveform, since they can be absorbed into a 5PN contribution to the
orbital phase evolution. The constant $\varpi$ in Eq.~(\ref{h22m}) depends on
whether radiation reaction terms are absorbed into a redefinition of the phase.
If we include radiation reaction terms in the waveform by using Eq.~(27) of
\cite{Kidder:2007gz} (as recommended by Kidder {\it et al.}) then
$\varpi=-24$. If instead we neglect the radiation reaction contribution by
using their Eq.~(32), we find $\varpi=-8/7$. In the following we present
analytical results including all known contributions to the waveform
(including the $\varpi$-dependent 2.5PN terms). Given the ambiguity due to the
inclusion of radiation reaction terms, we decided not to include 2.5PN
contributions in our comparisons with the $l=m=2$ numerical waveforms.

As stated earlier, to compute the projection of $\Psi_4$ onto spin-weighted
spherical harmonics we must take the second time derivative of expressions
like Eq.~(\ref{h22m}) above. Noticing that the logarithmic term in the phase
is of 4PN order \cite{Blanchet:1996pi}, we will simply neglect it when taking
the derivative\footnote{This is consistent with the PN order considered here.
  While it would be preferable to keep the logarithmic term, it introduces an
  extra unknown constant which we choose not to worry about in the present
  work.}. One can then show that, up to 2.5PN order,
\be \psi_{l\,,m}=-m^2\Omega^2(h_+-ih_{\times})_{l\,,m}\,, \ee
the only exception to this rule being the 2.5PN contribution to the amplitude
of the $l=m=2$ component.

In our calculation of the amplitudes we discard terms of order ${\cal
  O}(M\Omega)^{14/3}$ (i.e. we compute all terms in a 2.5PN expansion of the
gravitational wave amplitude, as given in \cite{Arun:2004ff}). We only list
the positive-$m$ components of the dominant multipoles, since negative-$m$
components are obtained by the symmetry property
\be \psi_{l\,,-m}=(-1)^l \psi_{l\,,m}^*\,. \ee
The small mass ratio limit of these results was obtained by Poisson
and by Tagoshi and Sasaki \cite{Poisson:1993vp}.
For comparable mass ratios, we find that the amplitudes of the dominant
components are:

\begin{subequations}
\label{dominant-psis}
\beq Mr\, \psi_{2\,, 2}e^{i\tilde{\phi}}&=& 
32\sqrt{\frac{\pi}{5}}\eta (M\Omega)^{8/3}
\left[1+\f{55\eta-107}{42}(M\Omega)^{2/3}+2\pi
(M\Omega)-\frac{2173+7483\eta-2047\eta^2}{1512}(M\Omega)^{4/3}\right.\nn\\
&+&\left. \left(\f{-107+34\eta}{21}\pi+\left(\varpi+\f{112}{5}\right)
i\eta\right)(M\Omega)^{5/3} \right]\,,
\label{c22m}\\
Mr\, \psi_{3\,, 3}e^{i\tilde{\phi}}&=&27\sqrt{\frac{6\pi}{7}}\eta \frac{\delta M}{M}(M\Omega)^{3} \left[
1-(4-2\eta)(M\Omega)^{2/3}+\left [3\pi-i\left(\f{21}{5}-6\ln{(3/2)}\right)\right ]M\Omega\right.\nn\\
&+&\left.\left(\frac{123}{110}-\frac{1838}{165}\eta+\frac{887}{330}\eta^2\right)(M\Omega)^{4/3} \right] \,,
\label{c33m}\\
Mr\, \psi_{4\,, 4}e^{i\tilde{\phi}}&=& \frac{1024}{9}\sqrt{\frac{\pi}{7}}\eta (M\Omega)^{10/3}\left\{(1-3\eta)
-\frac{1779-6365\eta+2625\eta^2}{330}(M\Omega)^{2/3}\right.\nn\\
&+&\left. \left[4\pi-i\left(\f{42}{5}-8\ln(2)\right)
-\eta\left(12\pi-i\left(\f{1193}{40}-24\ln(2)\right)\right) \right](M\Omega) \right\}\,,
\label{c44m}\\
Mr\, \psi_{2\,, 1}e^{i\tilde{\phi}}&=&\f{8}{3}\sqrt{\frac{\pi}{5}}\eta \frac{\delta M}{M} (M\Omega)^{3} \left[
1+\f{20\eta-17}{28}(M\Omega)^{2/3}+\frac{2\pi-i(1+\ln 16)}{2}M\Omega\right.
\nn\\
&+&\left. \left(-\frac{43}{126}-\frac{509}{126}\eta+\frac{79}{168}\eta^2\right )(M\Omega)^{4/3} \right]\,,
\label{c21m} \eeq
\end{subequations}
where $\eta$ is the symmetric mass ratio and we defined the phase
$\tilde{\phi}$ as
\be e^{i\tilde{\phi}}\equiv e^{im( \int \Omega dt-2M\Omega\ln{\Omega/\Omega_0})}\,.\ee
The complete expressions of all multipolar components are listed, for
reference, in Appendix \ref{app:multipoles}.  The leading order term in
(\ref{c22m}) is nothing but the quadrupole approximation: see eg.  Eq.~(24) of
BCP. As predictable from symmetry arguments the odd-$m$ multipoles, being
proportional to $\delta M/M$, are suppressed in the equal mass case. To
recover the spin-weighted expansion of the waveform $h_+-ih_{\times}$ one only
has to divide these expressions by $-m^2\Omega^2$. The only exception to this
rule is the $l=m=2$ component: the term proportional to $112/5$ in
(\ref{c22m}) is the lowest-order correction due to the fact that the orbital
angular velocity $\Omega$ is in fact a time dependent quantity.

Terms with $l>2$ and higher-order PN corrections provide a strong consistency
check on both the PN expansion and the numerical results. First, they tell us
if the PN expansion is a good approximation for higher multipolar components
of the radiation ($l>2$).  Secondly, they can be used to check convergence of
the PN expansion for any $(l\,,m)$. If the series is convergent, for example,
going beyond the so-called ``restricted PN approximation'' (i.e., including
higher powers of $(M\Omega)^{1/3}$ in the expansion for $\psi_{2\,,2}$) should
yield better agreement with the amplitude predicted by numerical simulations.

Notice also that multipoles which are {\em formally} of higher PN
order are not necessarily subdominant. For instance, $\psi_{2\,,1}$ is of
order $(M\Omega)^{3}$. Based on power counting, this term should be comparable
to $\psi_{3\,,3}$ and larger than $\psi_{4\,,4}$. However the amplitude of
these terms is proportional to $(8/3)\sqrt{\pi/5}\simeq 2.11$ for
$(l=2,~m=1)$, $27\sqrt{6\pi/7}\simeq 44.31$ for $(l=3,~m=3)$ and
$(1024/9)\sqrt{\pi/7}\simeq 76.22$ for $(l=4,~m=4)$, respectively. At the
maximum orbital frequency we are interested in (the ISCO frequency, which is
of order $M\Omega=M\Omega_{\rm ISCO}\simeq 0.1$), the $(2\,,1)$ amplitude is
much smaller than the $(4\,,4)$ amplitude: $\psi_{2\,,1}\simeq 0.06\, (\delta
M/M)\, \psi_{4\,,4}$. For this reason, in the following we will limit
consideration to terms with $l=m=2,~3,~4$.  Figure \ref{wf2} shows that the
dominance of these terms is quantitatively confirmed by numerical simulations
of the inspiral-merger transition.

\subsection{Estimates of the binary's orbital frequency from numerical simulations}
\label{freq-ests}

In the following, we estimate the orbital frequency $\Omega$ of a
binary at any given time by three different methods, that we list below.

\begin{itemize}
\item[(1)]{\em Orbital frequency from the gravitational wave frequency:
  $\Omega\simeq \omega_{Dm}$}

This estimate of $\Omega$ is based on the observation that the gravitational
wave frequency in a mode characterized by azimuthal number $m$ is $\omega_{\rm
  GW}=m\Omega$. In practice, the calculation can be carried out in two
equivalent ways:

(i) Decompose each mode into a real amplitude and a real phase,
$\psi_{l\,,m}={\cal A}_{l\,,m}\exp(i\phi_{l\,,m})$. Then compute:

\be\label{phidot}
\omega_{Dm}=\f{1}{m}\f{d\phi_{l\,,m}}{dt}\,.
\ee

(ii) Alternatively, observe that if some frequency dominates the Fourier
expansion of a signal, this frequency can be estimated by computing
\be\label{psidot}
\omega_{Dm}=-\f{1}{m}{\rm Im}\left[\f{\dot \psi_{l\,,m}}{\psi_{l\,,m}}\right]\,.
\ee

The latter method was used also in BCP, and it relies on the (implicit)
assumption that the modulus of the complex mode amplitudes $\psi_{l\,,m}$
changes slowly compared with their phase. However, we verified that methods
(i) and (ii) yield results which are basically indistinguishable from each
other.  In the following, when we refer to $\omega_{Dm}$ we always compute
Eq.~(\ref{psidot}) by finite differencing.  In any case, we verified for all
modes that using Eq.~(\ref{phidot}) would not produce appreciable differences.

\item[(2)]{\em Orbital frequency from the coordinate orbital motion of the
  punctures: $\Omega\simeq \omega_c$}

The idea here is to convert each puncture's motion in the $(x\,,y)$ plane into
polar coordinates $(r\,,\phi)$, then compute $\f{d\phi}{dt}$. There are two
problems with this estimate of the orbital frequency. The first is that this
definition obviously depends on the choice of coordinates, and we expect it to
get worse as we get closer to merger. In our particular set of coordinates,
the puncture motion agrees better with other estimates of the orbital
frequency for large mass ratio, when the system becomes more similar to a test
particle moving in Schwarzschild.
The second problem is that, to compare the puncture coordinate frequencies
against the other two, we need to take into account the finite time it takes
for the waves to reach the extraction sphere. We simply estimate this
propagation time to be $\Delta t\simeq r_{\rm ext}$, but since the propagation
speed may differ from unity and waves are not emitted from the origin, this
introduces a (small) additional uncertainty in the comparison.

\item[(3)]{\em Orbital frequency from the Post-Newtonian Quasi-Circular
  approximation: $\Omega\simeq \omega_{\rm PNQC}$}

BCP made the remarkable observation that, even very close to merger, the
$l=|m|=2$ modes of the inspiral waveform can be well approximated by the
standard quadrupole formula for a Newtonian binary in circular orbit. They
computed the leading-order term in Eq.~(\ref{c22m}):
\be\label{NQC}
Mr\, \psi_{2\,, \pm 2}= 32\sqrt{\frac{\pi}{5}}\eta (M\Omega)^{8/3}
e^{\mp 2i(\phi(t)-\phi_0)}\,,
\ee
where $\phi(t)$ is the accumulated phase of the orbit with respect to some
initial phase $\phi_0$. Ignoring logarithmic terms,
\be
\phi(t)=\int_0^t\Omega(t')\,dt'\,.
\ee
BCP pointed out that this simple {\em Newtonian Quasi-Circular} (NQC)
approximation can be used in two ways. First, given the orbital frequency
evolution $\Omega(t)$ we can compute (an approximation to) the wave amplitude.
Conversely, given the modulus of the wave amplitude, we can estimate the
orbital frequency $\Omega=\omega_{\rm NQC}$ by inverting the modulus of
Eq.~(\ref{NQC}), and check whether $\omega_{\rm NQC}$ agrees with the
estimates $\omega_{Dm}$ (computed from the gravitational wave frequency) and
$\omega_c$ (computed from the punctures' coordinate motion).

We will show below that using Eqs.~(\ref{c22m})--(\ref{c21m}) and, more
generally, the expressions listed in Appendix \ref{app:multipoles}, their
observation can be extended to all multipolar components of the radiation. Our
approximation improves on the simple NQC estimate in that we include all PN
terms in the expansion up to 2.5PN, but it still assumes that the orbits are
quasi-circular. For this reason, we call it a {\em Post-Newtonian
  Quasi-Circular} estimate of the frequency.

\end{itemize}

\begin{figure*}[ht]
\begin{center}
\begin{tabular}{cc}
\epsfig{file=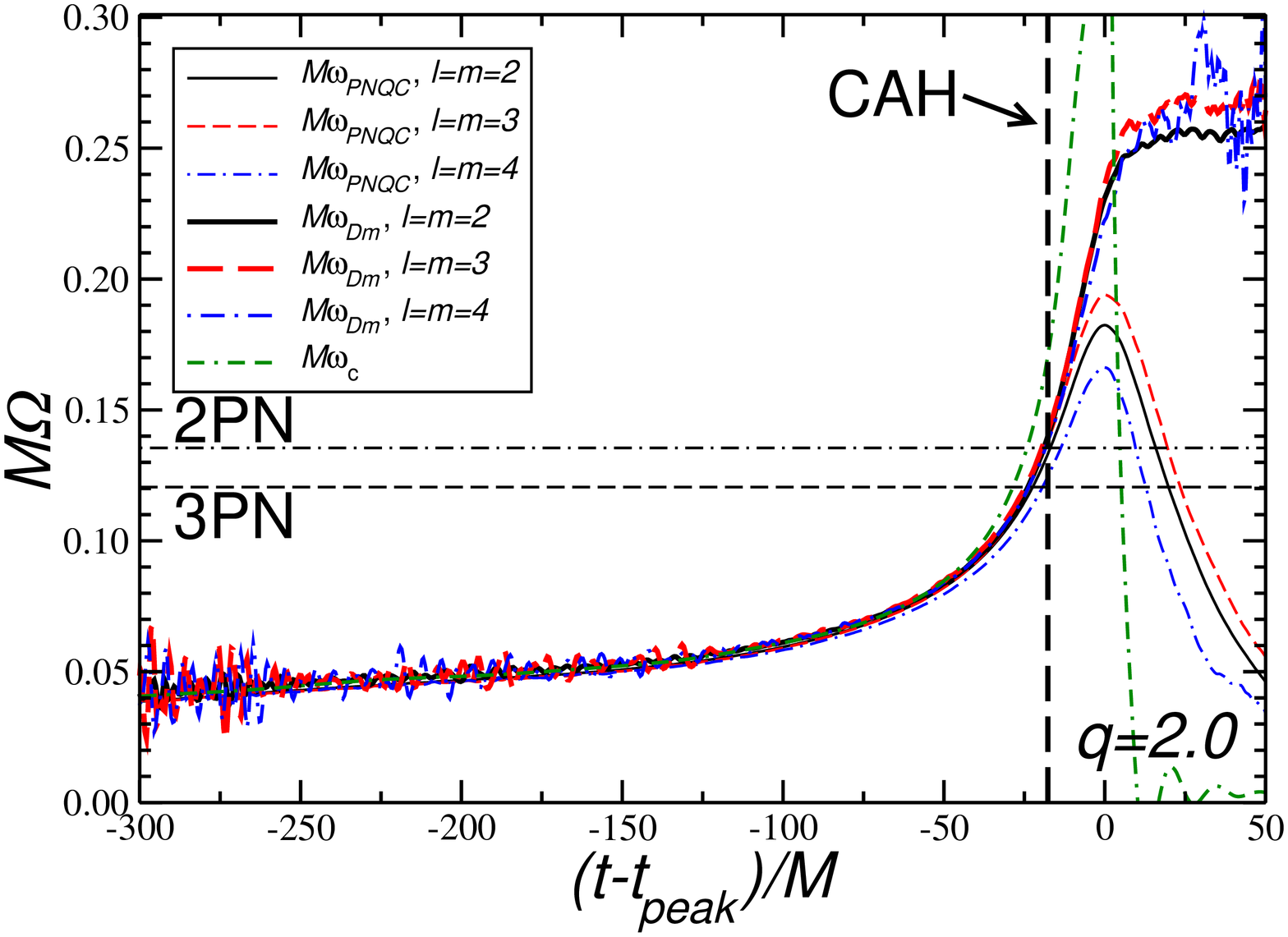,width=7cm,angle=-90} &
\epsfig{file=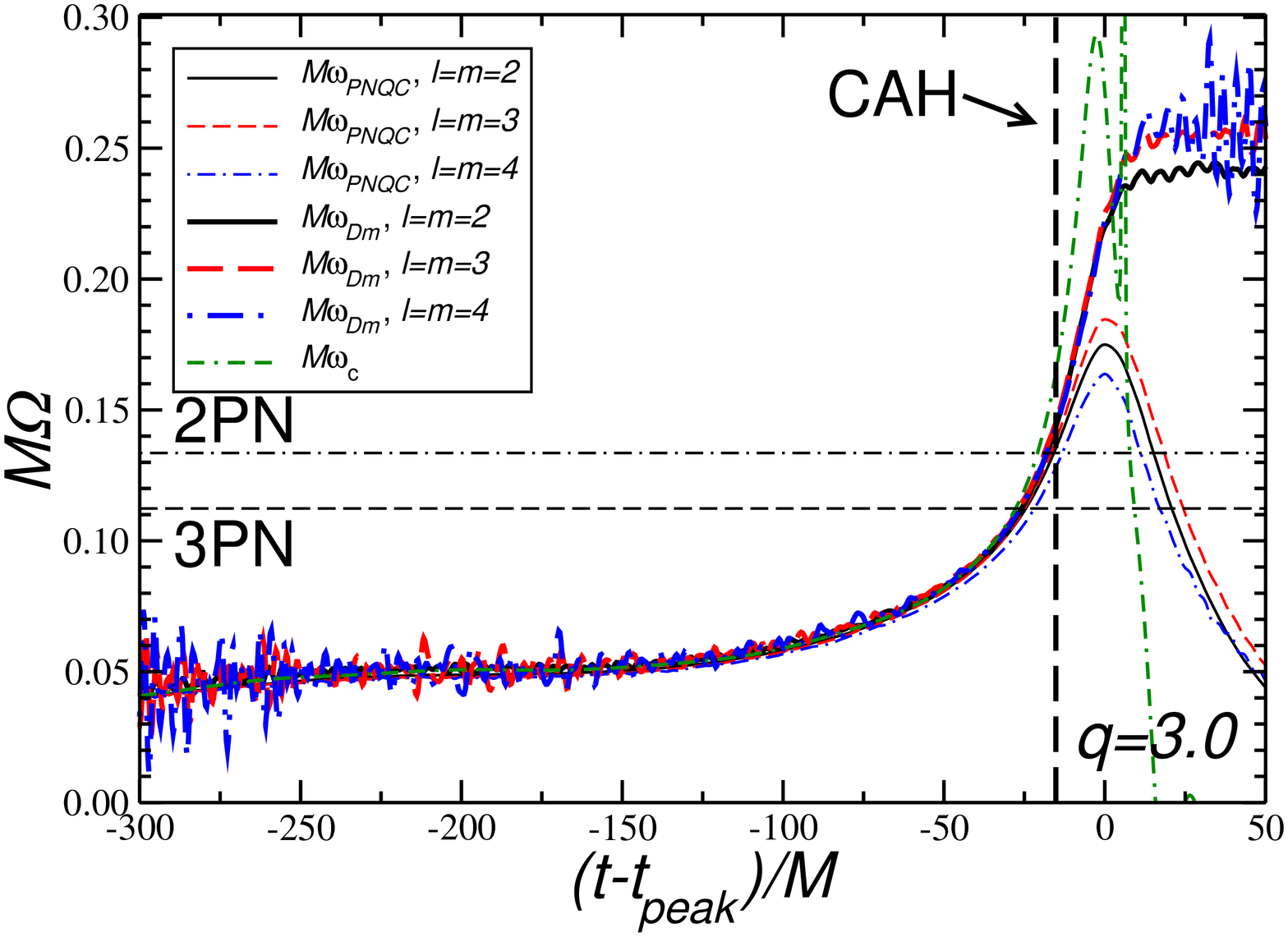,width=7cm,angle=-90} \\
\end{tabular}
\caption{Orbital frequencies from the puncture motion ($\Omega=\omega_c$),
  from the gravitational wave frequency ($\Omega=\omega_{Dm}$) and from the
  PNQC approximation ($\Omega=\omega_{\rm PNQC}$).  Here $\omega_{\rm PNQC}$
  is computed by inverting Eqs.~(\ref{c22m}), (\ref{c33m}) and (\ref{c44m}),
  and keeping only the leading order in the PN expansions on the right hand
  side. Times are measured starting from $t_{\rm peak}$, the peak of the
  $l=m=2$ mode amplitude for the given mass ratio (see Table
  \ref{tab:pars}). Horizontal lines mark 2PN and 3PN estimates of the ISCO
  frequency (as listed in Tables \ref{tab:ISCO2pn} and \ref{tab:ISCO3pn}); the
  vertical dashed line marks (an estimate of) the CAH formation. The plots
  refer to runs D8 and two different mass ratios ($q=2.0$ on the left, and
  $q=3.0$ on the right).
  \label{omegas}}
\end{center}
\end{figure*}

The three different estimates are shown in Fig.~\ref{omegas}, to be compared
with Fig.~7 in BCP. For this plot we choose runs D8 (since the inspiral part
lasts longer) and we compare two values of the mass ratio: $q=2.0$ and
$q=3.0$. After the final black hole's formation (roughly, for $t>t_{\rm CAH}$)
the system is no longer a binary, and therefore different estimates of the
orbital frequency disagree with each other. The most physically meaningful
quantity after merger is the gravitational wave frequency $\omega_{Dm}$. This
frequency levels off to a constant, which is roughly proportional to the
fundamental $l=m$ quasinormal frequency of the final black hole (see Section
\ref{fitmerger} for a detailed analysis of the merger-ringdown transition).

The puncture coordinate frequency $\omega_c$ is a reliable estimate at early
times and large separations, whereas $\omega_{Dm}$ is initially noisy, being
contaminated by spurious initial data radiation or noise from boundary
reflections. However, the puncture coordinates provide a bad estimate of the
orbital frequency already $\sim 30M$ before the peak of the radiation. In this
sense, when we are close to merger our coordinates are not as good as the
generalized harmonic coordinates used in BCP. Coordinate choices having such a
big impact, some care is required if we want to attach physical meaning to
quantities like $\omega_c$. For example, BCP use the ``decoupling point''
where $\omega_c$ separates from $\omega_{Dm}$ to mark the beginning of the
merger phase. With our particular choice of coordinates this decoupling point
would occur much earlier, clearly invalidating the estimate. We argue that
comparing $\omega_{Dm}$ and (our best PN guess for) $\omega_{\rm PNQC}$ should
provide a coordinate-independent, more reliable estimate of the decoupling
point.

In Fig.~\ref{omegas} the PNQC frequency $\omega_{\rm PNQC}$ is computed by
inverting Eqs.~(\ref{dominant-psis}) and keeping only the leading order. This
simple leading-order approximation is in excellent agreement with the other
estimates ($\omega_c$ and $\omega_{Dm}$) until $\sim 20M$ before the radiation
peak.  At this point the orbit transitions from inspiral to plunge, and we
cannot expect the PN inspiral calculation to provide the correct orbital
frequency anymore.

We expect the transition from inspiral to plunge to happen, roughly, when the
binary's orbital frequency crosses the ISCO frequency. To estimate the ISCO we
look for extrema of the 2PN and 3PN Taylor expansions of the binding energy
(see Appendix \ref{app:postplunge} for numerical values of $M \Omega_{\rm
  ISCO}$, and Section IIIA of Ref.~\cite{Buonanno:2002ft} for a discussion of
this and alternative methods of estimating the ISCO).  The corresponding
estimates are marked by horizontal lines in Fig.~\ref{omegas}.  Around the
transition region, $\omega_{\rm PNQC}$ (which is computed {\it assuming} that
the motion is a slow, quasi-circular inspiral) should deviate more and more
from $\omega_c$ and $\omega_{Dm}$.

\begin{figure*}[ht]
\begin{center}
\begin{tabular}{cc}
\epsfig{file=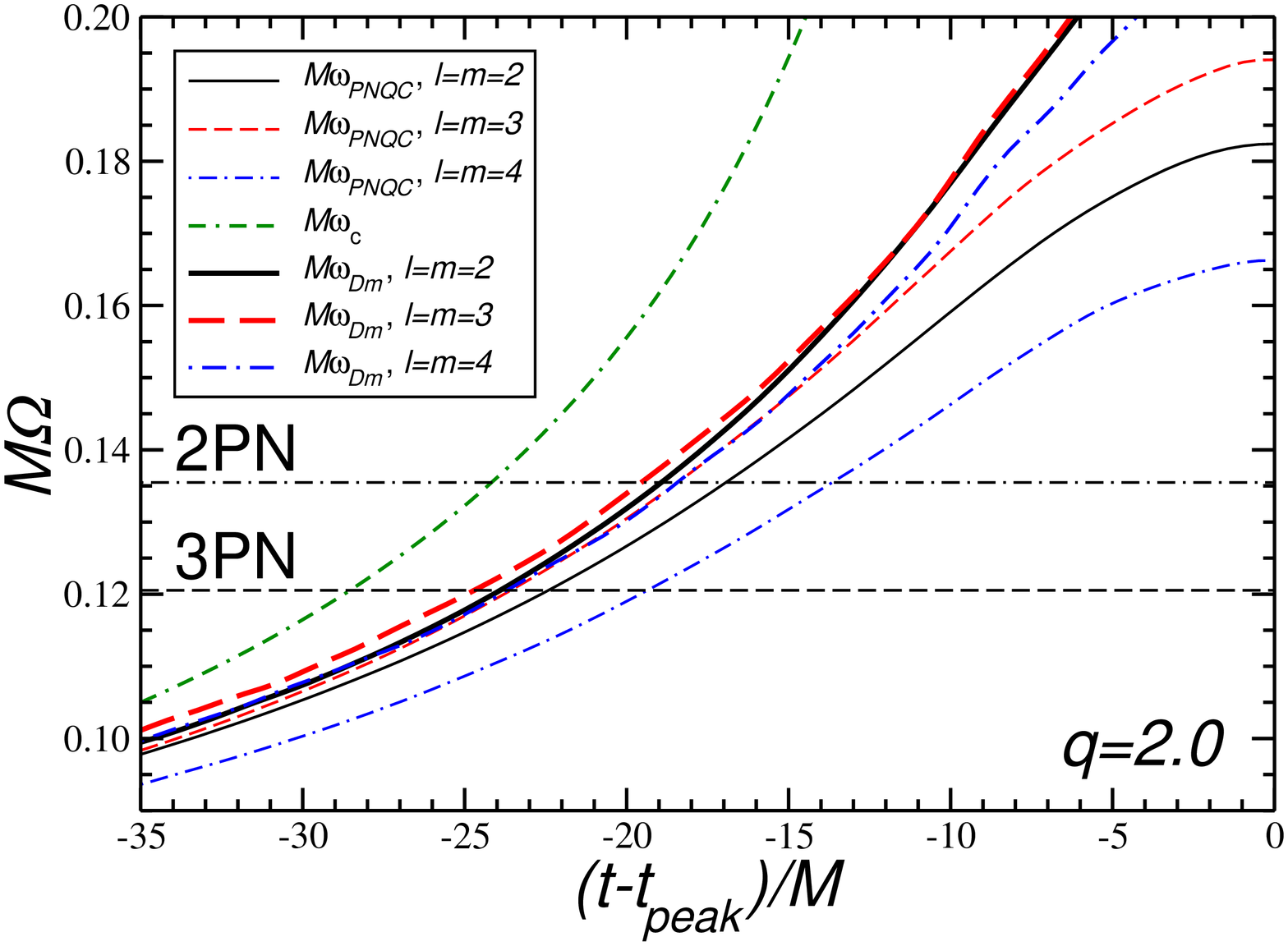,width=7cm,angle=-90} &
\epsfig{file=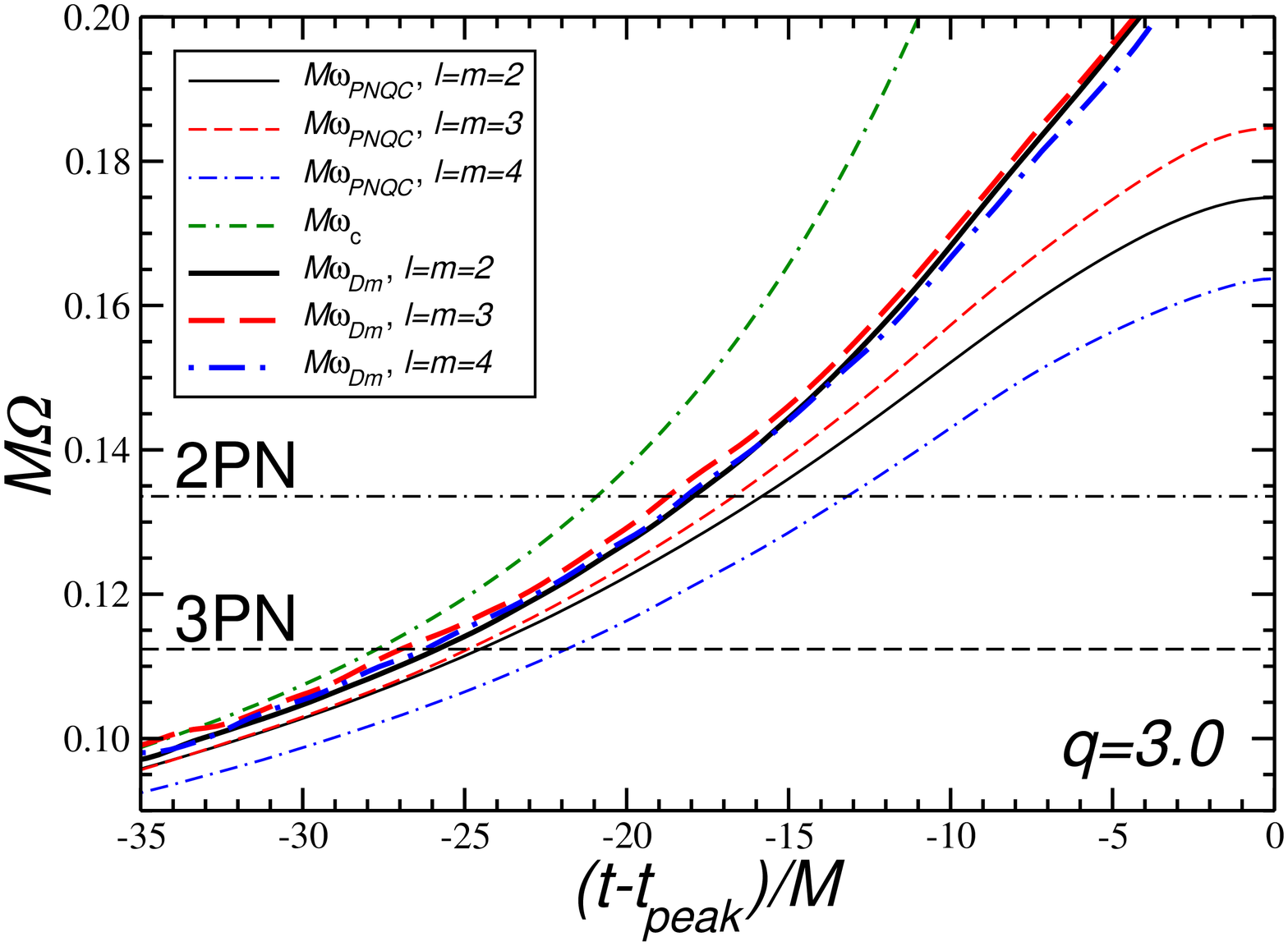,width=7cm,angle=-90} \\
\end{tabular}
\caption{Orbital frequencies from the puncture motion ($\Omega=\omega_c$),
  from the gravitational wave frequency ($\Omega=\omega_{Dm}$) and from the
  PNQC approximation ($\Omega=\omega_{\rm PNQC}$) in the region around the
  ISCO. Here $\omega_{\rm PNQC}$ is computed inverting Eqs.~(\ref{c22m}),
  (\ref{c33m}) and (\ref{c44m}), and keeping only the leading order in the PN
  expansions on the right hand side.
  \label{omegas-ISCO}}
\end{center}
\end{figure*}

This statement is made more quantitative in Fig.~\ref{omegas-ISCO}, where we
zoom in around the ISCO region. Thick lines (the actual gravitational wave
frequencies of the system in a multipole with $l=m$, divided by $m$) and thin
lines (the PNQC estimates) are almost parallel to each other before the ISCO,
and they deviate significantly as the orbit crosses the ISCO. The pre-ISCO
agreement is better, and the post-ISCO deviation larger, for the $q=3.0$
binary than for the $q=2.0$ binary. This is in agreement with the physical
expectation that, as the binary's masses become comparable, the very notion of
an ISCO becomes less and less significant: roughly speaking, the system cannot
be described anymore by the simple-minded picture of a ``small'' particle
orbiting a larger black hole.

An interesting question is if the agreement between the PNQC frequency
$\omega_{\rm PNQC}$ and other estimates of the orbital frequency, namely
$\omega_c$ and $\omega_{Dm}$, improves if we include additional terms in
Eqs.~(\ref{c22m}), (\ref{c33m}) and (\ref{c44m}). In other words, can we use different estimates of
the PNQC orbital frequency $\omega_{\rm {PNQC}}$ to estimate the convergence
rate of the PN approximation? Conversely, if we substitute $\omega_{Dm}$ into
Eqs.~(\ref{dominant-psis}) to compute some PN approximation to the amplitude,
does the agreement with the modulus of the numerical amplitude
$|Mr\,\psi_{l\,,m}|$ get better as we increase the PN order?  We address these
issues below.

\subsubsection{Convergence of the Post-Newtonian quasi-circular approximation}
\label{PNQC-conv}

\begin{figure*}[ht]
\begin{center}
\begin{tabular}{cc}
\epsfig{file=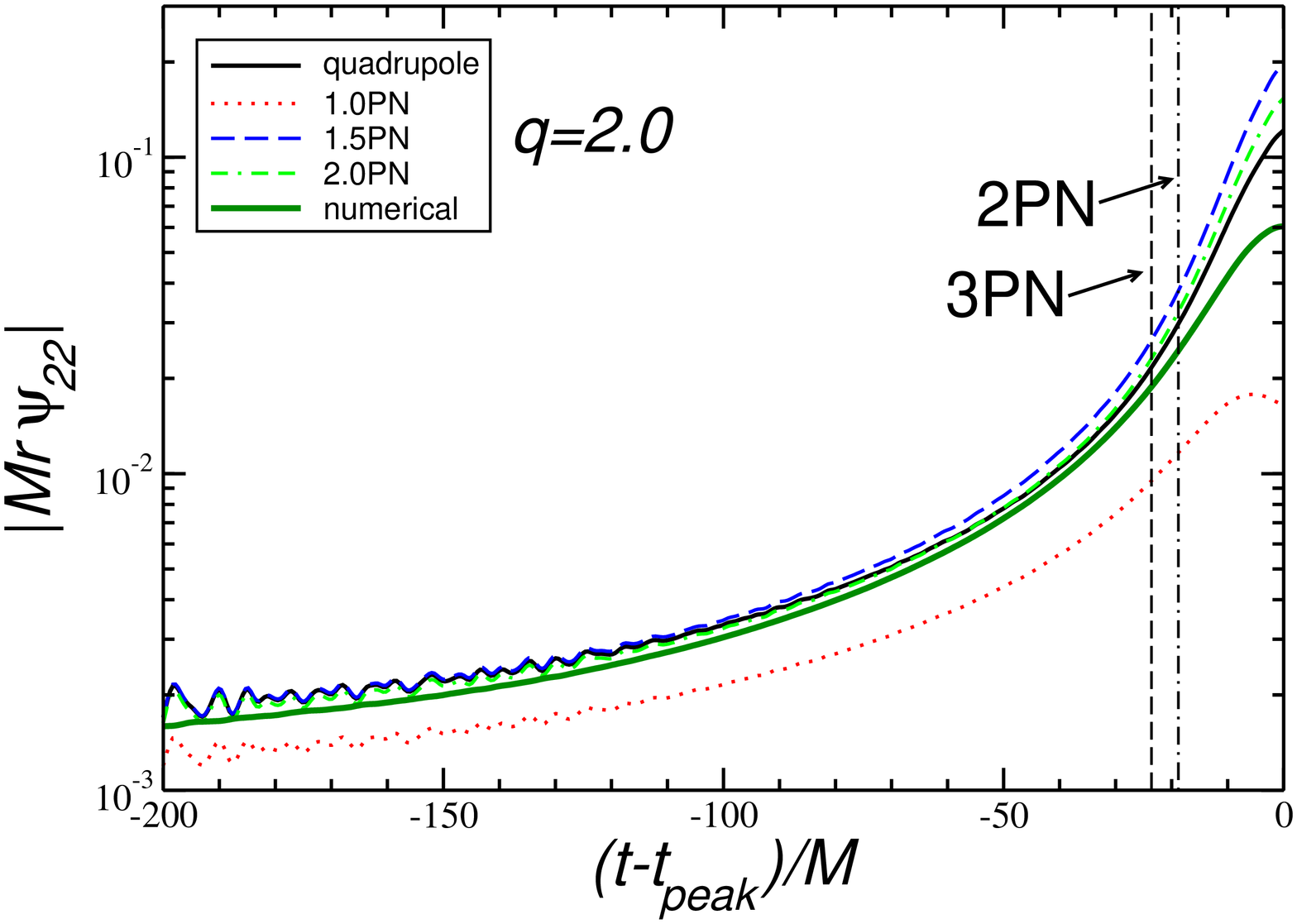,width=7cm,angle=-90} &
\epsfig{file=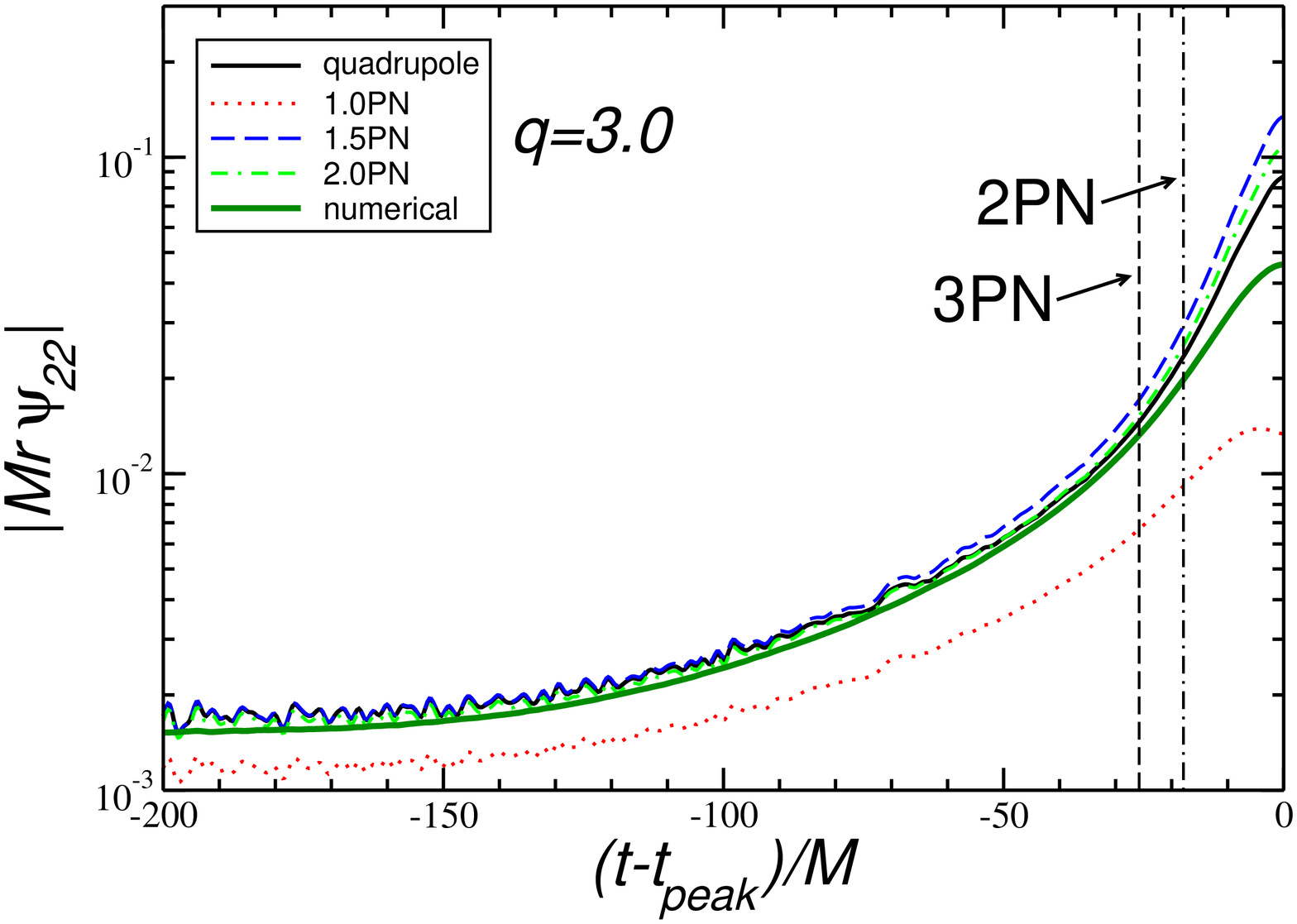,width=7cm,angle=-90} \\
\epsfig{file=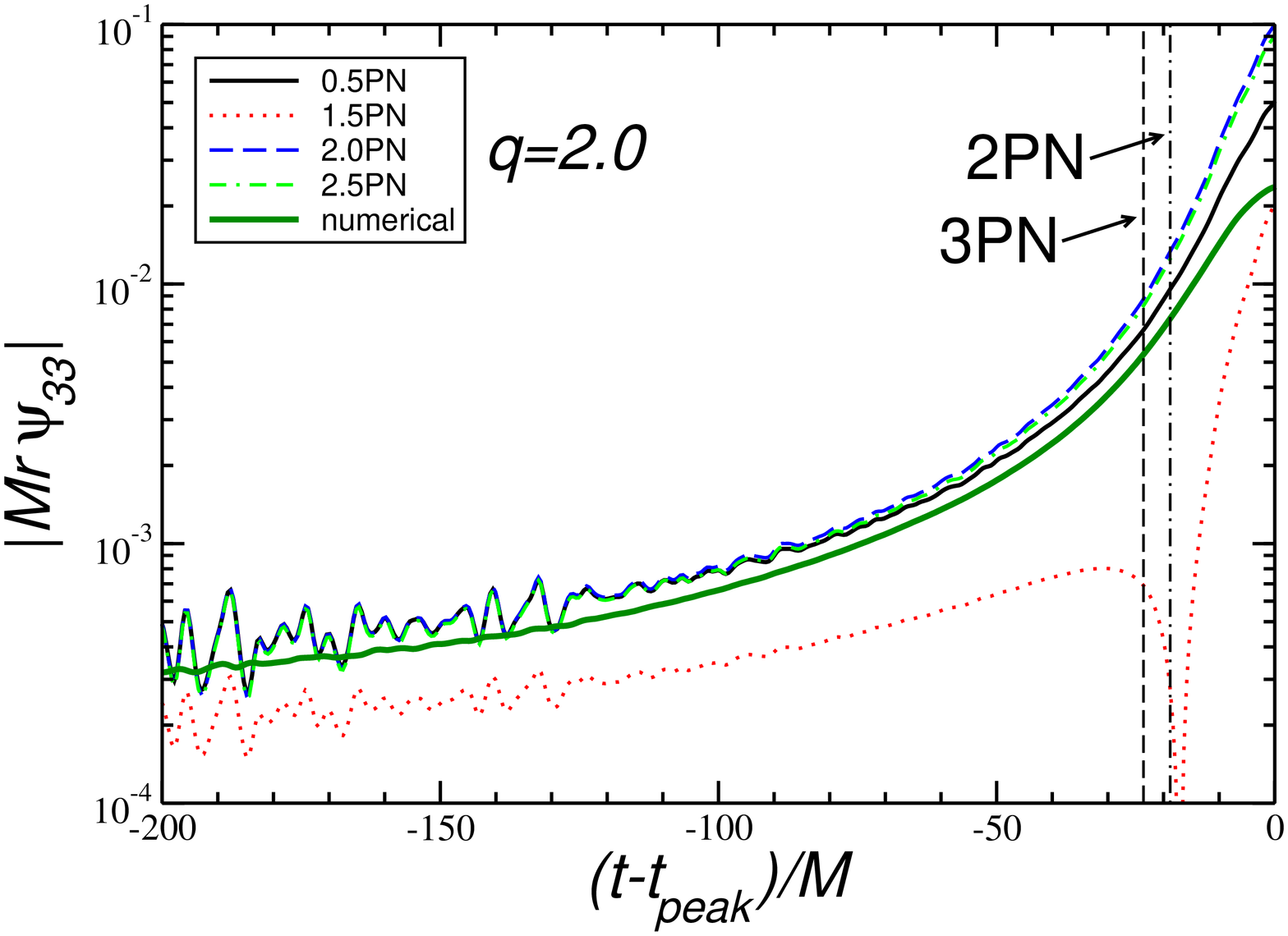,width=7cm,angle=-90} &
\epsfig{file=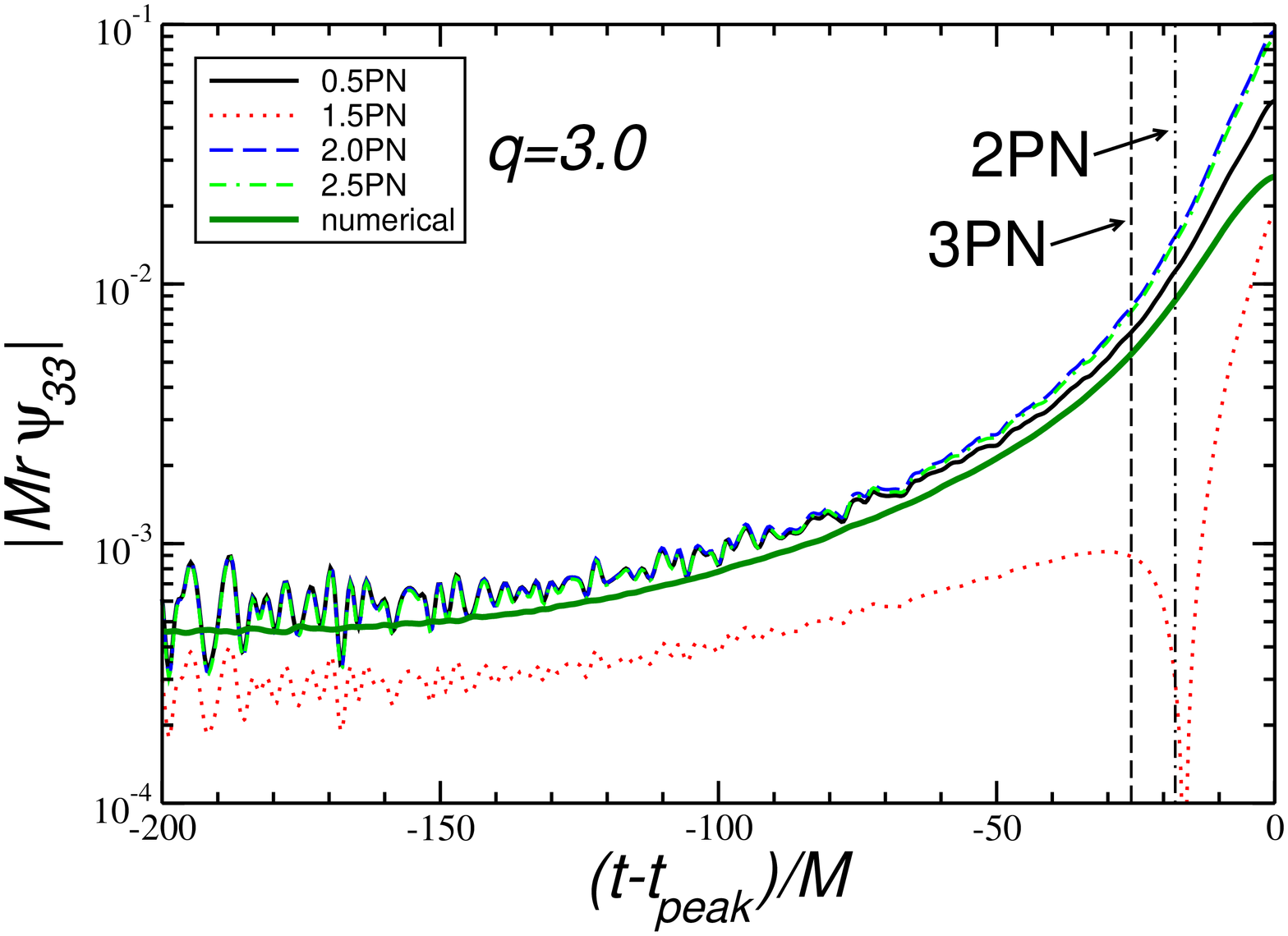,width=7cm,angle=-90} \\
\end{tabular}
\caption{Amplitudes obtained by substituting $\omega_{Dm}$ into the PNQC
  equations are compared with the numerical amplitude. All plots refer to runs
  D8. Figures on the left refer to $q=2.0$, those on the right to $q=3.0$. The
  top row shows amplitudes for $l=m=2$, the bottom row for $l=m=3$. Vertical
  lines mark 2PN and 3PN estimates of the ISCO.
  \label{amplPNQC}}
\end{center}
\end{figure*}

In Fig.~\ref{amplPNQC} we show a simple ``visual'' convergence test of the PN
approximation. We substitute $\omega_{Dm}$ into Eqs.~(\ref{c22m}) and
(\ref{c33m}) and we compute successive PN approximations to the wave amplitude
as functions of time; then we compare the results with the modulus of the
actual numerical amplitude $|Mr\,\psi_{l\,,m}|$.

At early times we clearly see oscillations in the PN estimates of the
amplitude, that damp away as the binary evolves.  These oscillations are due
to $\omega_{D2}$ being very noisy near the beginning of the simulation
(compare the early portion of Fig.~\ref{omegas}), and they would not be
present if we used as a reference $\omega_c$, which is much smoother at early
times\footnote{The reason why we do {\it not} choose $\omega_c$ as a baseline
  for comparison is that this frequency significantly deviates from the others
  close to merger, where the comparison between PN theory and numerical
  simulations is most interesting.}.  The first PN correction is seen to
deviate significantly from all other PN approximations. This is a general
feature of PN expansions. The poor convergence properties of the PN
approximation have long been known in the point-particle limit
\cite{Poisson:1995vs}, where exact results can be obtained by simply
integrating the Zerilli equation.  Fortunately, in our case higher-order PN
expansions (of order higher than 1PN for $l=m=2$) are reasonably consistent
with each other.

A comparison of the PNQC orbital frequencies computed at different PN orders
is also instructive. Let us assume that, within the accuracy of our numerical
simulations, $\omega_{Dm}$ is a good representation of the ``true'' orbital
frequency of the binary\footnote{In practice, of course, this is only true in
  an approximate sense. For example, in the early inspiral $\omega_{Dm}$ is
  heavily contaminated by the initial data burst and boundary reflection
  noise; finite differencing effects will introduce errors in the calculation
  of $\omega_{Dm}$ given the computed waveform; and finally, the waveform
  itself is obtained at finite extraction radius, introducing additional
  uncertainties. What we are really assuming is that, taken together, all of
  these effects are smaller than the errors introduced by the PN approximation
  to general relativity.}. If by increasing the PN order we find that
$\omega_{\rm PNQC}$ gets closer and closer to $\omega_{Dm}$, this would
provide an indication that the PN expansion is converging to the actual
solution of the full, nonlinear problem.

\begin{figure*}[ht]
\begin{center}
\begin{tabular}{cc}
\epsfig{file=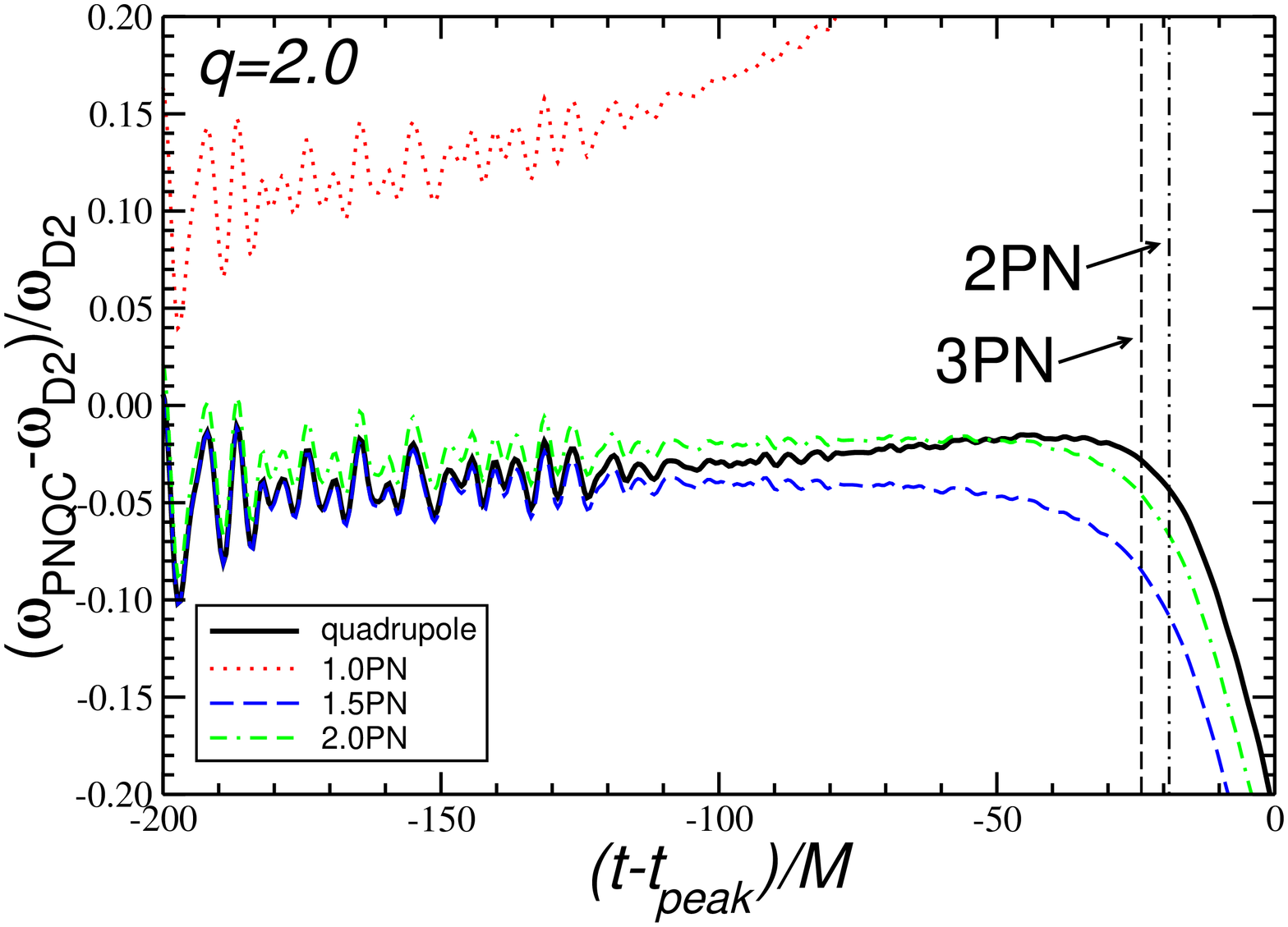,width=7cm,angle=-90} &
\epsfig{file=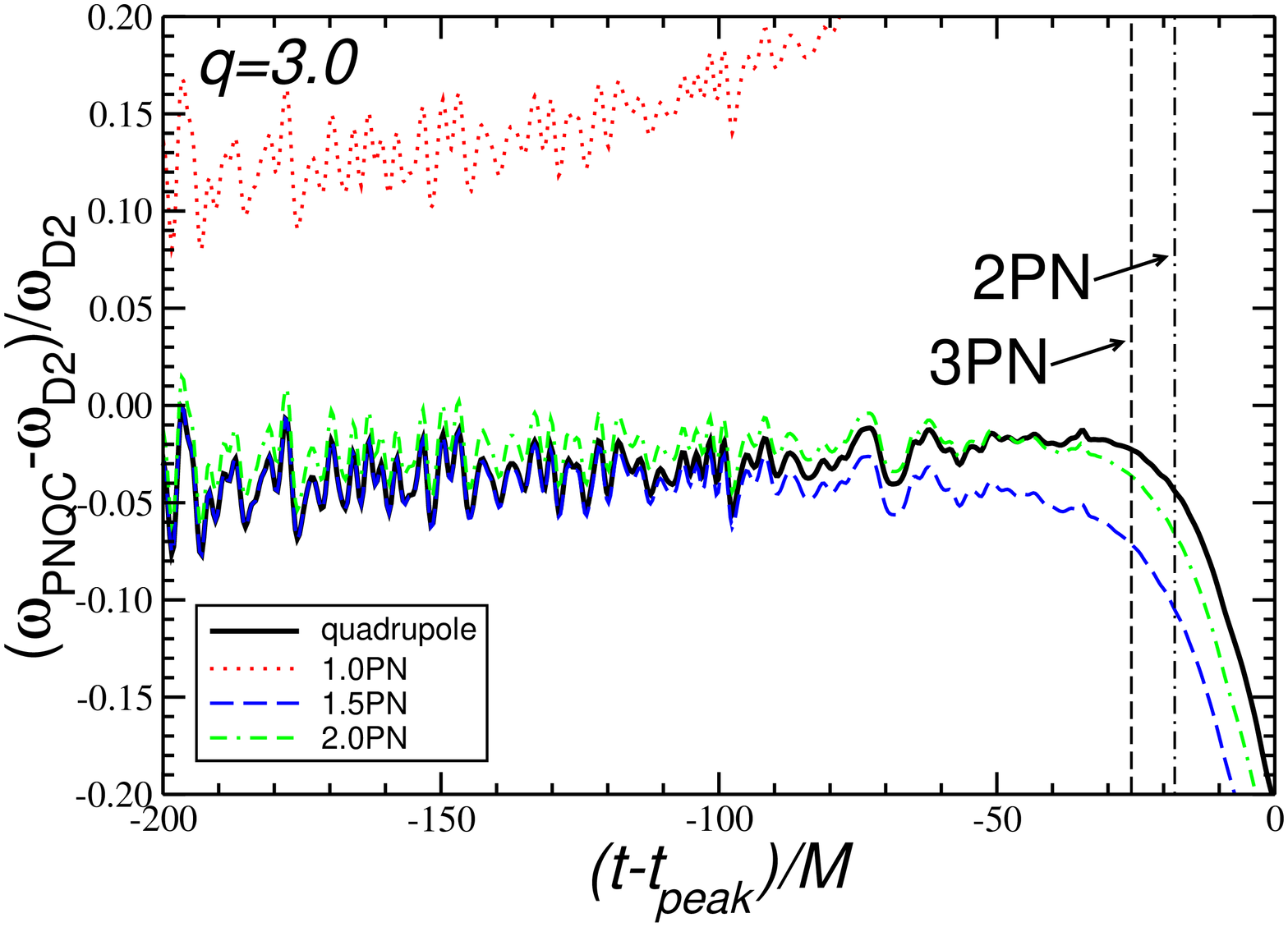,width=7cm,angle=-90} \\
\end{tabular}
\caption{Relative deviation between the orbital frequency obtained from the
  PNQC inspiral formulae and the ``true'' frequency of the signal
  $\omega_{D2}$. Plots refer to runs D8 with $q=2.0$ (left), $q=3.0$
  (right).\label{domegas-ISCO}}
\end{center}
\end{figure*}

Fig.~\ref{domegas-ISCO} shows the relative deviation between $\omega_{\rm
  PNQC}$, as computed by keeping more and more terms in
Eqs.~(\ref{dominant-psis}), and the supposedly more accurate orbital frequency
$\omega_{D2}$.  Once again, at early-times we see oscillations in the relative
deviation, that damp away as the binary evolves. The magnitude of the relative
deviation $|(\omega_{\rm PNQC}-\omega_{D2})/\omega_{D2}|$ can be taken as an
indicator of the accuracy of the PN approximation.  These plots confirm, from
a slightly different perspective, the non-monotonic convergence of the PN
series. After the transition from inspiral to plunge (very roughly
corresponding to the vertical lines, marking the estimated location of the
ISCO at 2PN and 3PN) the PNQC frequency, which is only valid for the inspiral
phase, clearly decouples from $\omega_{D2}$, and the relative error becomes
much larger.  Perhaps in the future, as the accuracy of numerical simulations
increases and higher-order PN calculations become available, it will be
possible to use the change in slope of $|(\omega_{\rm
  PNQC}-\omega_{D2})/\omega_{D2}|$ to monitor the occurrence of an orbital
instability (the ``plunge phase'') in full general relativity.

\begin{figure*}[ht]
\begin{center}
\begin{tabular}{ccc}
\epsfig{file=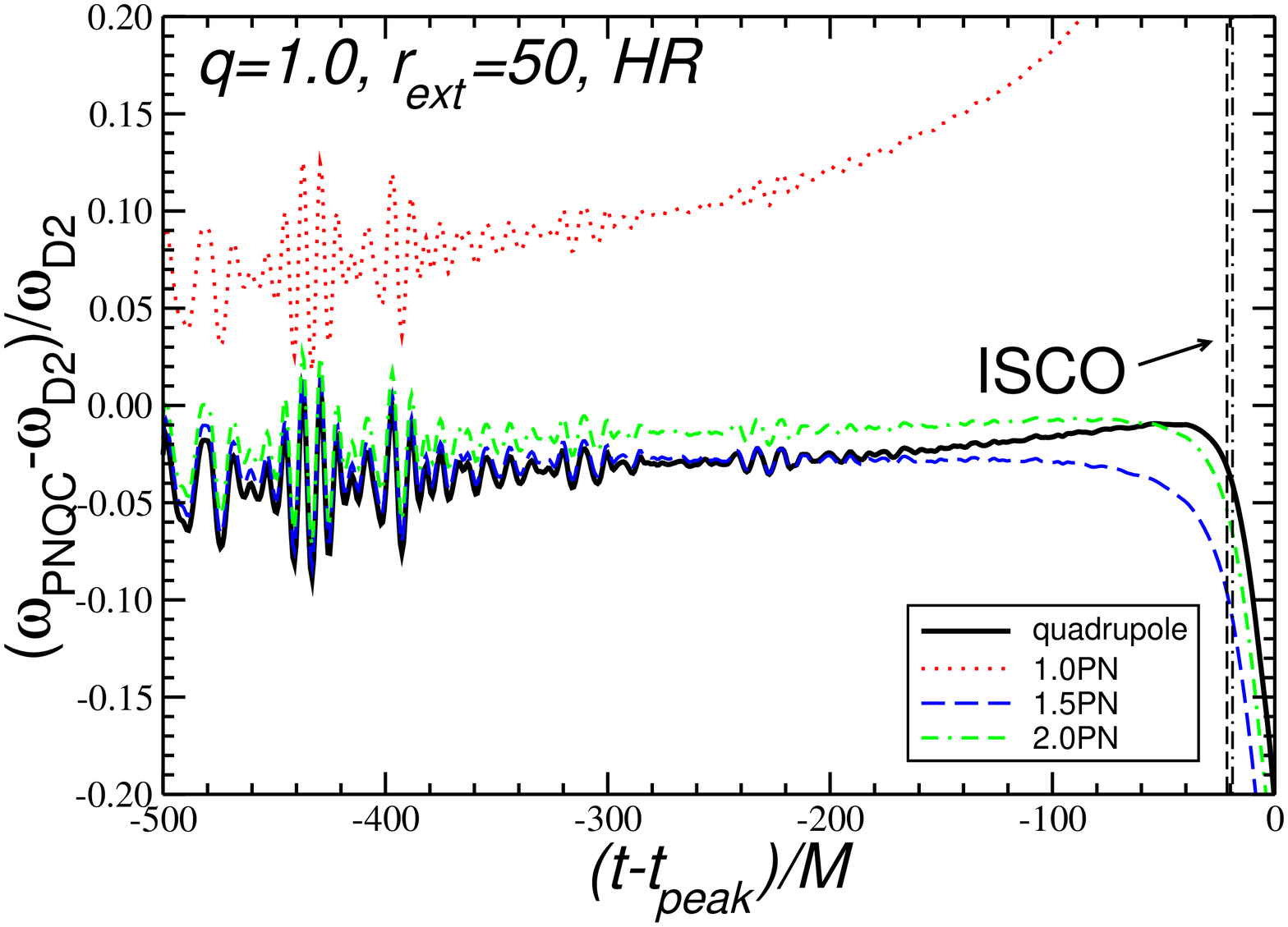,width=4.5cm,angle=-90} &
\epsfig{file=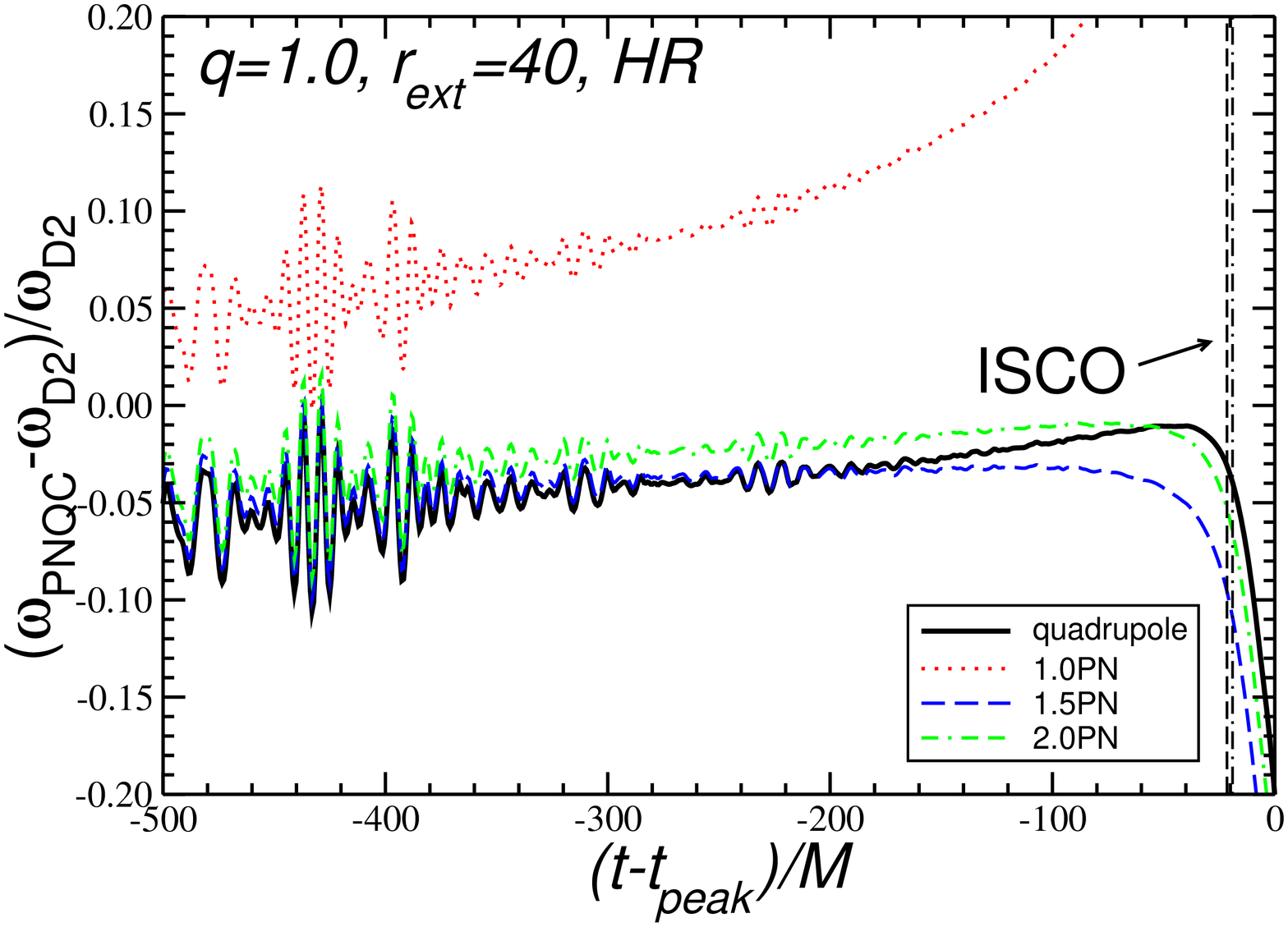,width=4.5cm,angle=-90} &
\epsfig{file=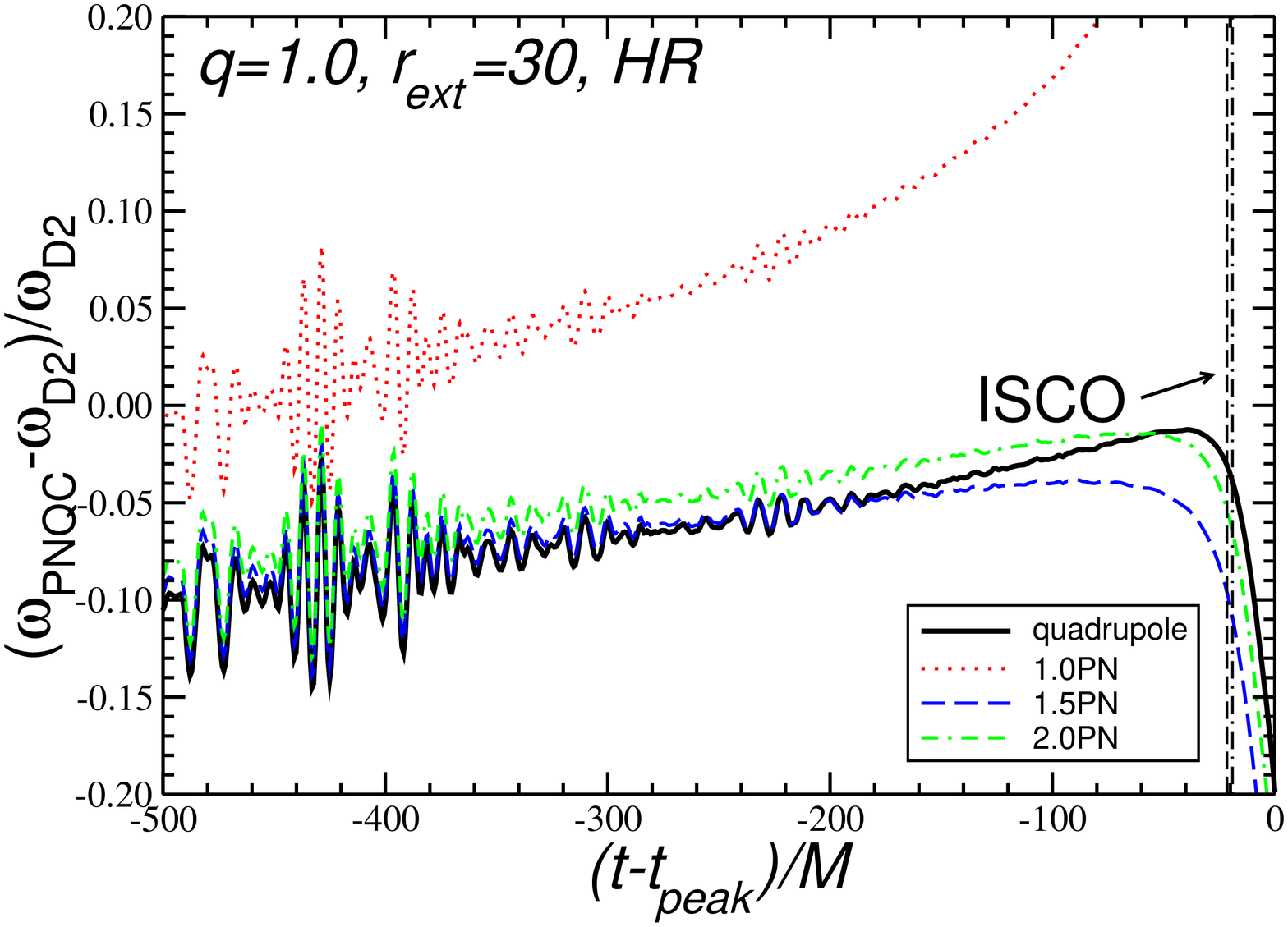,width=4.5cm,angle=-90} \\
\epsfig{file=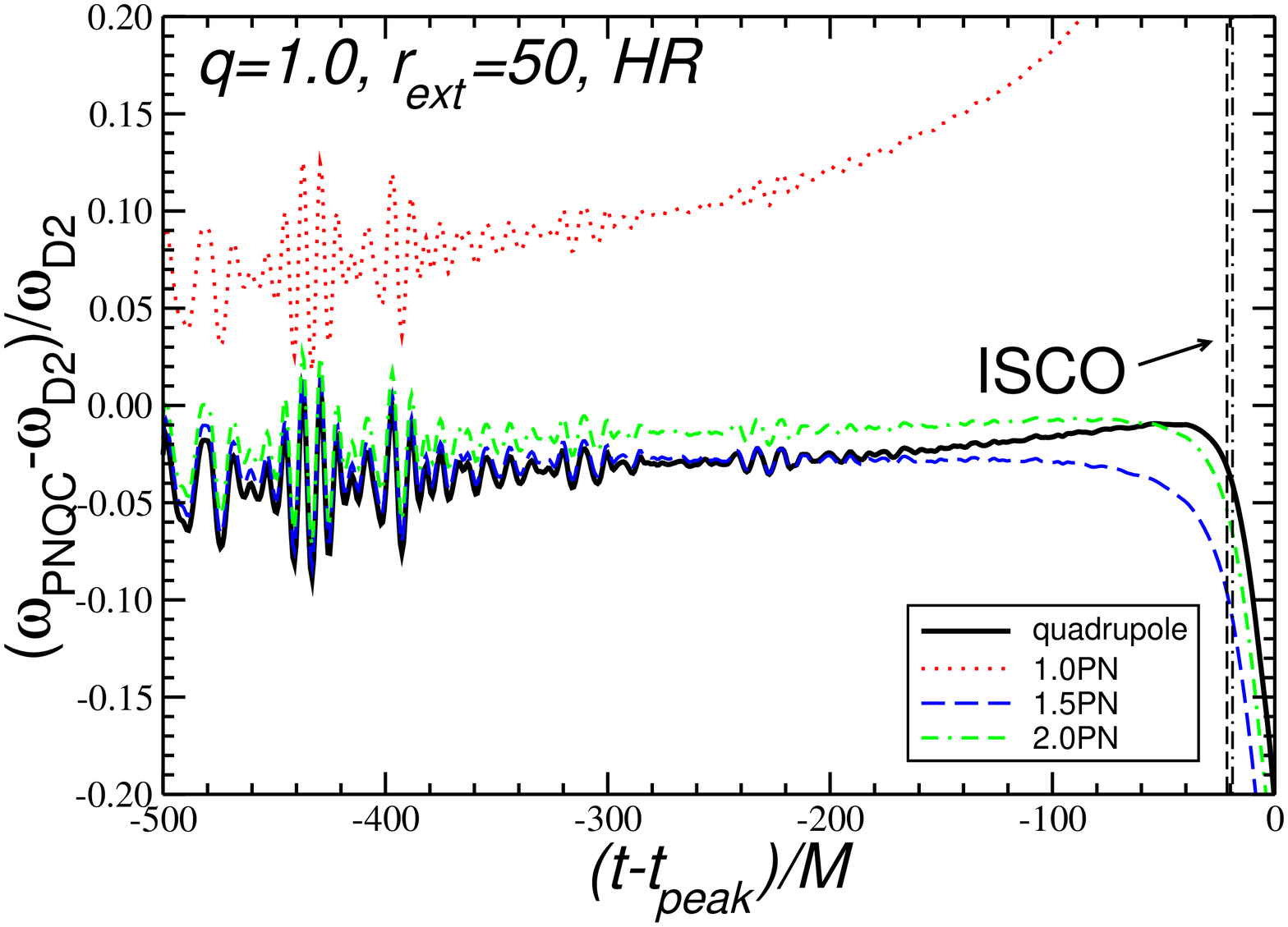,width=4.5cm,angle=-90} &
\epsfig{file=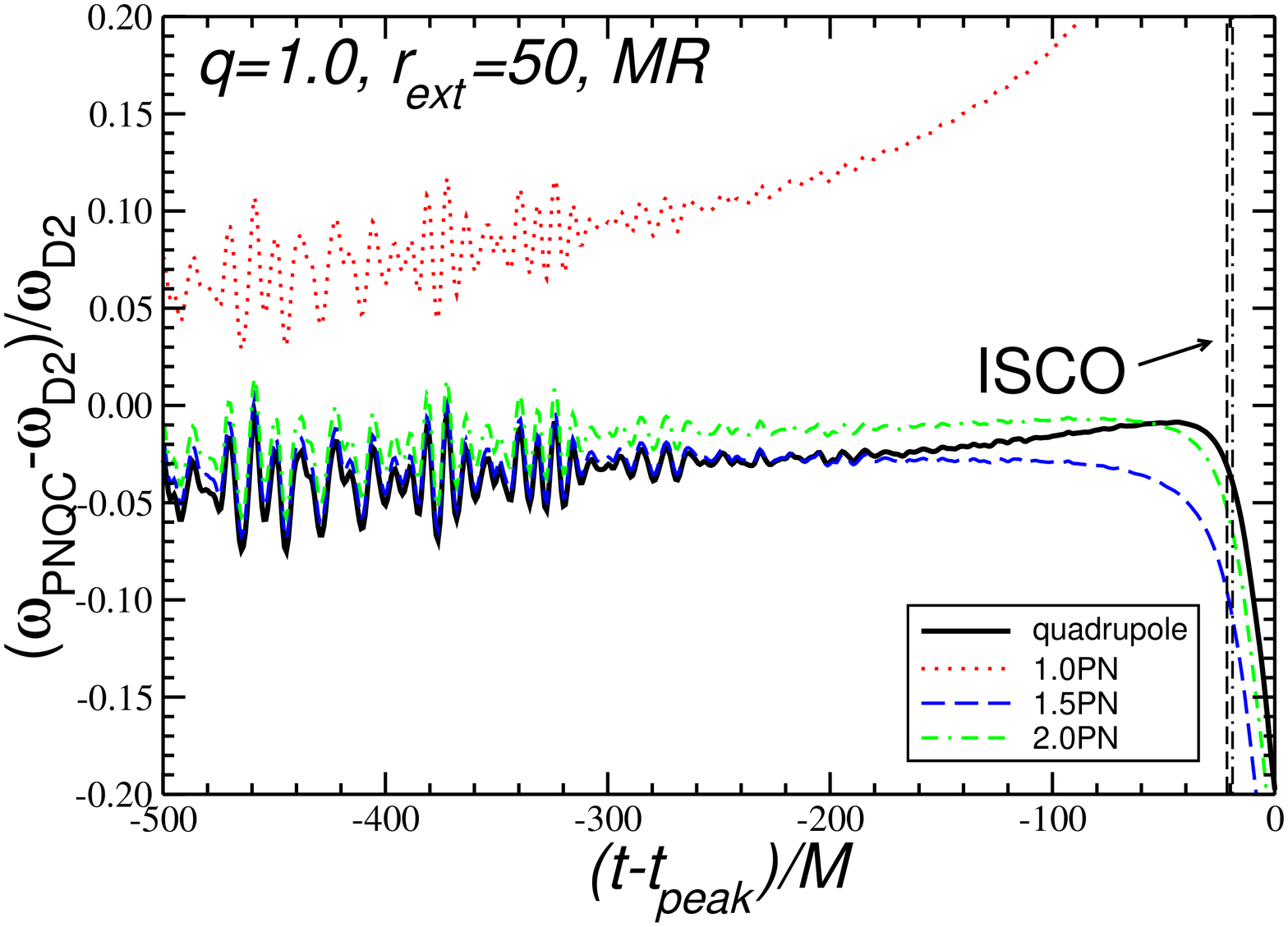,width=4.5cm,angle=-90} &
\epsfig{file=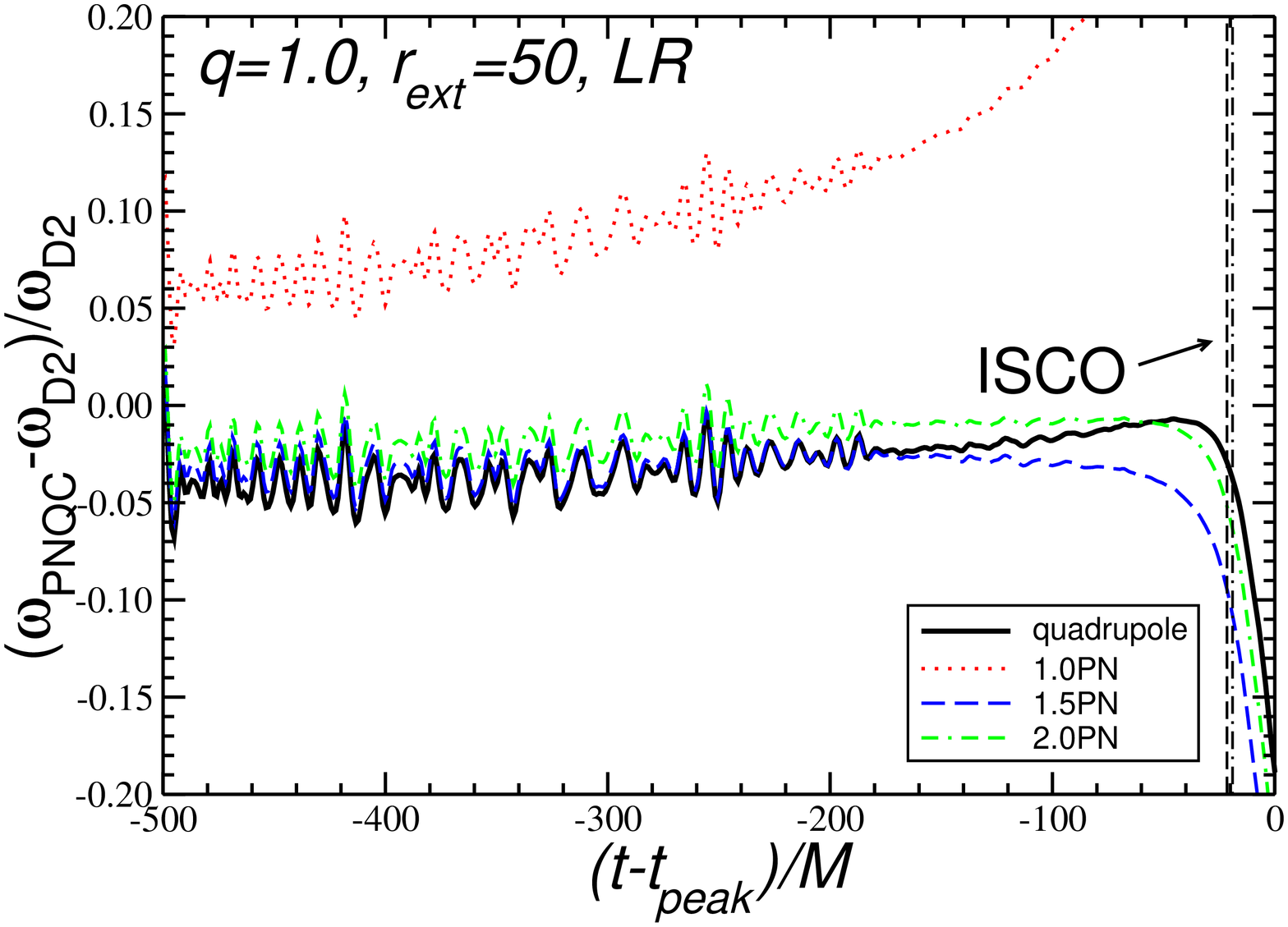,width=4.5cm,angle=-90} \\
\end{tabular}
\caption{The effect of changing extraction radius (top) and resolution
  (bottom) on the errors. These plots refer to mass ratio $q=1.0$, run D10. The
  extraction radii ($r_{\rm ext}=30,~40$ and $50$) and resolutions (HR, MR, LR
  stands for high, medium and low resolution, respectively) are indicated in
  the inset.
  \label{omegas-res-rext}}
\end{center}
\end{figure*}

We already pointed out that our assumed ``exact'' orbital frequency,
$\omega_{D2}$, is in practice affected by various sources of numerical error:
finite-differencing errors, the finite extraction radius and the initial data
burst all introduce uncertainties. To bracket these uncertainties, in
Fig.~\ref{omegas-res-rext} we choose one of our longest runs (D10 for $q=1.0$)
and we study the effect of resolution and extraction radius on $(\omega_{\rm
  PNQC}-\omega_{D2})/\omega_{D2}$. 

Two remarkable features emerge from this plot. First of all, at lower
resolution the ``wiggles'' induced by initial data are still visible at later
times. The second effect is perhaps the most important for matching numerical
waveforms to the PN approximation, and for building template banks for
gravitational wave detection. We see that {\it small extraction radii produce
  a systematic bias (i.e., a larger deviation of $\omega_{\rm PNQC}$ from
  $\omega_{Dm}$) at large orbital separations}.  This effect is easily
understandable: the typical wavelength in the ``early'' inspiral part is of
order $\lambda\sim 100M$, which is actually {\it larger} than the typical
extraction radii used in the present simulations. Such small extraction radii
inevitably produce a bias in the waveform. We observed a similar, and probably
related problem in the context of what we called the memory effect (Section
\ref{memory}).

\subsection{Radiated energy and angular momentum}
\label{energy-j}

\subsubsection{Total radiated energy}

\begin{figure*}[ht]
\begin{center}
\begin{tabular}{cc}
\epsfig{file=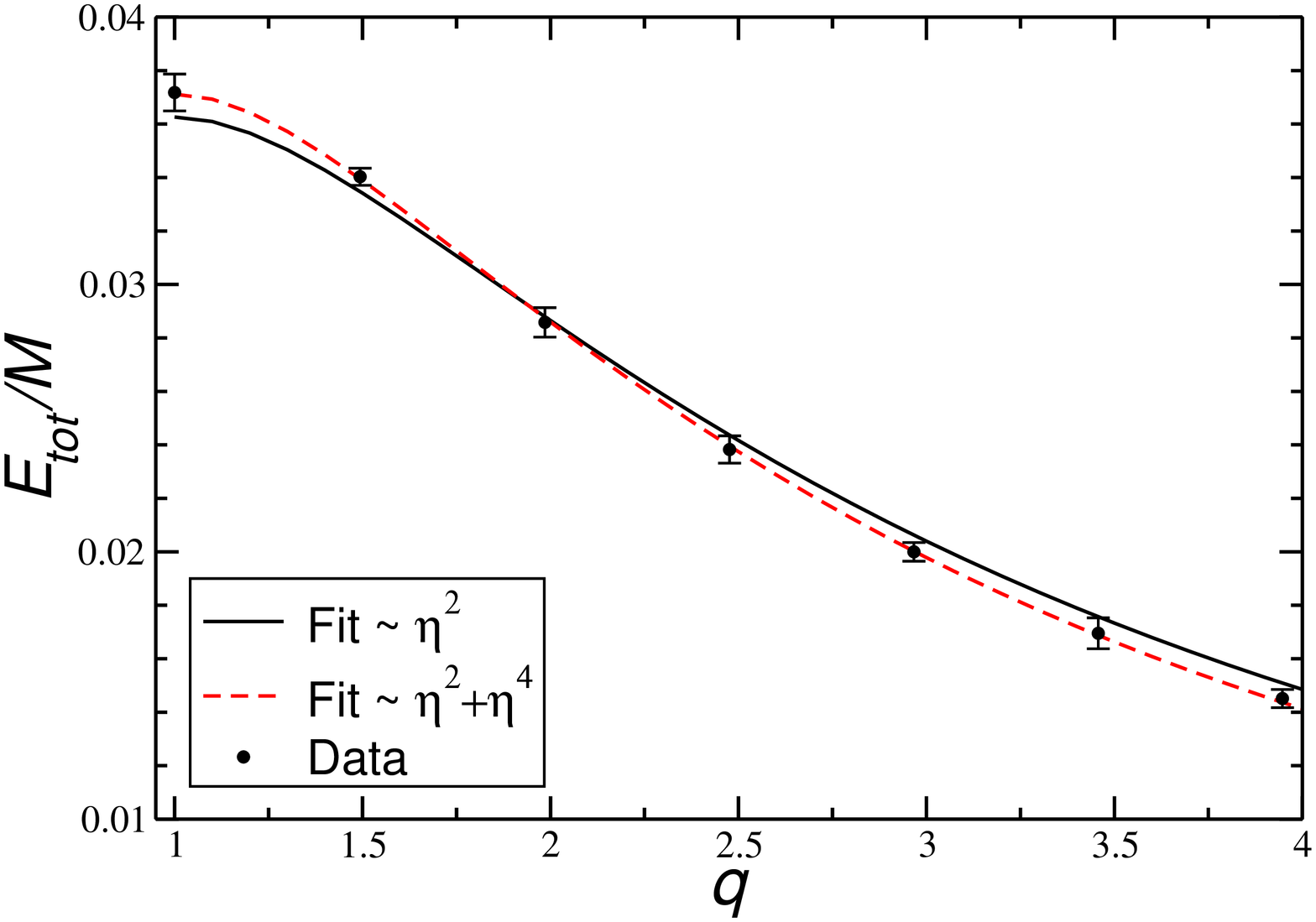,width=7cm,angle=-90} &
\epsfig{file=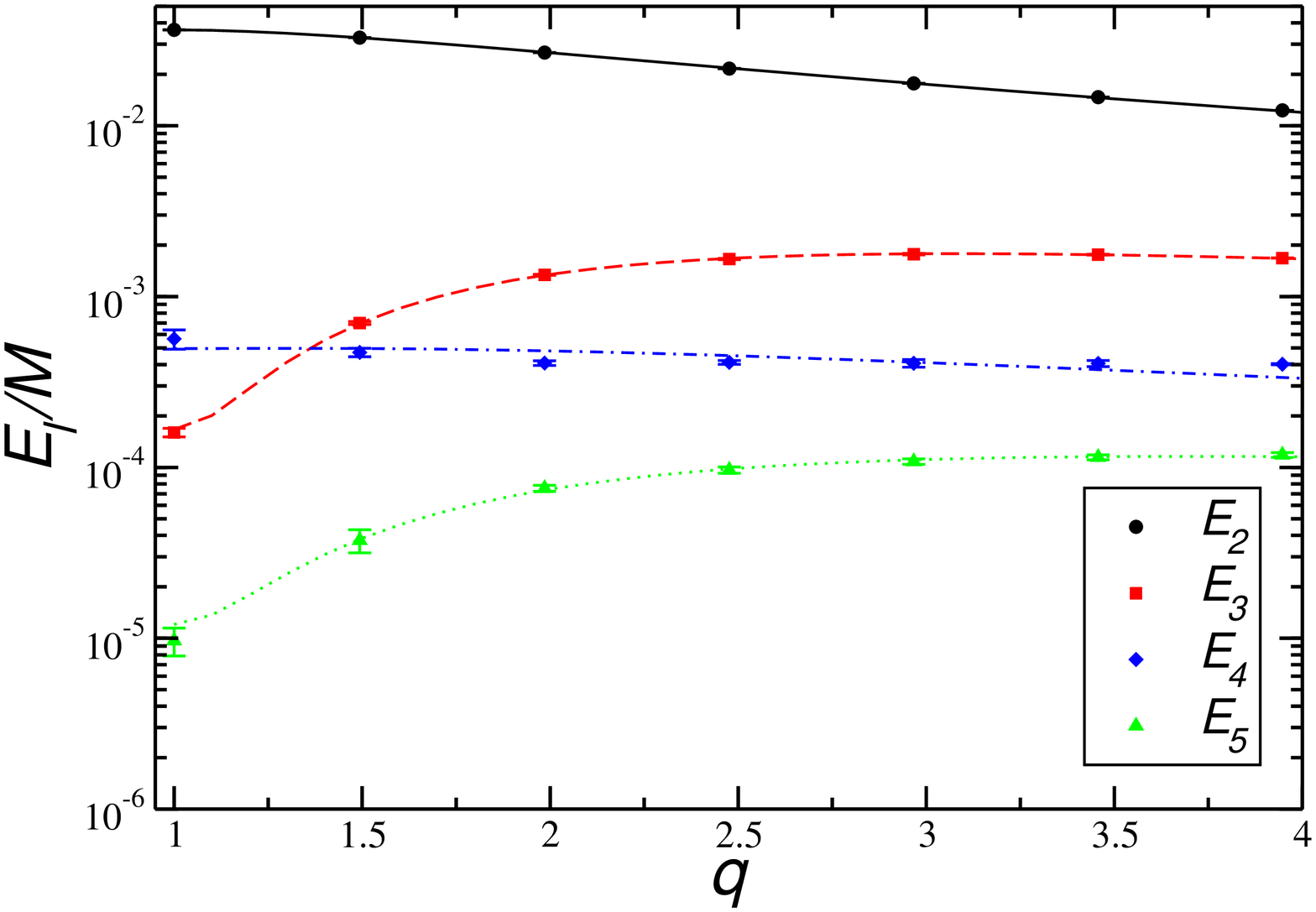,width=7cm,angle=-90} \\
\end{tabular}
\caption{Left: total energy $E_{\rm tot}/M$ radiated in the merger (including
  the initial data burst) fitted by Eq.~(\ref{Etot-lin}) and
  Eq.~(\ref{Etot-quad}), respectively. Right: numerical data for the energy
  $E_l/M$ emitted into each multipole $l$ are compared with the fitting
  functions (\ref{El-fits}). Error bars are estimated by Richardson
  extrapolation. Notice the suppression of odd multipoles as $q\to 1$.
  \label{fig:test-fits}}
\end{center}
\end{figure*}

The total radiated energy computed from wave-extraction methods, and the
energy radiated into each multipole $l$, are listed in Table \ref{tab:El-high}
and plotted in Fig.~\ref{fig:test-fits}.  The error estimates listed in the
table are obtained from Richardson extrapolation as described in
Sec.~\ref{setup} (cf.~also Table \ref{tab:FitResults-HR} below, where we also
list the radiated energies and the final angular momenta computed using QNM
fits).

\begin{table}[ht]
  \centering \caption{\label{tab:El-high} Total energy radiated in merger
    simulations of unequal mass black holes, and percentage of energy in the
    different multipoles (normalized to the total energy radiated in
    $l=2,\dots,5$). The numbers refer to high-resolution D7 runs, and error
    estimates are obtained using Richardson extrapolation. In parentheses we
    list numbers obtained eliminating the initial data burst (in practice, we
    remove all data for $t<t_0=75M$).
  }
\begin{tabular}{cccccc}
\hline \hline
$q$  &$E_{\rm tot}/M (\%)$ &$l=2$ &$l=3$ &$l=4$ &$l=5$ \\
\hline
1.0& $3.718\pm0.069$  & $98.02\pm0.22$  & $0.428\pm0.025$   & $1.521\pm0.197$  & $0.026\pm0.005$ \\
   &($3.651\pm0.065$) &($98.08\pm0.20$) &($0.368\pm0.008$)  &($1.526\pm0.208$) &($0.024\pm0.015$)\\
1.5& $3.403\pm0.032$  & $96.43\pm0.03$  & $2.070\pm0.042$   & $1.384\pm0.079$  & $0.110\pm0.017$ \\
   &($3.340\pm0.025$) &($96.51\pm0.04$) &($2.014\pm0.036$)  &($1.369\pm0.075$) &($0.109\pm0.007$)\\
2.0& $2.858\pm0.055$  & $93.62\pm0.10$  & $4.693\pm0.035$   & $1.426\pm0.047$  & $0.264\pm0.011$ \\
   &($2.802\pm0.055$) &($93.73\pm0.10$) &($4.648\pm0.040$)  &($1.388\pm0.053$) &($0.236\pm0.006$)\\
2.5& $2.383\pm0.051$  & $90.87\pm0.12$  & $6.991\pm0.060$   & $1.730\pm0.046$  & $0.405\pm0.017$ \\
   &($2.334\pm0.053$) &($91.00\pm0.13$) &($6.957\pm0.075$)  &($1.679\pm0.052$) &($0.362\pm0.002$)\\
3.0& $2.000\pm0.035$  & $88.55\pm0.05$  & $8.877\pm0.083$   & $2.036\pm0.103$  & $0.541\pm0.020$ \\
   &($1.958\pm0.036$) &($88.68\pm0.05$) &($8.854\pm0.075$)  &($1.975\pm0.106$) &($0.487\pm0.019$)\\
3.5& $1.695\pm0.058$  & $86.56\pm0.02$  & $10.374\pm0.058$  & $2.393\pm0.101$  & $0.676\pm0.023$ \\
   &($1.659\pm0.059$) &($86.70\pm0.02$) &($10.359\pm0.084$) &($2.326\pm0.096$) &($0.615\pm0.006$)\\
4.0& $1.451\pm0.034$  & $84.84\pm0.02$  & $11.573\pm0.018$  & $2.770\pm0.021$  & $0.815\pm0.027$ \\
   &($1.419\pm0.036$) &($84.99\pm0.04$) &($11.565\pm0.044$) &($2.700\pm0.012$) &($0.746\pm0.012$)\\
\hline \hline
\end{tabular}
\end{table}

Table \ref{tab:El-high} and Fig.~\ref{fig:test-fits} clearly illustrate that
the relative contribution of higher multipoles becomes more relevant as the
mass ratio increases. As expected from symmetry considerations (and from the
calculations in Appendix \ref{app:multipoles}), odd values of $m$ are
suppressed as the mass ratio $q\to 1$.  The dominant components for all mass
ratios are $(l,~m)=(2,~2),~(3,~3),~(4,~4)$. We often observe a non-negligible
contribution (partly due to spurious initial data radiation) also in
$(l,~m)=(2,~1),~(2,~0),~(4,~0),~(5,~5)$.  The initial data radiation burst can
be eliminated by starting the integration of the energy flux after the initial
burst has passed.  In Table \ref{tab:El-high} we decided, somewhat
arbitrarily, to start the integration at $t=75M$. Changes in the starting time
have a marginal impact on the results: at the level of $0.1\%$ for the
($2\,,2$) modes, and of about a few percent for the weakest modes (which have
higher errors anyway).

In practical applications it may be useful to have some fitting formulas for
the total energy radiated and for the contribution of different multipoles.
Since the energy is proportional to $|\dot \Psi_4|^2$, and the $l=m=2$
component $\psi_{2\,,2}\sim \eta$ dominates the radiation, we expect the total
radiated energy to be roughly
proportional to $\eta^2$ (recall that the symmetric mass ratio
$\eta=q/(1+q)^2$ tends to 1/4 in the equal mass limit).  Indeed, it turns out
that the total radiated energy {\em in the merger} $E_{\rm tot}$ is fitted
extremely well (deviations from the data being $\lesssim 4\%$) by the function
\be\label{Etot-lin}
\frac{E_{\rm tot}}{M}=0.036262 \left[\f{4q}{(1+q)^2}\right]^2 \,. 
\ee
Fitting by a higher-order function, eg.
\be\label{Etot-quad}
\frac{E_{\rm tot}}{M}=0.032661 \left[\f{4q}{(1+q)^2}\right]^2 +
0.004458 \left[\f{4q}{(1+q)^2}\right]^4\,, 
\ee
marginally improves the quality of the fit, bringing the agreement with the
data to the level of $\sim 1\%$ (see the left panel of
Fig.~\ref{fig:test-fits}).

The different multipolar components are slightly harder to fit. Since again we
expect the energy in each component to be proportional to the square of the
amplitudes, the even components with $l=m$ should be proportional to
$\left[4q/(1+q)^2\right]^2$, and the odd components should scale with
$\left[q(q-1)/(1+q)^3\right]^2$. After some experimentation with including
higher-order corrections in $\eta$, we found that the following functions
provide a satisfactory fit of the data:

\begin{subequations}
\label{El-fits}
\beq
\frac{E_{l}}{M}&=&
c_1 \left[\f{4q}{(1+q)^2}\right]^2 + c_2 \left[\f{4q}{(1+q)^2}\right]^4\,,
\qquad
(l~\mathrm{even})\,,\\
\frac{E_{l}}{M}&=&
d_1 +
\left[\f{q(q-1)}{(1+q)^3}\right]^2 \left(d_2
+ d_3 \left[\f{4q}{(1+q)^2}\right]^2\right)\,,
\qquad
(l~\mathrm{odd})\,.
\eeq
\end{subequations}

For $l=2$, we find the best-fit coefficients to be $(c_1=0.024231\,,
c_2=0.012163)$; for $l=4$, they are $(c_1=0.0010294\,,c_2=-0.0005328)$. For
$l=3$ we find $(d_1=0.00017\,, d_2=0.10509\,, d_3=0.13990)$, and for $l=5$
$(d_1=0.000012\,,d_2=0.011020\,,d_3=0.000463)$. Numerical data for different
multipolar contributions and the corresponding fitting functions are shown in
the right panel of Fig.~\ref{fig:test-fits}.

\subsubsection{Final angular momentum}

\begin{figure*}[ht]
\begin{center}
  \epsfig{file=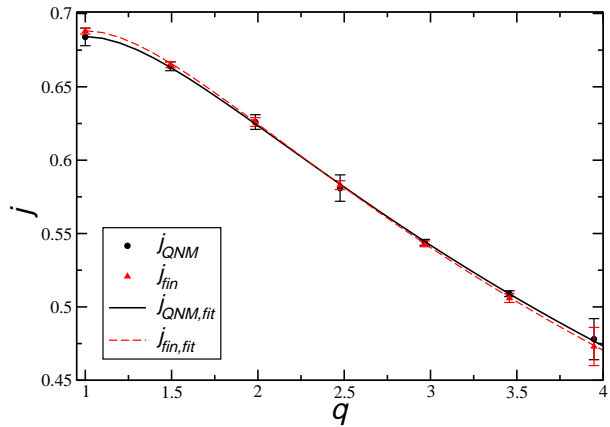,width=7cm,angle=-90} \caption{Angular momentum
    estimated from a QNM fit and from wave extraction, and corresponding fits.
    Error bars are estimated by Richardson extrapolation. \label{j-fit.ps}}
\end{center}
\end{figure*}

A good fit to the final angular momentum, for small mass ratios, was found in
Ref.~\cite{Gonzalez:2006md}:
\be j \sim 0.089+2.4\frac{q}{(1+q)^2}\,. \ee
In this paper we compute the final angular momentum in two ways. One estimate,
that we denote by $j_{\rm fin}$, is obtained by subtracting the total radiated
angular momentum (as computed by wave extraction) from the total angular
momentum at the beginning of the simulation. A second estimate $j_{\rm QNM}$
is based on QNM fits, and will be described in detail in Section
\ref{fitmerger} below. The actual data, together with error estimates based on
Richardson extrapolation, can be found in Table \ref{tab:FitResults-HR}.  We
find our estimates to be accurate within a few percent for $q>3$, and even
more accurate as $q\to 1$.  We found that very good fits (accurate to within
$\sim 1\%$ of the numerical data) are given by
\begin{subequations}
\label{bestfit-qnm}
\beq
j_{\rm QNM} &\simeq&
3.352\frac{q}{(1+q)^2}-2.461\frac{q^2}{(1+q)^4}\,,
\\
j_{\rm fin} &\simeq&
3.272\frac{q}{(1+q)^2}-2.075\frac{q^2}{(1+q)^4}\,.
\eeq
\end{subequations}
The quality of these fits, and the very good agreement between the two
different estimates of the final angular momentum, is illustrated in
Fig.~\ref{j-fit.ps}.

The functional form of Eq.~(\ref{bestfit-qnm}) can be justified by a simple
physical argument.  Consider an extreme-mass ratio binary with the smaller
body orbiting near the ISCO. The orbital angular momentum at the ISCO, in the
small mass ratio limit and for non-spinning bodies, is given by
\be L_{\rm ISCO}=2\sqrt{3}M_1M_2\,. \ee
In Appendix \ref{app:postplunge} we show that the numerical results for the
angular momentum radiated after the ISCO are well fitted by
Eq.~(\ref{JISCOfit}), that we reproduce here:
\be \frac{\Delta J_{\rm ISCO}}{M^2}\approx 2.029 \frac{q^2}{(1+q)^4}\,.\ee
Therefore the angular momentum of the final hole should be well described by
\be j\approx
2\sqrt{3}\frac{q}{(1+q)^2}-2.029\frac{q^2}{(1+q)^4}=3.464\frac{q}{(1+q)^2}-2.029\frac{q^2}{(1+q)^4}\,,\label{iscofinalj}
\ee
which is remarkably close to the best fits (\ref{bestfit-qnm}). 

\subsubsection{Energy and angular momentum fluxes}
\label{sec:flux}

The purpose of this Section is to compare analytical PN estimates of the
energy and angular momentum flux for a quasi-circular, unequal mass inspiral
against the corresponding numerical calculations. To begin with, we summarize
how to compute these quantities in PN theory and in numerical relativity.

The gravitational wave energy flux emitted by a binary moving along an
adiabatic sequence of circular orbits is currently known analytically through
3.5PN order for non-spinning BHs in circular orbits~\cite{Blanchet:2004ek}. It
reads
\begin{eqnarray}
\label{flux}
 F_{\rm E}^{\rm PNQC}  &=& {32\over 5}\eta^2 (M \Omega)^{10/3} \biggl\{ 1 +
\left(-\frac{1247}{336}-\frac{35}{12}\eta \right) (M\,\Omega)^{2/3} + 4\pi
(M\,\Omega)\nonumber \\ &+&
\left(-\frac{44711}{9072}+\frac{9271}{504}\eta+\frac{65}{18} \eta^2\right)
(M\,\Omega)^{4/3}
+\left(-\frac{8191}{672}-\frac{583}{24}\eta\right)\pi
(m\,\Omega)^{5/3}\nonumber \\
&+&\left(\frac{6643739519}{69854400}+\frac{16}{3}\pi^2-\frac{1712}{105}\gamma_E
-\frac{856}{105}\ln (16~\!(M\,\Omega)) \right.\nonumber\\
&+&\left.\left[-\frac{134543}{7776}+\frac{41}{48}\pi^2\right]\eta-\frac{94403}{3024}\eta^2
-\frac{775}{324}\eta^3\right) (M\,\Omega)^2\nonumber\\ &+&
\left(-\frac{16285}{504}+\frac{214745}{1728}\eta
+\frac{193385}{3024}\eta^2\right)\pi (M\,\Omega)^{7/3} \biggr\}\;,
\end{eqnarray}
where $\gamma_E$ is Euler's number and $\eta$ denotes, as usual, the symmetric
mass ratio. The numerical energy flux can be obtained from the mode amplitudes
$Mr\,\psi_{l\,, m}(t)$ as
\begin{equation}
  \label{fluxlm}
  F_{\rm E}=\frac{dE}{dt} =
  \sum_{l=2}^\infty F_{{\rm E}\,,l}=
  \sum_{l=2}^\infty \sum_{m=-l}^l F_{{\rm E}\,,lm}=
  \frac{1}{16\pi}\sum_{l\,, m}{|D_{l\,, m}(t)|^2},
\end{equation}
where $D_{l\,, m}(t)$ is a dimensionless first time integral of $\psi_{l
  m}(t)$ defined by
\begin{equation}
   D_{l\,, m}(t) \equiv \frac{1}{M}\int_0^t{dt^\prime
        Mr\,\psi_{l\,,m}(t^\prime)}\,.
\end{equation}

Each term in the sum (\ref{fluxlm}) represents the multipolar contribution of
a different mode. A PN estimate for the flux in each $(l\,,m)$ mode can be
obtained by using the expansion of $\Psi_4$ in spin-weighted spherical
harmonics. Using the same approximation discussed in Section \ref{sec:pnqc},
namely neglecting the logarithmic term in the phase, we get the folllowing PN
estimate for the $(l\,,m)$ component of the energy flux:
\be\label{Flm}
F_{{\rm E}\,,lm}^{\rm PNQC}=\frac{dE_{l\,,m}}{dt} =
\frac{1}{16\pi m^2 \Omega^2}|Mr\,\psi_{l\,,m}|^2\,,
\ee
where we use the best available PN expansions of $Mr\,\psi_{l\,,m}$, as listed
in Appendix \ref{app:multipoles}. The 2.5PN term in the $l=m=2$ waveform
suffers from the usual ambiguity in $\varpi$ related with the inclusion of
radiation reaction terms \cite{Kidder:2007gz}. For this reason we only
consider $l=m=2$ corrections up to and including 2PN terms.

For quasi-circular orbits, the PN angular momentum flux is simply
\be\label{FJPNQC}
F_{\rm J}^{\rm PNQC}=\f{1}{\Omega}F_{\rm E}^{\rm PNQC}\,.
\ee
The numerical angular momentum flux $F_{\rm J}$ can be computed using
Eq.~(\ref{FJnum}).

\begin{figure*}[ht]
\begin{center}
\begin{tabular}{cc}
\epsfig{file=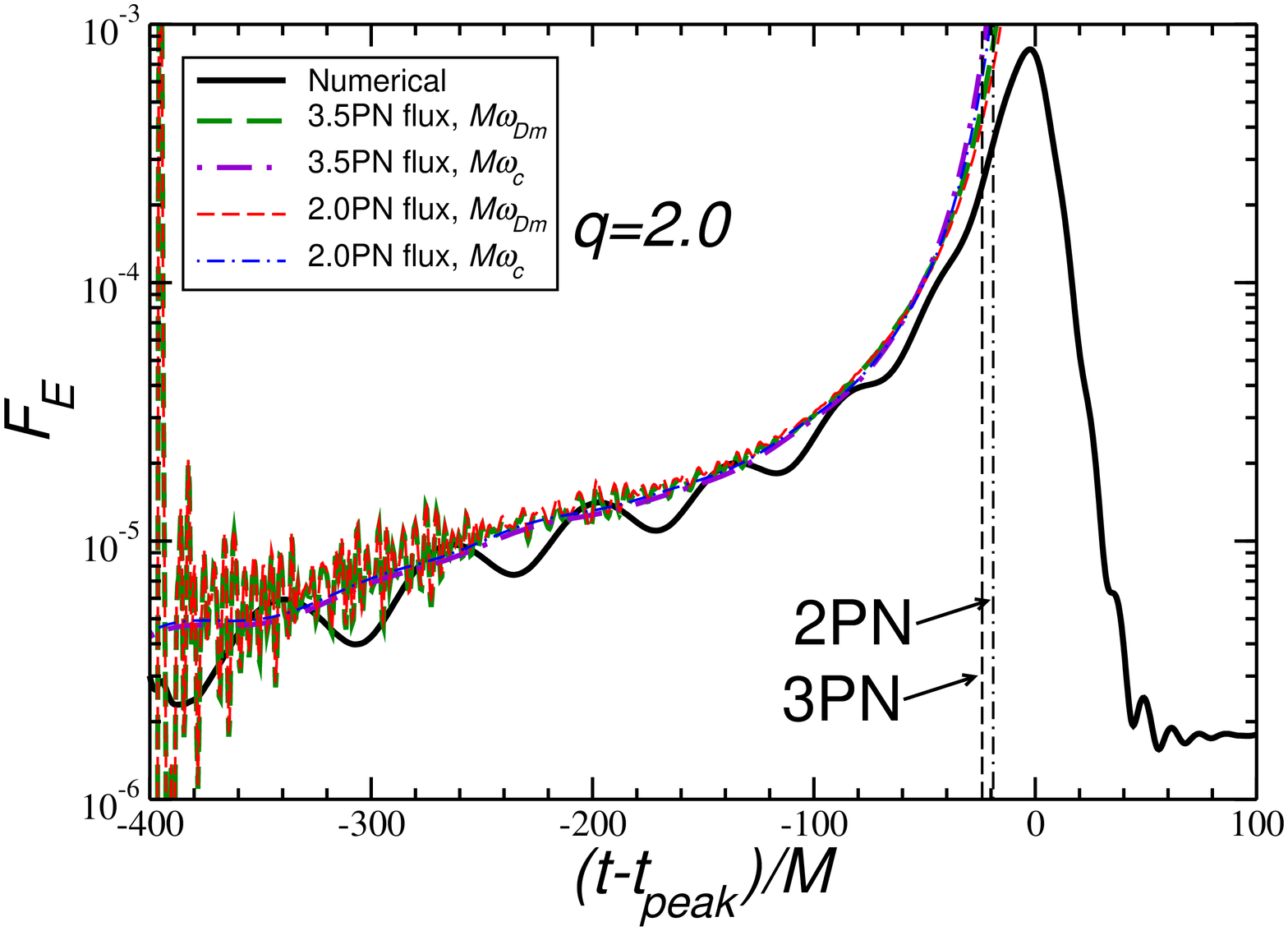,width=7cm,angle=-90} &
\epsfig{file=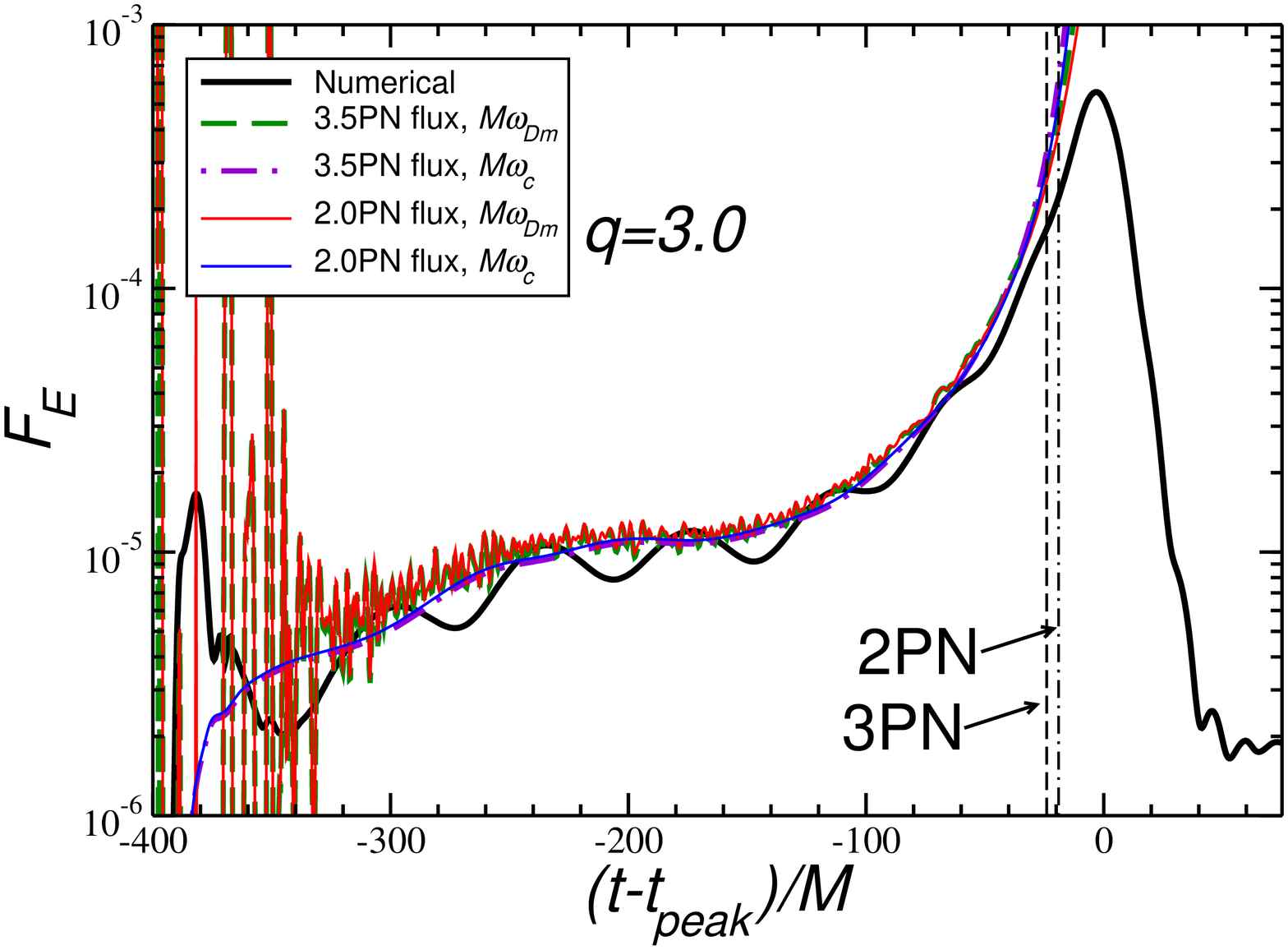,width=7cm,angle=-90} \\
\end{tabular}
\caption{Total energy flux computed numerically (thick solid line), and
  substituting two different estimates of the orbital frequency ($\omega_{Dm}$
  and $\omega_c$) into the 2PN and 3.5PN energy fluxes, i.e., keeping
  different numbers of terms in Eq.~(\ref{flux}). The 2PN and 3.5PN fluxes are
  so close they cannot be resolved ``by eye'' on the scale of this plot. 
  The plots refer to runs D8. On the left we show results for $q=2.0$, on the
  right for $q=3.0$.
  \label{fig:flux}}
\end{center}
\end{figure*}

In Fig.~\ref{fig:flux} we compare the total energy flux (as computed from our
numerical simulations) with the PNQC energy flux (\ref{flux}) evaluated at 2PN
and 3.5PN. The 2PN, 3PN and 3.5PN expansions are very close to each other, and
to improve readability we decided not to display the 3PN results\footnote{The
  2.5PN expansion of the flux (not shown in the plots) has a physically
  unreasonable zero crossing $\sim 20-40M$ before the radiation peak.  The
  poor quality of the 2.5PN flux is in agreement with well-known results in
  the extreme mass ratio limit (see eg. Fig.~1 in \cite{Poisson:1995vs}).}.
At each different PN order we evaluate Eq.~(\ref{flux}) by using two different
estimates of the orbital frequency ($\omega_{Dm}$ and $\omega_c$).

Some features of Fig.~\ref{fig:flux} should be quite familiar from the
discussion in Section \ref{freq-ests}. First of all, because of the (small)
orbital eccentricity, the numerical flux oscillates around a ``mean'' value
given by the PNQC estimate.  
The numerical flux starts deviating quite clearly from the 2PN and 3.5PN
fluxes $\sim 20-40M$ before the ISCO, and the agreement between numerical and
analytical fluxes gets slightly better for larger mass ratio.  A remarkable
feature of the numerical flux is that it does {\it not} reduce to zero after
the exponentially decaying ringdown phase. We believe this to be, at least in
part\footnote{Some contribution to the non-zero flux at late times may come
  from numerical noise.}, an artifact of the memory effect discussed in
Section \ref{memory}.

\begin{figure*}[ht]
\begin{center}
\begin{tabular}{cc}
\epsfig{file=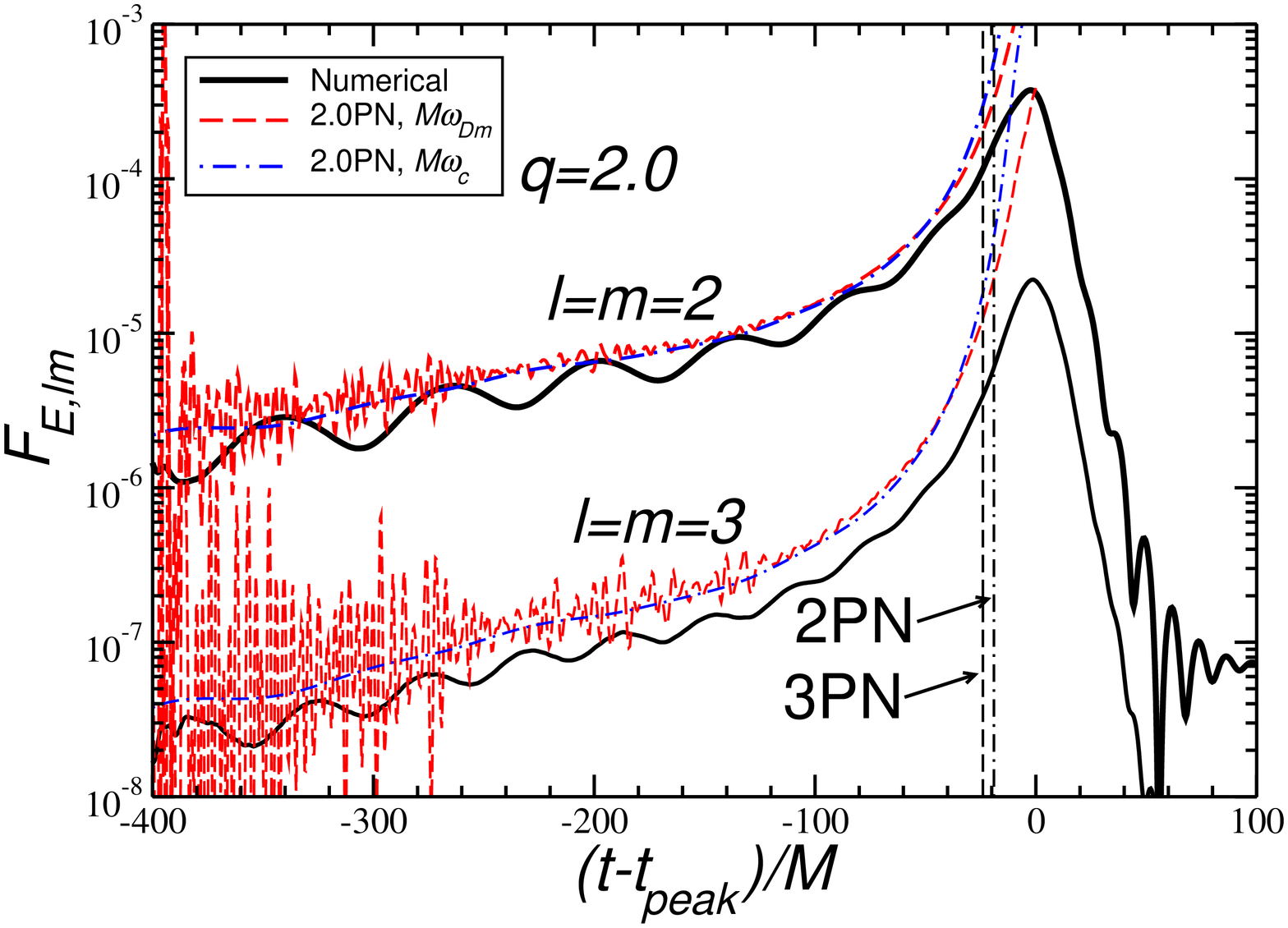,width=7cm,angle=-90} &
\epsfig{file=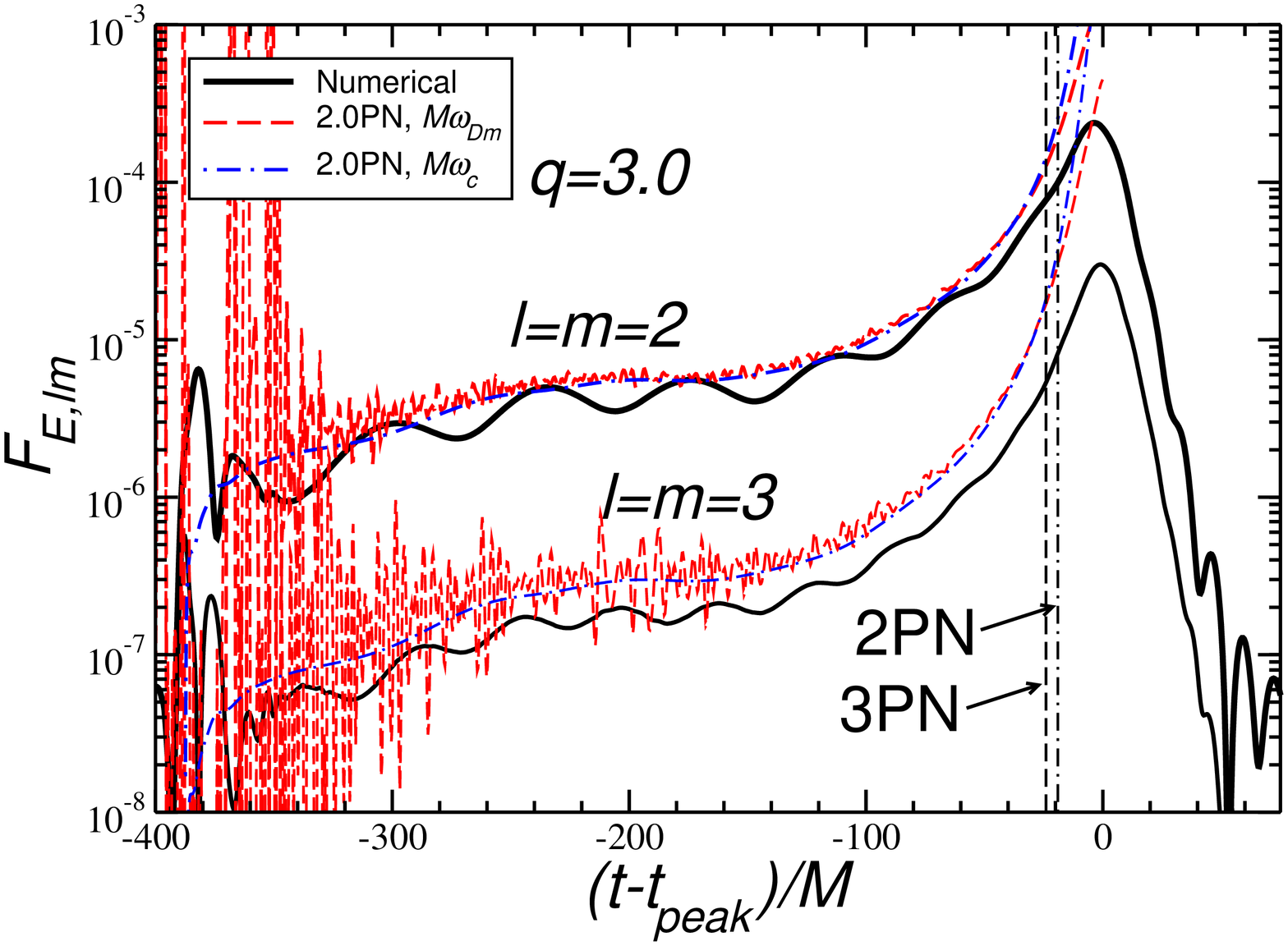,width=7cm,angle=-90} \\
\end{tabular}
\caption{Dominant multipolar components of the energy flux computed
  numerically (thick solid lines), and substituting two different estimates of
  the orbital frequency ($\omega_{Dm}$ and $\omega_c$) into the 2PN multipolar
  contribution to the energy flux, Eq.~(\ref{Flm}). The plots refer to runs
  D8. On the left we show results for $q=2.0$, on the right for $q=3.0$.
  \label{fig:flux-lm}}
\end{center}
\end{figure*}

In Fig.~\ref{fig:flux-lm} we look at the dominant multipoles contributing to
the total energy flux: $l=m=2$ and $l=m=3$. The available analytical
expansions of the mode amplitudes $Mr\,\psi_{l\,,m}$ are only 2.5PN accurate.
Since the 2.5PN term in the $l=m=2$ waveform depends on our particular choice
of $\varpi$, we dropped all terms of order higher than 2PN in Eq.~(\ref{Flm}).

By comparing Fig.~\ref{fig:flux} and Fig.~\ref{fig:flux-lm}, we can easily
check that the $l=m=2$ component carries almost half of the total energy flux
(as long as we compare numerical results with numerical results, or consider
the same truncation order in the PN approximation). The agreement between the
2PN estimate and the numerical flux is quite good for the dominant ($l=m=2$)
component. However, the 2PN approximation seems to systematically
overerestimate the flux in $l=m=3$.  Given the present accuracy of our
numerical code (and the poor convergence properties of the total energy flux
at 2.5PN) it is hard to tell whether this is an artifact of eccentricity in
the simulations, or a genuine indicator of the convergence properties of the
PN approximation for higher multipolar components.

%

Whether we consider numerical results or the PNQC approximation, the relative
contribution of higher-$l$ multipoles to the flux can be seen to increase with
mass ratio. It also increases (for fixed mass ratio) as we get close to
merger. To leading order in a PNQC expansion we can show that the ratio of the
dominant multipolar components of the flux as a function of frequency (or
alternatively, as a function of time to coalescence) is given by
\begin{subequations}
\beq
\f{F_{E, 33}}{F_{E, 22}}&\simeq&
\f{40}{21}\left (\f{27}{32}\right )^2\left (\f{\delta M}{M}\right)^2
(M\Omega)^{2/3}\simeq
\f{10}{21}\left (\f{27}{32}\right )^2\left (\f{\delta M}{M}\right)^2
\left(\f{5}{\eta}\right)^{1/4}(t_c-t)^{-1/4}\,,\\
\f{F_{E,44}}{F_{E,22}}&\simeq&
\f{1280}{567} (1-3\eta)^2(M\Omega)^{4/3} \simeq
\f{80}{567}(1-3\eta)^2\left(\f{5}{\eta}\right)^{1/2}(t_c-t)^{-1/2}\,,
\eeq
\end{subequations}
where $t_c$ is the coalescence time.

\begin{figure*}[ht]
\begin{center}
\begin{tabular}{cc}
\epsfig{file=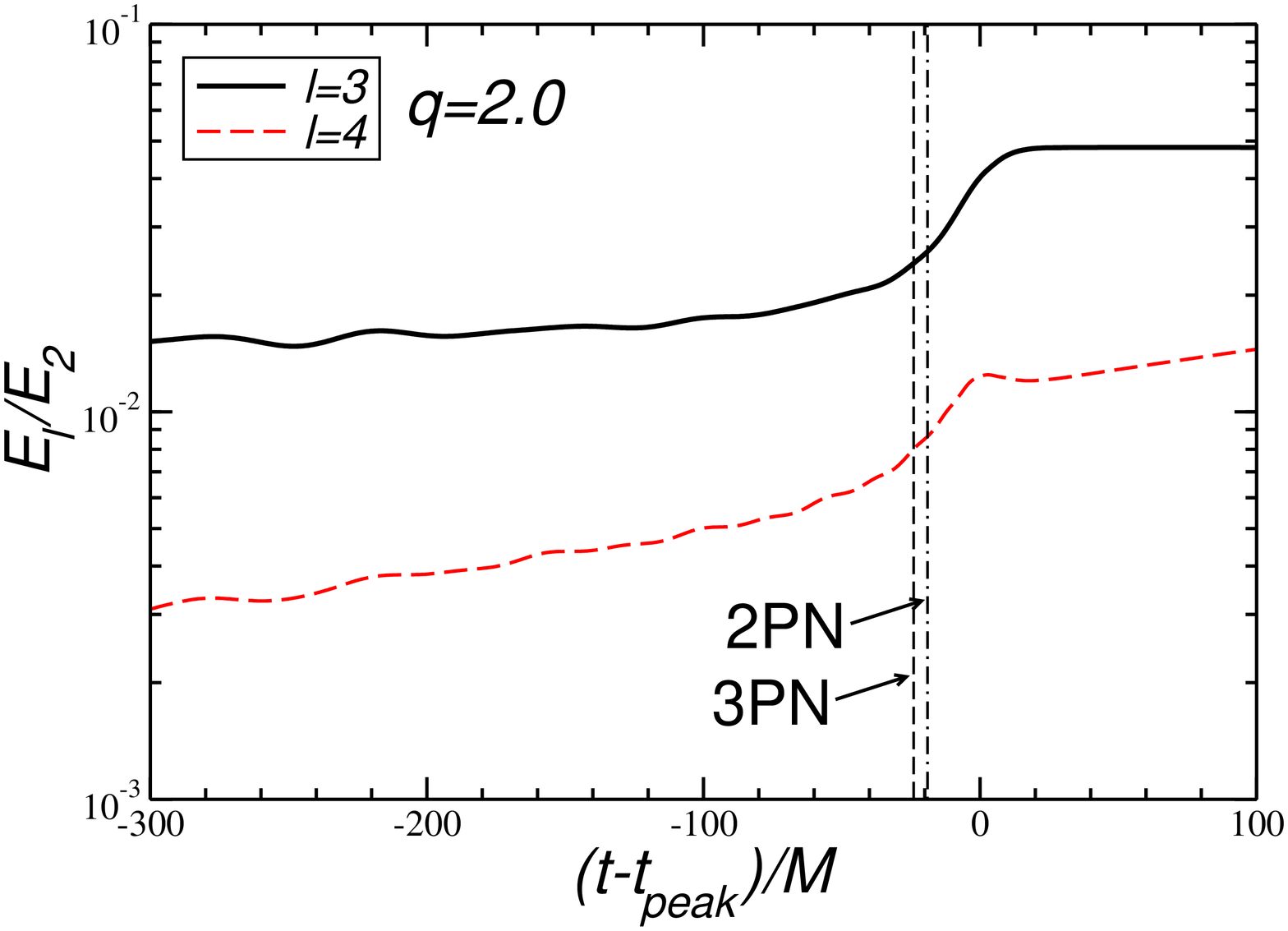,width=7cm,angle=-90} &
\epsfig{file=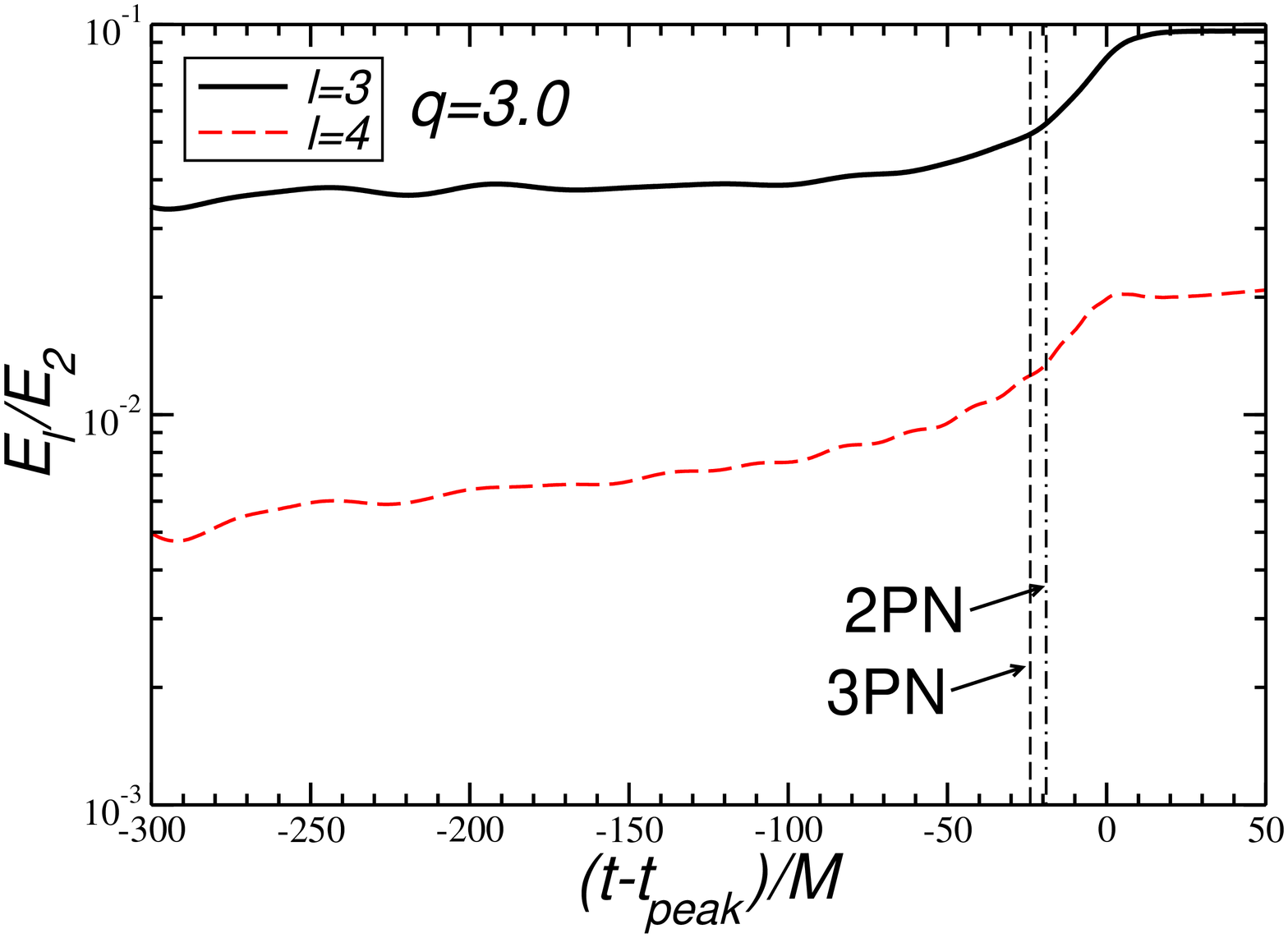,width=7cm,angle=-90} \\
\end{tabular}
\caption{Relative contribution of different multipolar components to the
  integrated energy flux as a function of time. The plots refer to runs D8. On
  the left we show results for $q=2.0$, on the right for $q=3.0$.
  \label{fig:energyratios}}
\end{center}
\end{figure*}

In Fig.~\ref{fig:energyratios} we show the ratio of the {\it integrated}
(numerical) energy flux in different multipolar components as a function of
time. This plot confirms that the relative contribution of higher multipoles
increases for large mass ratio, and (for given mass ratio) it increases as we
get close to merger.

\begin{figure*}[ht]
\begin{center}
\begin{tabular}{cc}
\epsfig{file=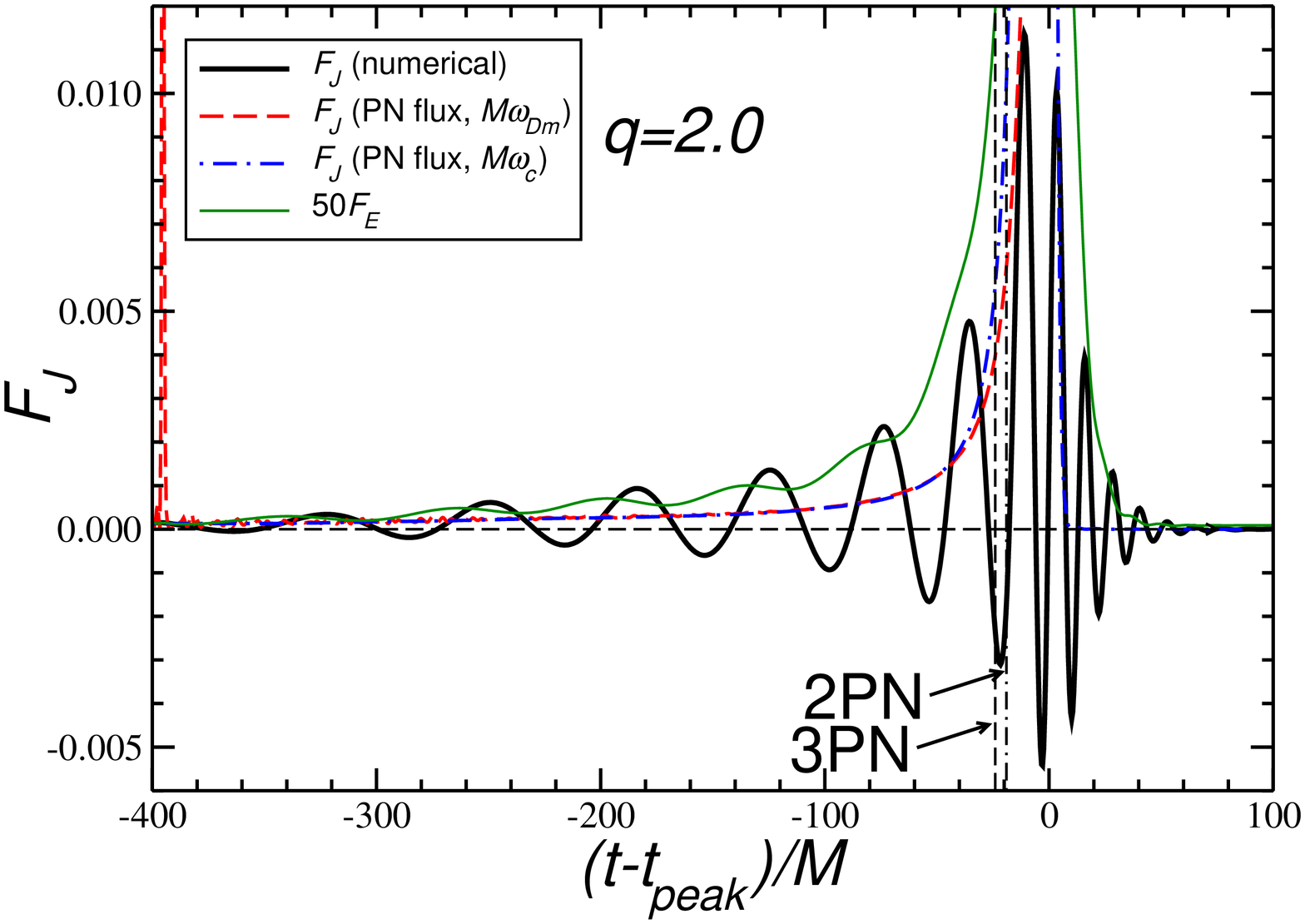,width=7cm,angle=-90} &
\epsfig{file=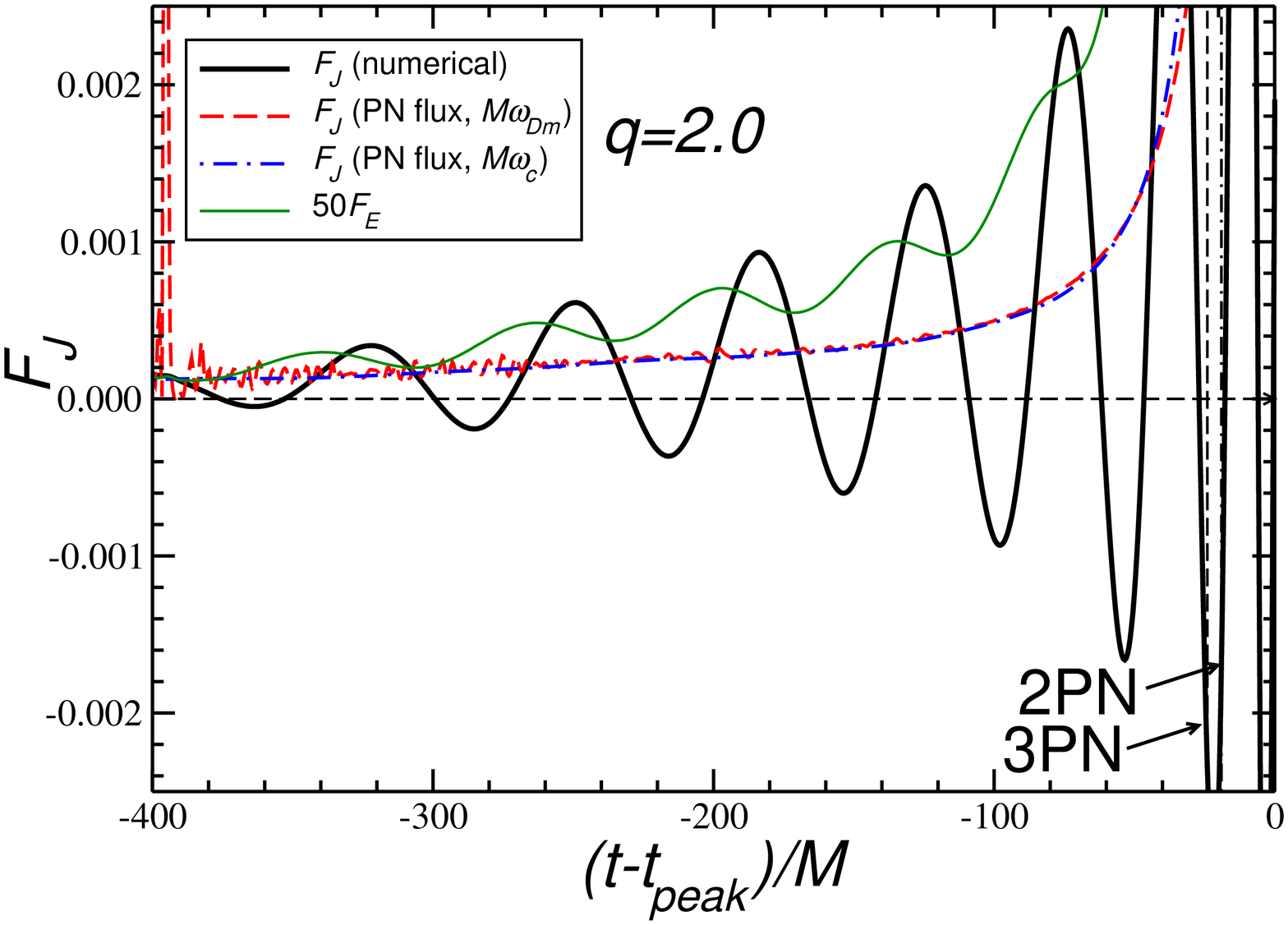,width=7cm,angle=-90} \\
\end{tabular}
\caption{Total angular momentum flux computed numerically (solid line) and
  substituting two different estimates of the orbital frequency, $\omega_{Dm}$
  and $\omega_c$, into Eq.~(\ref{FJPNQC}). We overplot the energy flux
  (multiplied by 50) for run D8 with $q=2.0$, to show that zero-crossings of
  the angular momentum flux correspond (roughly) to local extrema of the
  energy flux.
  \label{fig:fluxj}}
\end{center}
\end{figure*}

Finally, in Fig.~\ref{fig:fluxj} we compare the numerical angular momentum
flux with the PNQC prediction. The oscillations in the numerical flux seem to
be a general feature: compare eg. Fig.~27 of BCP, where they are attributed in
part to improper initial conditions in the time integrals required to obtain
the flux from $\Psi_4$. Our results confirm that, as remarked by BCP,
Eq.~(\ref{FJPNQC}) seems to hold {\it on average} throughout the whole
inspiral (possibly with larger deviations close to merger). In addition we
point out an interesting correlation between the energy and angular momentum
fluxes.  In Fig.~\ref{fig:fluxj}, besides the angular momentum flux, we also
plot the {\it energy} flux $F_{\rm E}$ (multiplied by 50 for scale). The plots
clearly show that oscillations in $F_{\rm J}$ have (roughly) the same period
as oscillations in $F_{\rm E}$. Perhaps this could be evidence that the
observed oscillations are somehow related with the orbital eccentricity,
minima and maxima corresponding to periastron and apastron. A detailed study
of this correlation is beyond the scope of this paper.

\clearpage

\section{The merger-ringdown transition}
\label{fitmerger}

The goal of this Section is to study the ringdown phase, and to explore the
properties of the final black hole formed after merger. We will compare
different fitting methods to extract information from the ringdown waveforms.
As discussed in \cite{Berti:2007dg}, such a comparison can help us resolve
real physical effects (such as, for example, time variations of the ringdown
frequencies) from systematic parameter estimation errors due to the variance
and bias of each particular fitting algorithm. In particular, here we consider
two classes of fitting algorithms: the matrix pencil (MP) and Kumaresan-Tufts
(KT) methods, which are modern variants of the so-called Prony
linear-estimation algorithms for damped exponentials in noise
\cite{matrixpencil,kumaresantufts}; and a standard non-linear least-squares
technique, the Levenberg-Marquardt (LM) algorithm \cite{LM}.

In \cite{Berti:2007dg} we pointed out that Prony methods have a number of
advantages with respect to standard non-linear least-squares techniques: (i)
They do not require an {\em initial guess} of the fitting parameters; (ii)
They provide us with a simple, efficient way to estimate QNM frequencies for
the {\em overtones}, and even to estimate how many overtones are present in
the signal; (iii) Statistical properties of Prony-based methods in the
presence of noise (such as their variance and bias) are well studied and under
control. When compared with the LM algorithm, Prony methods seems to have
comparable variance but slightly smaller bias.

In BCP, the real and imaginary parts of $\psi_{l\,,m}$ were fitted {\it
  separately} using standard non-linear least-squares methods. Prony-like
methods allow us to fit the ``full'', complex signal by a function of the form
\be\label{fitfunc} \psi^{\rm fit}_{l\,,m}= \sum_{l'm'n} {\cal A}_{l'm'n} e^{-i[\hat \omega_{l'm'n}(j,~M_{\rm
fin}) (t-t_{\rm peak}) +\phi_{l'm'n}]}  \,, \ee
where $\hat \omega_{l'm'n}(j,~M_{\rm fin})$ denotes a complex QNM frequency.
In BCP the final black hole's mass and spin ($j\,,M_{\rm fin}$) are taken as
the independent fitting parameters, and the different QNM frequencies $\hat
\omega_{lmn}$ are obtained, for given $(j,~M_{\rm fin})$, either by using
fitting relations or by interpolating numerical tables \cite{bcw}.

BCP allow for general mode-mixing due to the expansion of {\it spherical}
harmonics in terms of {\it spheroidal} harmonics. We will assume that each
spherical $(l\,,m)$ mode is well described by a single $(l\,,m)$ ringdown
mode. Another difference is that BCP include overtones in the QNM expansion.
Adding overtones provides a good fit of the strong-field phase by effectively
{\em increasing} the number of fitting parameters (mode amplitudes ${\cal
  A}_{lmn}$ and phases $\phi_{lmn}$ of the overtones). This idea is perfectly
consistent with QNM expansions in the context of linear black hole
perturbation theory. An obvious drawback of the idea is that it {\it assumes}
the validity of linear perturbation theory to extend the QNM fit before the
peak of the radiation.  Another potential problem is that, by using many
fitting parameters, we can always get very good agreement with the numerical
waveforms, but we do not necessarily get a better physical description of QNM
excitation. For simplicity, in this paper we do not attempt to include
overtones in the fit, but we only assess the accuracy of fits of the
fundamental QNM. In \cite{Berti:2007dg} we have shown that the QNM frequency
and damping time evolve quite rapidly right after merger. This evolution could
be interpreted as a bias in the fitted frequencies induced by the omission of
higher overtones; or, alternatively, it could mean that the mass and angular
momentum of the newly formed, dynamical black hole spacetime really are {\em
  evolving} on timescales much smaller than the QNM timescales, producing an
effective redshift in the QNM frequencies
\cite{Zlochower:2003yh,Papadopoulos:2001zf}.  Issues such as the inclusion of
overtones and the detailed study of nonlinearities will be addressed in the
future.

\subsection{Choice of the fitting window}
\label{fitwindow}

Independently of the chosen fitting method, there is some arbitrariness in
choosing the time interval $[t_0,~t_f]$ to perform the fit. A well-known
problem with the merger-ringdown transition is that we do not know {\it a
  priori} when the ringdown starts
\cite{Buonanno:2006ui,Berti:2006wq,Dorband:2006gg}. This problem is discussed
at length in Section \ref{sec:rdstart} below.  Ideally, the starting time for
the fit $t_0$ should be determined by a compromise between the following
requirements: (i) $t_0$ should be small enough to include the largest possible
number of data points: in particular, we do not want to miss the large
amplitude, strong-field part of the waveform after merger; (ii) $t_0$ should
be large enough that we do not include parts of the waveform which are {\em
  not} well described by a superposition of complex exponentials: the
inclusion of inspiral and merger in the ringdown waveform would produce a bias
in the QNM frequencies.

A judicious choice of $t_f$ is also necessary. Usually we would like the time
window to be as large as possible, but Fig.~\ref{wf1} and Fig.~\ref{wf2}
clearly show that the low amplitude, late-time signal is usually dominated by
numerical noise (mainly caused by reflection from the boundaries).  This noise
can reduce the quality of the fit, especially for the subdominant components
with $l>2$ and for large values of $t_0$.  A practical criterion for the
choice of $t_f$ is suggested by a look at Fig.~\ref{wf2}. If the ringdown
waveform were not affected by noise from boundary reflections,
$|Mr\,\psi_{l\,,m}|$ should decay linearly on the logarithmic scale of the
plots\footnote{With larger resolution and longer running times, eventually the
  exponential decay should turn into the well known power-law tail induced by
  backscattering of the radiation off the spacetime curvature
  \cite{Price:1971fb}. In the simulations we consider, noise produced by
  boundary effects is large enough that this effect is not visible.}. At low
signal amplitudes, we see boundary noise-induced wiggles superimposed to this
linear decay. The first occurrence of these wiggles is a good indicator of the
time $t_f$ at which numerical results cannot be trusted anymore. To test the
robustness of fitting results to late-time numerical noise, while at the same
time keeping the largest number of data points in the waveform, we decided to
use two different ``cutoff criteria'':

\begin{itemize}

\item[1)] ``Relative'' cutoff: remove from the waveforms all data for times
  $t>t_f=t_{\rm rel}$, where $t_{\rm rel}$ is the time when the amplitude of
  each multipolar component $|Mr\,\psi_{l\,,m}|$ becomes less than some factor
  $\psi_{\rm cutoff}$ times the peak amplitude $|\psi_{l\,,m}(t_{\rm peak})|$
  (values of $t_{\rm peak}$ for $l=m=2$ are listed in Table \ref{tab:pars}):

  \be
  \f{|Mr\,\psi_{l\,,m}(t_{\rm rel})|}{|Mr\,\psi_{l\,,m}(t_{\rm peak})|}<
  \psi_{\rm cutoff}\,.
  \ee

\item[2)] ``Absolute'' cutoff: remove from the fit all data with $t>t_f=t_{\rm
    abs}$, where $t_{\rm abs}$ is the time at which the {\em absolute} value
  of the amplitude $|Mr\,\psi_{l\,,m}|<\psi_{\rm cutoff}/10$.

\end{itemize}

The choice of the cutoff amplitude is somewhat arbitrary. We chose $\psi_{\rm
  cutoff}=10^{-3}$ for low resolution, and $\psi_{\rm cutoff}=10^{-4}$ for
high resolution.

For each chosen $t_f$, we compare the different fitting routines as we let
$t_0$ vary in the range $[t_{\rm peak},~t_f]$. By monitoring the convergence
of the QNM frequencies to some ``asymptotic'' value as $t_0\to \infty$, we can
tell if the black hole settles down to a stationary Kerr state, or if, on the
contrary, non-linearities and mode coupling are always present.  Notice that
as $t_0$ grows the signal amplitude decreases exponentially, and we
effectively reduce the signal-to-noise ratio (SNR) in our fitting window.
Robust fitting methods should give reasonable results even for large values of
$t_0$ (that is, modest values of the SNR).

\subsection{From ringdown frequencies to black hole parameters}
\label{wrjm}

In \cite{Berti:2007dg} we fitted the frequency $\omega$ and quality
factor\footnote{We recall that the quality factor $Q\equiv
  |\omega/(2\alpha)|$, where $\alpha=1/\tau$ is the imaginary part of the QNM
  frequency, i.e. the inverse of the damping time.} $Q$ of the $l=m=2$
fundamental mode of the newly-formed black hole as a function of $t_0$.  The
results show that the QNM frequencies evolve quite rapidly in the first
$10M-20M$ after merger: see in particular the bottom panels of Fig.~7 in
\cite{Berti:2007dg}, where a rapid decrease of $Q(t_0)$ is clearly visible for
simulation times $240\lesssim t_0/M\lesssim 260M$.  Assuming linear
perturbation theory to be valid, the real and imaginary parts of each QNM
frequency are unique functions of the mass $M_{\rm fin}$ and of the
(dimensionless) angular momentum $j=J_{\rm fin}/M_{\rm fin}^2$ of the final
black hole: say,
$M_{\rm fin}\omega_{lmn}=f_{lmn}(j)$,
$M_{\rm fin}\alpha_{lmn}=M_{\rm fin}/\tau_{lmn}=g_{lmn}(j)$
\cite{bcw}. The quality factor of the oscillations $Q_{lmn}$, being
dimensionless, must be a function of $j$ only. A numerical calculation shows
that for the dominant modes ($l=m=2,~3,~4$) this function is monotonic and
invertible (see eg. Fig.~5 in Ref.~\cite{bcw}). Therefore we can easily invert
$Q_{lmn}(j)$ to compute $j(t_0)$, either by using fitting relations or by
interpolating QNM tables.

\begin{figure*}[ht]
\begin{center}
\begin{tabular}{cc}
\epsfig{file=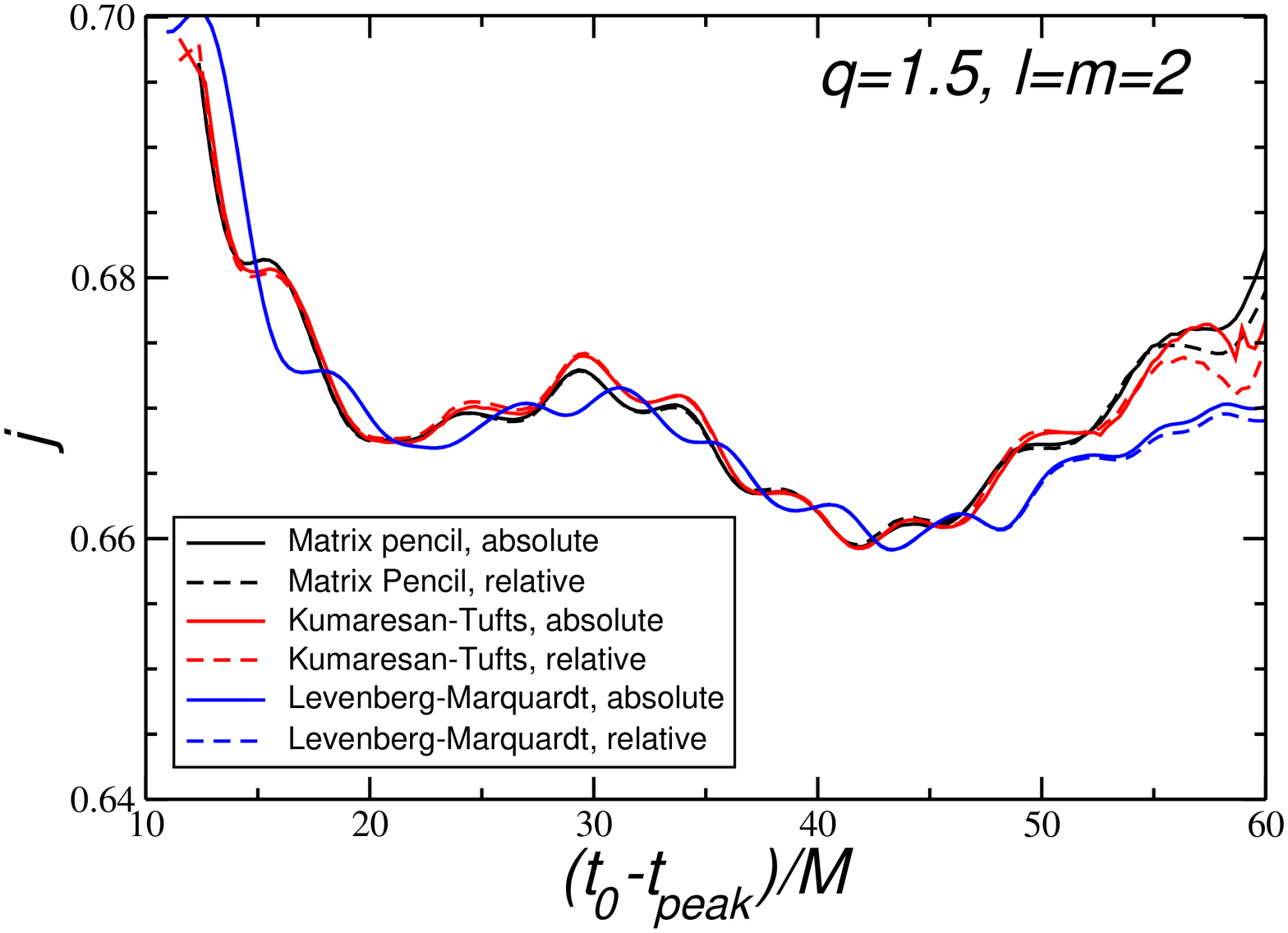,width=7cm,angle=-90}&
\epsfig{file=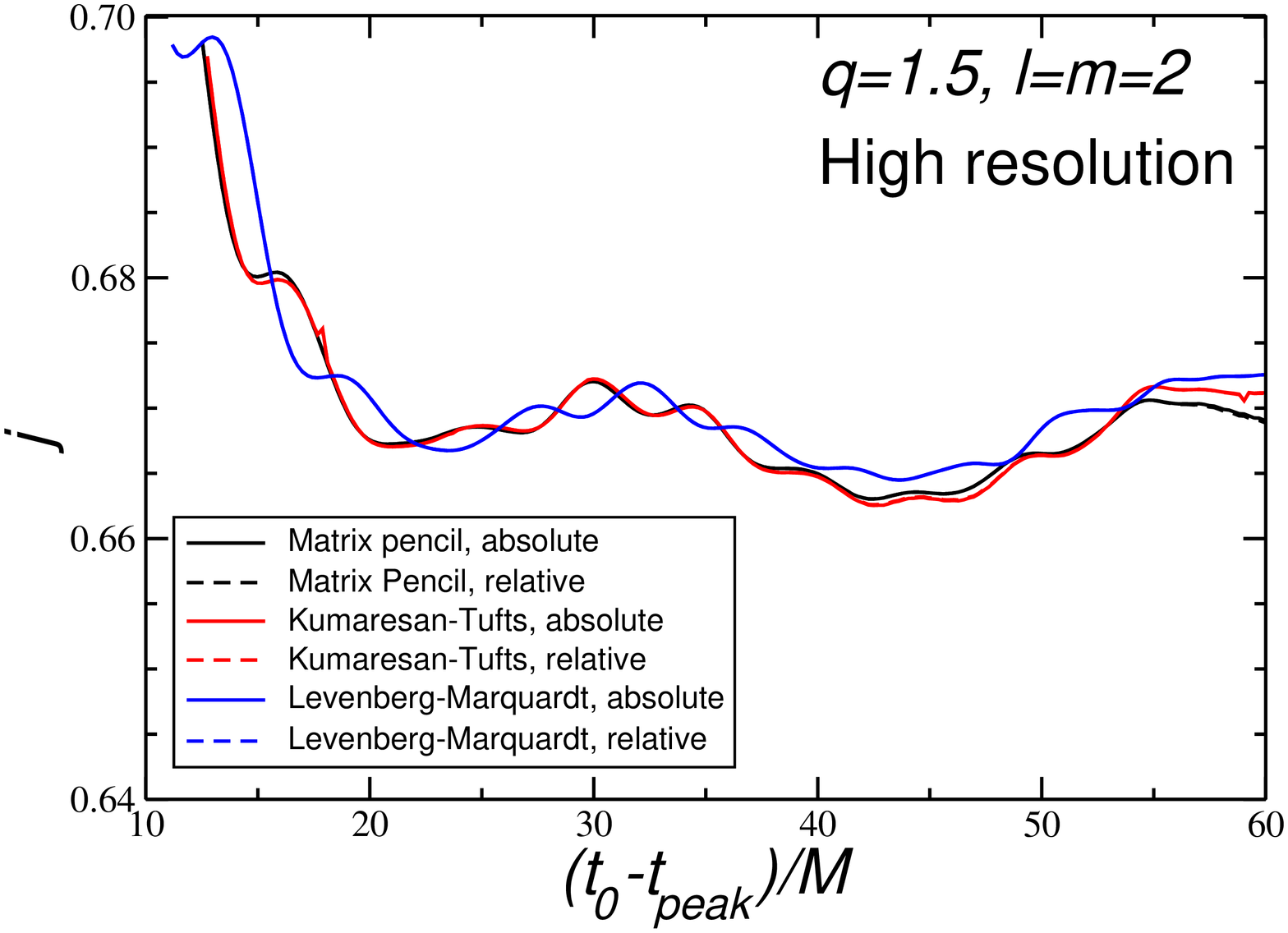,width=7cm,angle=-90}\\
\epsfig{file=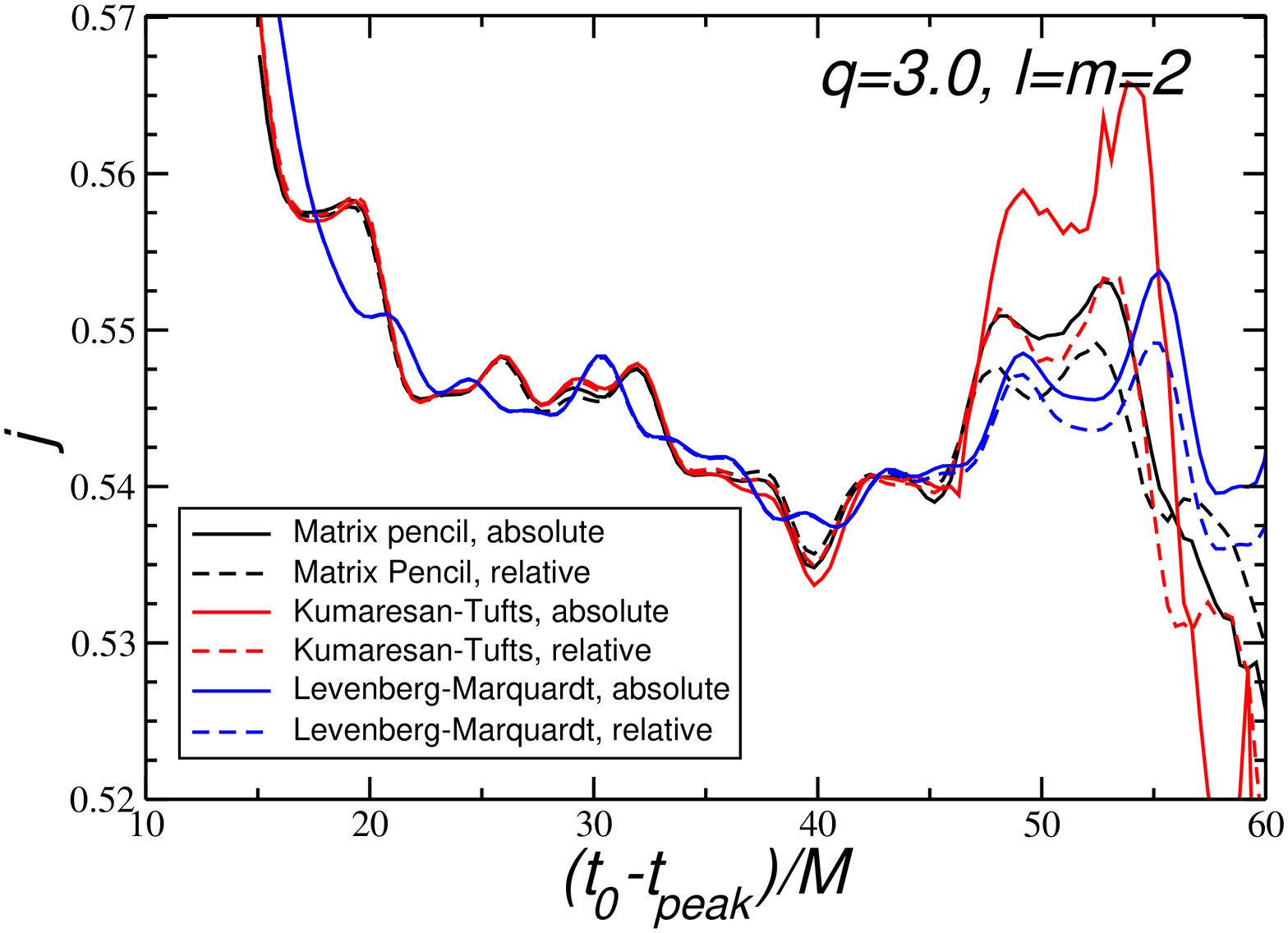,width=7cm,angle=-90}&
\epsfig{file=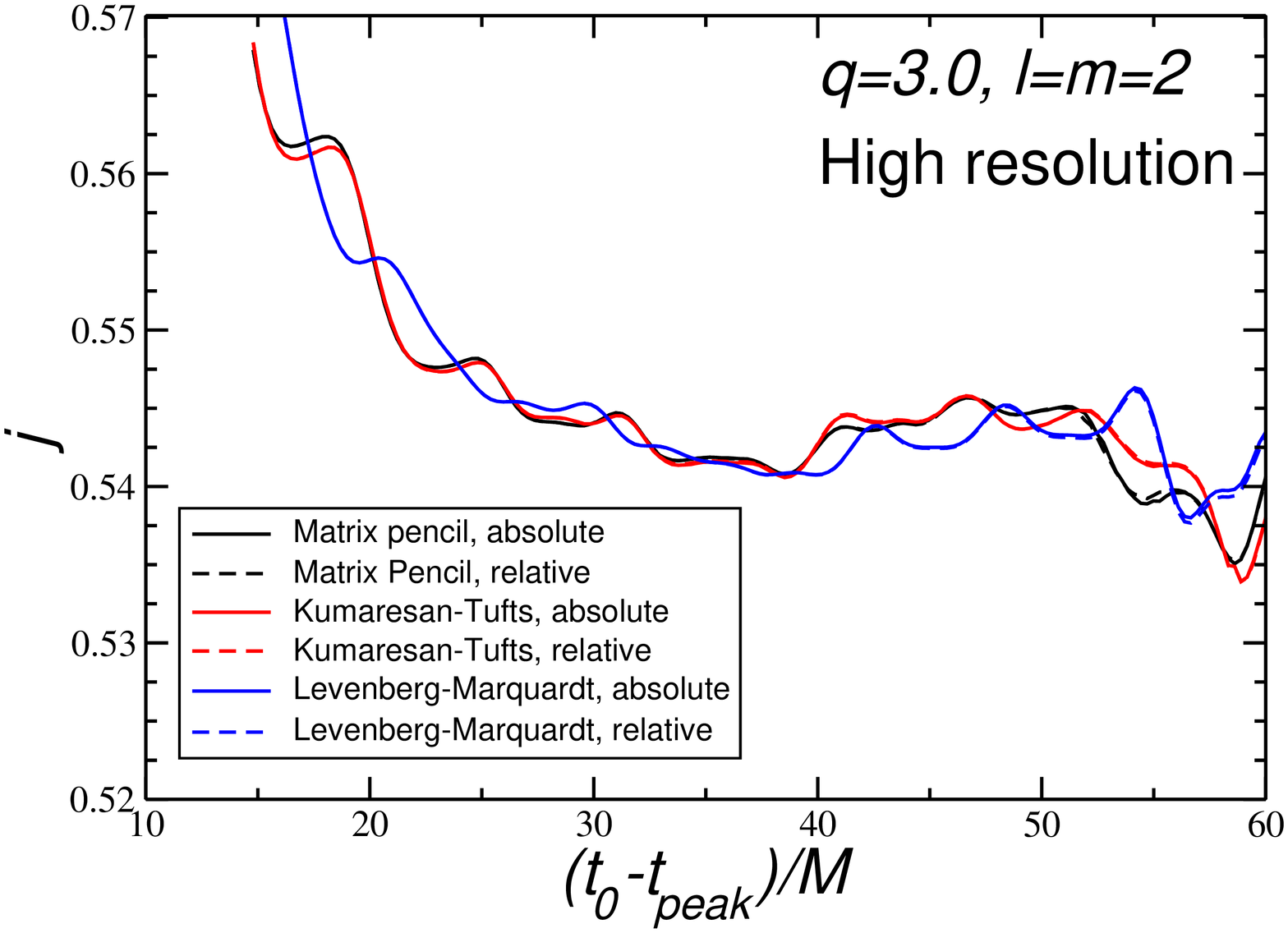,width=7cm,angle=-90}\\
\end{tabular}
\caption{Estimate of the angular momentum from a fit of the $l=m=2$ waveform
  using different methods. Top panels refer to a merger with $q=1.5$,
  bottom panels to a merger with $q=3.0$. Results on the left were obtained
  from low-resolution D7 runs, and those on the right from high-resolution
  D7 runs.
  \label{jestimate}}
\end{center}
\end{figure*}

The results of this inversion for the fundamental $l=m=2$ mode of the black
hole formed as a result of inspirals with $q=1.5$ and $q=3.0$ are shown in
Fig.~\ref{jestimate}. As the origin of the time axis we choose the time
$t_{\rm peak}$ at which the $l=m=2$ amplitude has a maximum (see Table
\ref{tab:pars}). Solid lines refer to the ``absolute'' truncation criterion,
and dashed lines to the ``relative'' truncation criterion (see Section
\ref{fitwindow}).  On the scale of these plots, different truncation criteria
affect the estimated parameters only for low-resolution simulations and at
relatively late starting times ($t_0/M\gtrsim 50$), when the signal amplitude
becomes comparable to numerical noise.  Not surprisingly, there is remarkable
agreement between KT and MP methods. The main difference when we use the
non-linear least-squares LM method is a {\it systematic time-shift in the
  angular momentum}: the blue lines would be in excellent agreement with the
prediction from Prony methods if shifted backwards in time by $\Delta t_0 \sim
2-3M$. This time shift can easily be understood. In the non-linear least
squares fit we are ignoring the imaginary part of the waveform. Since the real
and imaginary parts of the waveforms are essentially time-shifted copies of
each other, this produces a constant dephasing in the predicted physical
parameters of the final black hole.

\begin{figure*}[ht]
\begin{center}
\epsfig{file=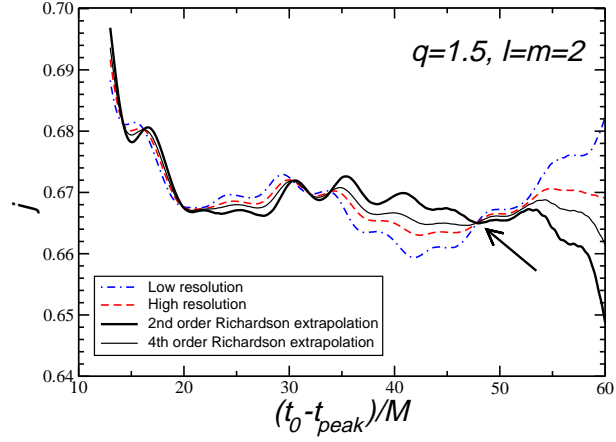,width=7cm,angle=-90}
\caption{Richardson extrapolation of the estimated angular momentum for
  $l=m=2$ and $q=1.5$, assuming second-order (thick line) and fourth-order
  (thin line) convergence.
  \label{jrichardson}}
\end{center}
\end{figure*}

In the absence of numerical errors and mode coupling $j(t_0)$ should
monotonically decrease, approaching a constant as $t_0\to\infty$.
Fig.~\ref{jestimate} clearly shows that this is not the case.  All fitting
routines consistently predict non-trivial time variations (roughly of order a
percent) in $j$. Increasing the resolution reduces the amplitude of these
variations, and produces a flattening of $j(t_0)$ for $40\lesssim
t_0/M\lesssim 60$.  The angular momentum increase that can be seen for $q=1.5$
and $(t_0-t_{\rm peak})\gtrsim 45 M$, and the oscillations in $j$ for $q=3.0$
in the same time range, are clearly artifacts of insufficient resolution. We
tried to perform a Richardson extrapolation of the results assuming
second-order and fourth-order convergence, to determine if angular momentum
oscillations (which could be a sign of ``new'' physics) disappear in the limit
of infinite resolution. Our results are shown in Fig.~\ref{jrichardson}. They
are compatible with the possibility that oscillations disappear in the limit
of infinite resolution, but more simulations and better control of the errors
are required to reach a firm conclusion.

Fig.~\ref{jestimate} clearly illustrates that resolution plays a role in the
accuracy with which we can estimate black hole parameters from ringdown fits,
especially at late times, when the signal is very weak and affected by
numerical noise (likely caused by reflections off refinement and outer
boundaries). Fortunately, changing the extraction radius does {\it not} affect
the quality of the fits. We checked this by fitting the $l=m=2$ and $l=m=4$
modes for equal mass ($q=1$), large separation (D10) binary mergers with
different extraction radii $r_{\rm ext}=30,~40$ and $50$.  The functional form
of $j(t_0)$ is exactly the same at different extraction radii. Changing
$r_{\rm ext}$ only produces a trivial shift of the time axis by $\Delta
t_0\simeq \Delta r_{\rm ext}$, due to the finite propagation speed of the
waves.

\begin{figure*}[ht]
\begin{center}
\begin{tabular}{cc}
\epsfig{file=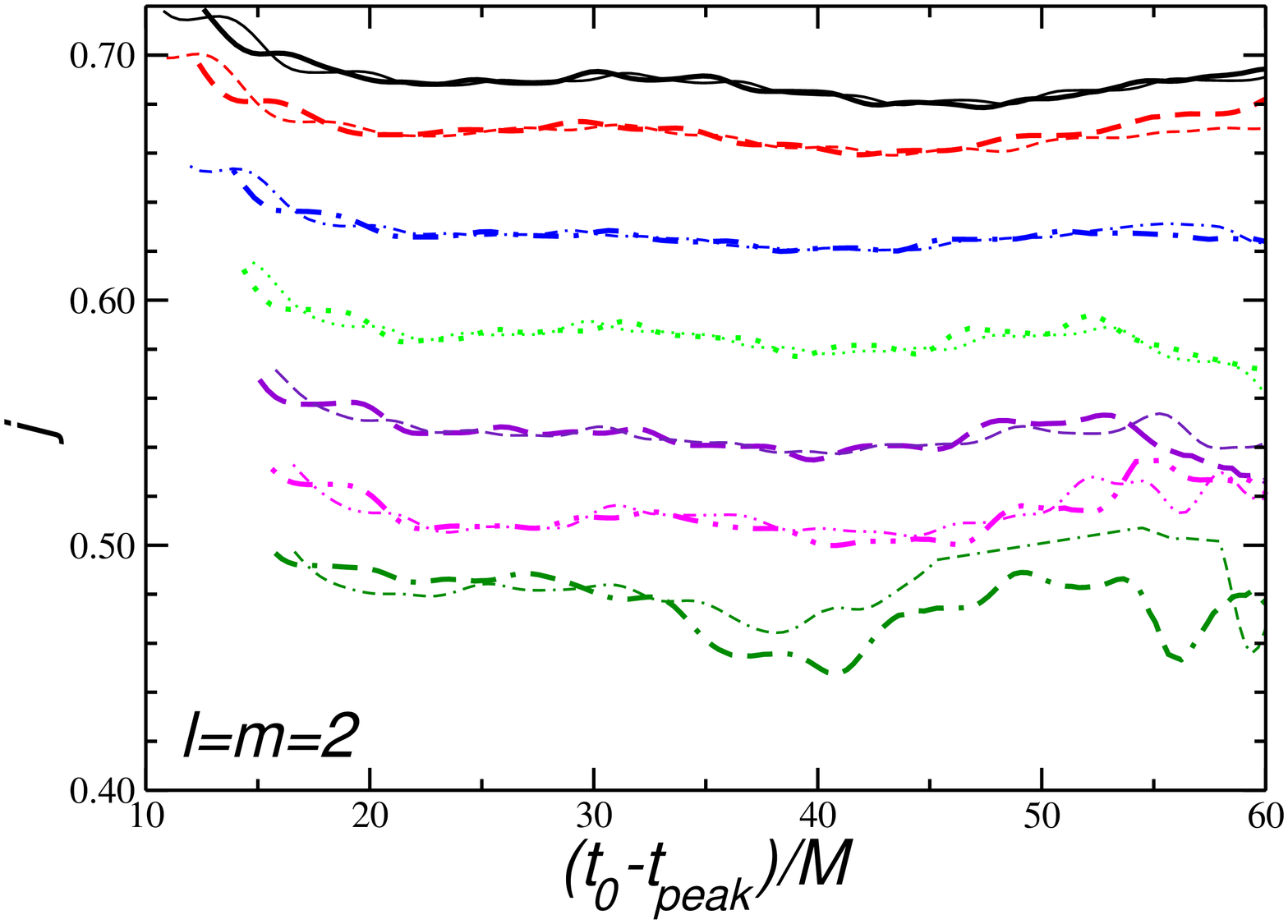,width=7cm,angle=-90} &
\epsfig{file=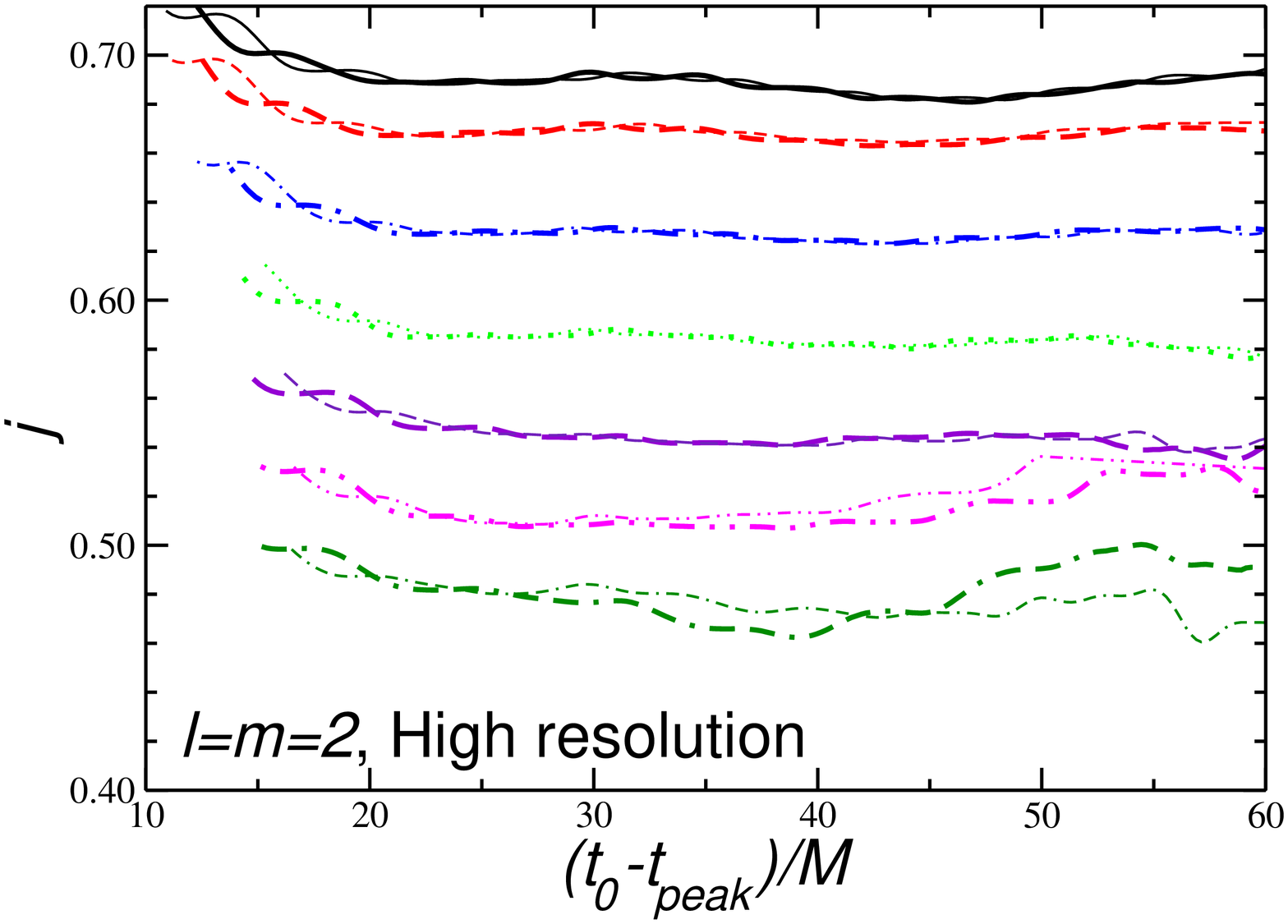,width=7cm,angle=-90} \\
\end{tabular}
\caption{\label{jq22} Angular momentum estimated applying the MP method (thick
  lines) and the LM method (thin lines) to the $l=m=2$ waveforms. Lines from
  top to bottom refer to different mass ratios:
  $q=1.0,~1.5,~2.0,~2.5,~3.0,~3.5,~4.0$.}
\end{center}
\end{figure*}

Estimates of the angular momentum as a function of $t_0$, obtained by fitting
the dominant mode ($l=m=2$) for different values of $q$, are shown in
Fig.~\ref{jq22}. The angular momentum is constant within about $\sim 1\%$, but
the quality of the estimates rapidly degrades with mass ratio. Even with
high-resolution runs, the estimated angular momenta have errors $\sim 10\%$
for $q\gtrsim 3$. In the next Section we will show that improved estimates are
possible if we {\em cross-correlate} information from different multipoles of
the radiation, making use of the no-hair theorem of general relativity.

\subsection{\label{sec:crosscorrelate}Cross-correlating information from different multipoles to
determine the black hole parameters}
We already pointed out that the quality factor of each QNM $Q_{lmn}$, being
dimensionless, must be a function of $j$ only; and for the dominant modes
($l=m=2,~3,~4$) this function is monotonically increasing, so we can easily
invert $Q_{lmn}(j)$ to compute $j(t_0)$ by using fitting relations or by
interpolating QNM tables \cite{bcw}. If linear perturbation theory were an
{\it exact} description of the final black hole's dynamics, the value of the
angular momentum obtained from different QNMs -- that is, from different
values of $(l,~m,~n)$ -- should be the same for all modes and all values of
$t_0$. In practice this is only approximately true. First of all, non-linear
effects should be present close to merger, so that linear perturbation theory
provides only an approximation to the ``true'' oscillation frequencies (if the
definition of QNMs makes sense at all in the non-linear regime). Secondly,
mode mixing induced by the use of spin-weighted spherical harmonics with some
given given $(l\,,m)$ rather than spin-weighted {\it spheroidal} harmonics,
will produce additional QNM frequencies\footnote{This mode mixing was actually
  observed by BCP: when fitting the $l=3$, $m=2$ waveform they also found the
  $l=m=2$ QNM frequency.}  with different $l$'s and the same $m$. Finally,
numerical error and the omission of overtones will inevitably produce some
bias in the estimation of the frequencies, whatever fitting routine we use to
extract them.

All of these effects should be reasonably small, especially at late times,
since linear perturbation theory can be expected to be a good approximation
once the final black hole is ``reasonably close'' to a Kerr state (where
``reasonably close'' is here a loosely defined concept that can be made more
precise, for example, through the use of quantities such as the ``speciality
index'' $S$ \cite{Baker:2000zm}). Conversely, if we estimate angular momenta
by fitting different multipolar components of the radiation, we can determine
when perturbation theory is a good description of the system by looking for
points (or intervals) in time when the angular momenta obtained from different
fits agree with each other.

\begin{figure*}[ht]
\begin{center}
\begin{tabular}{cc}
\epsfig{file=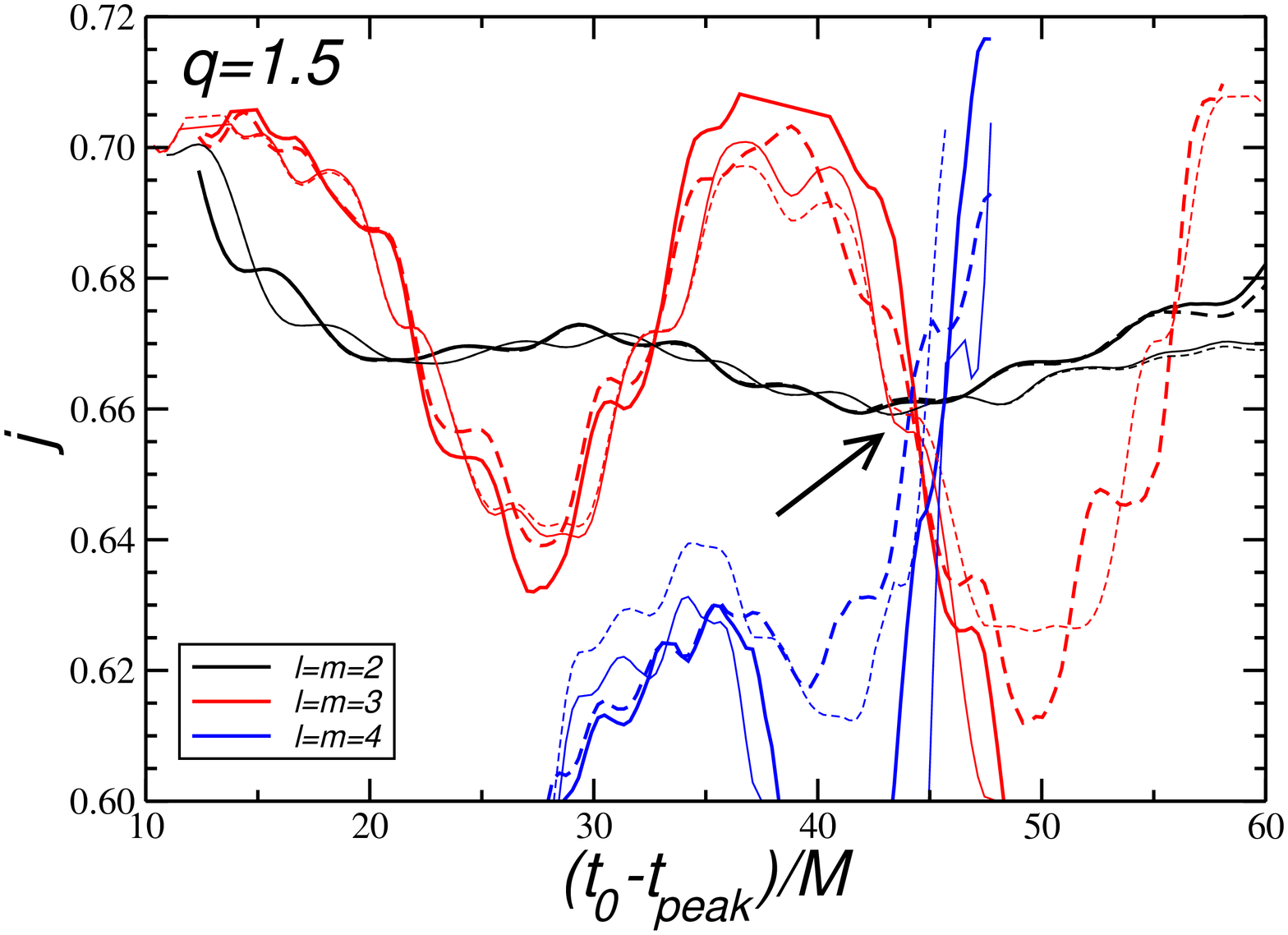,width=7cm,angle=-90}&
\epsfig{file=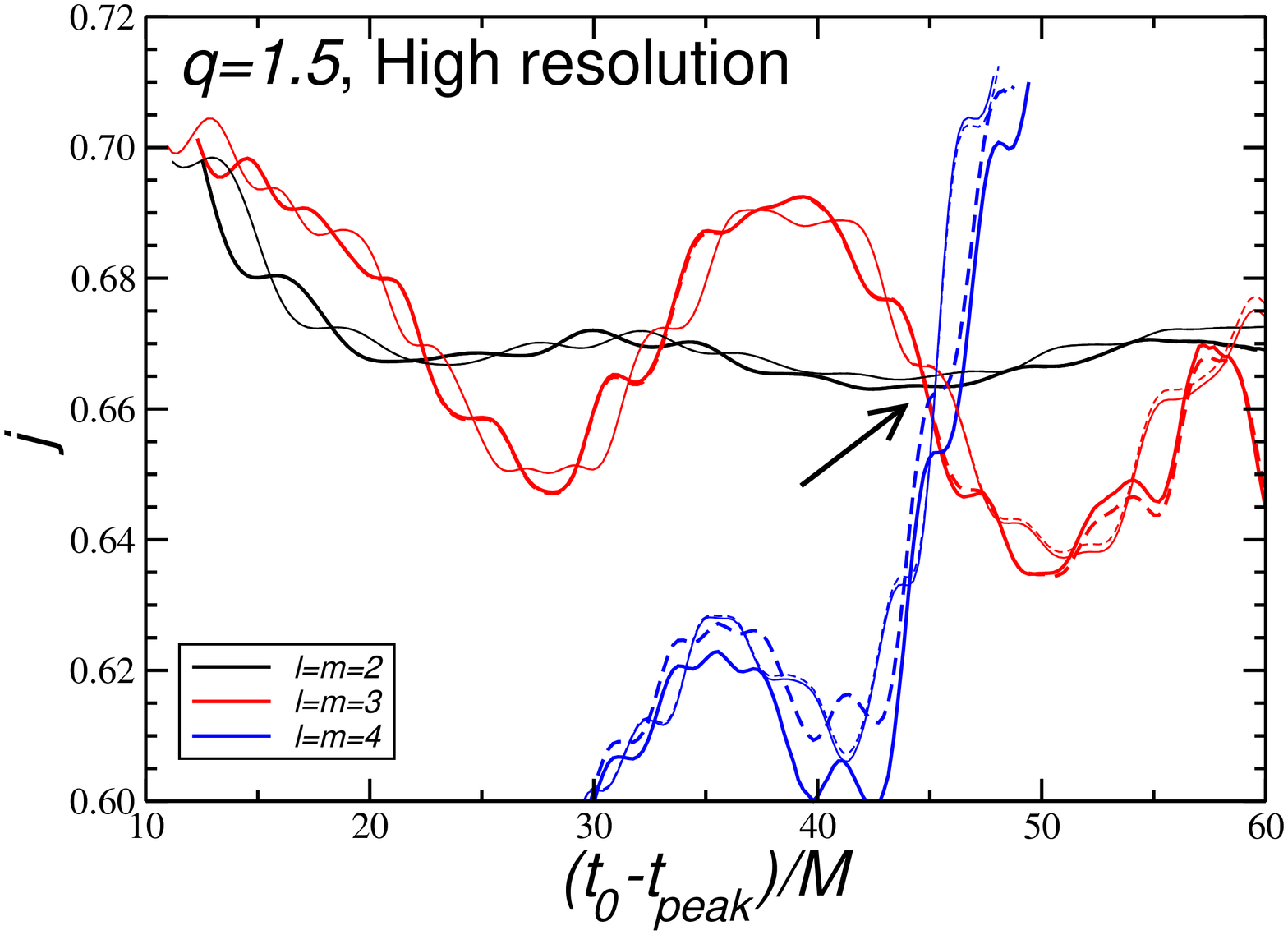,width=7cm,angle=-90}\\
\end{tabular}
\caption{Consistency in the radiated angular momentum, as predicted
 by different multipoles, for a $q=1.5$, D7 run. Thick lines use the MP
 method, thin lines use the LM method. Solid lines are obtained by removing
 all points for which the absolute amplitude drops below $10^{-4}$, and dashed
 lines by removing all points for which the relative amplitude drops below
 $10^{-3}$ the peak value. The choice of fitting method and truncation
 criterion has negligible influence on the results. Arrows mark the last point
 in time when the fit can still be considered reliable, {\it and} different
 multipoles agree with linear perturbation theory of Kerr black holes.
 \label{jconsistency}}
\end{center}
\end{figure*}

In Fig.~\ref{jconsistency} we plot the angular momenta estimated by fitting
the dominant multipolar components of the radiation emitted in a $q=1.5$, D7
merger. Angular momenta from $l=m=2$ and $l=m=3$ are generally in good
agreement, but they display oscillations around some mean value. The magnitude
of the oscillations is larger for $l=3$, and it also gets larger for coarse
resolutions. However, there are discrete points in time when the angular
momenta predicted by different multipolar components agree with each other.

\begin{figure*}[ht]
\begin{center}
\begin{tabular}{cc}
\epsfig{file=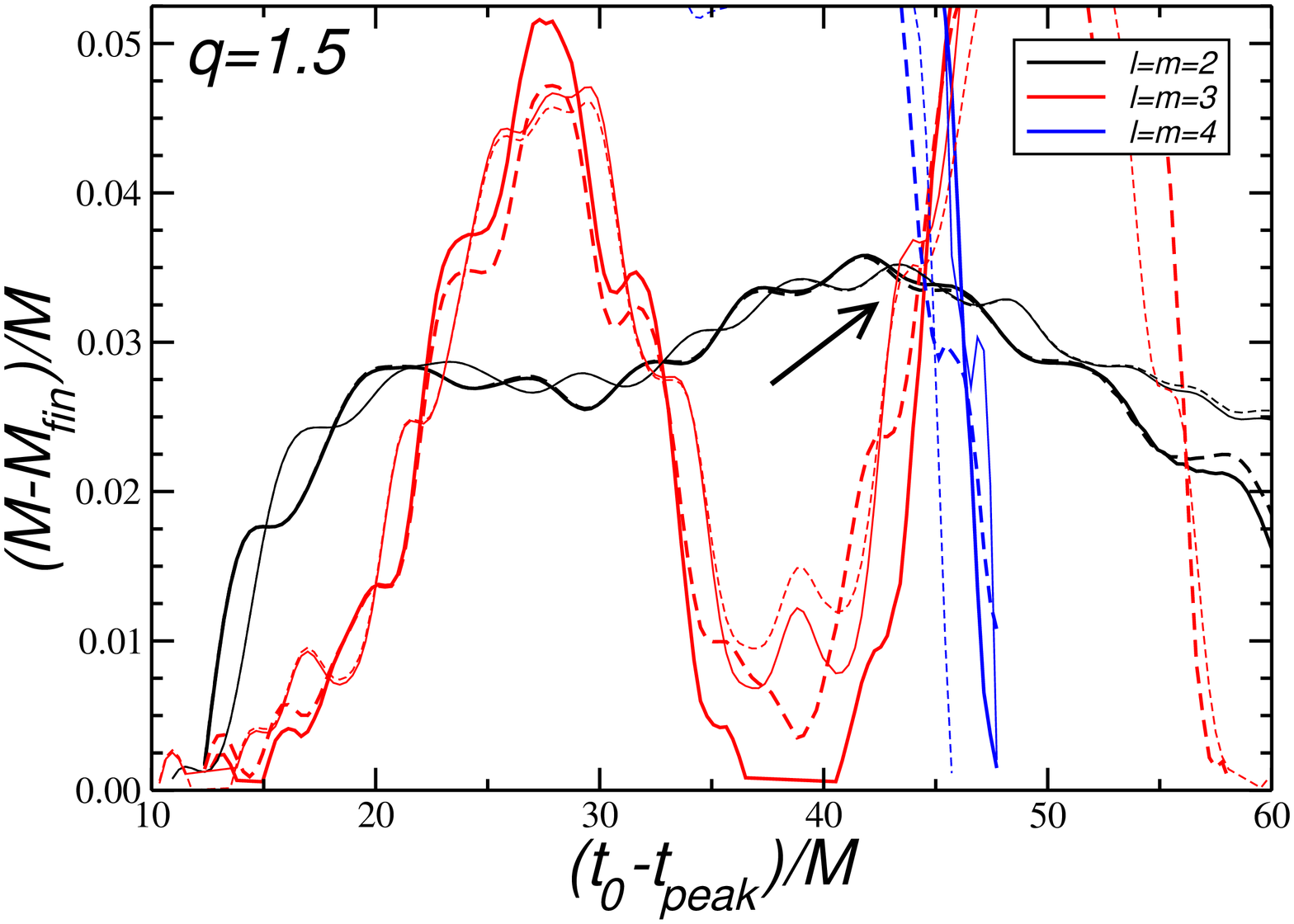,width=7cm,angle=-90}&
\epsfig{file=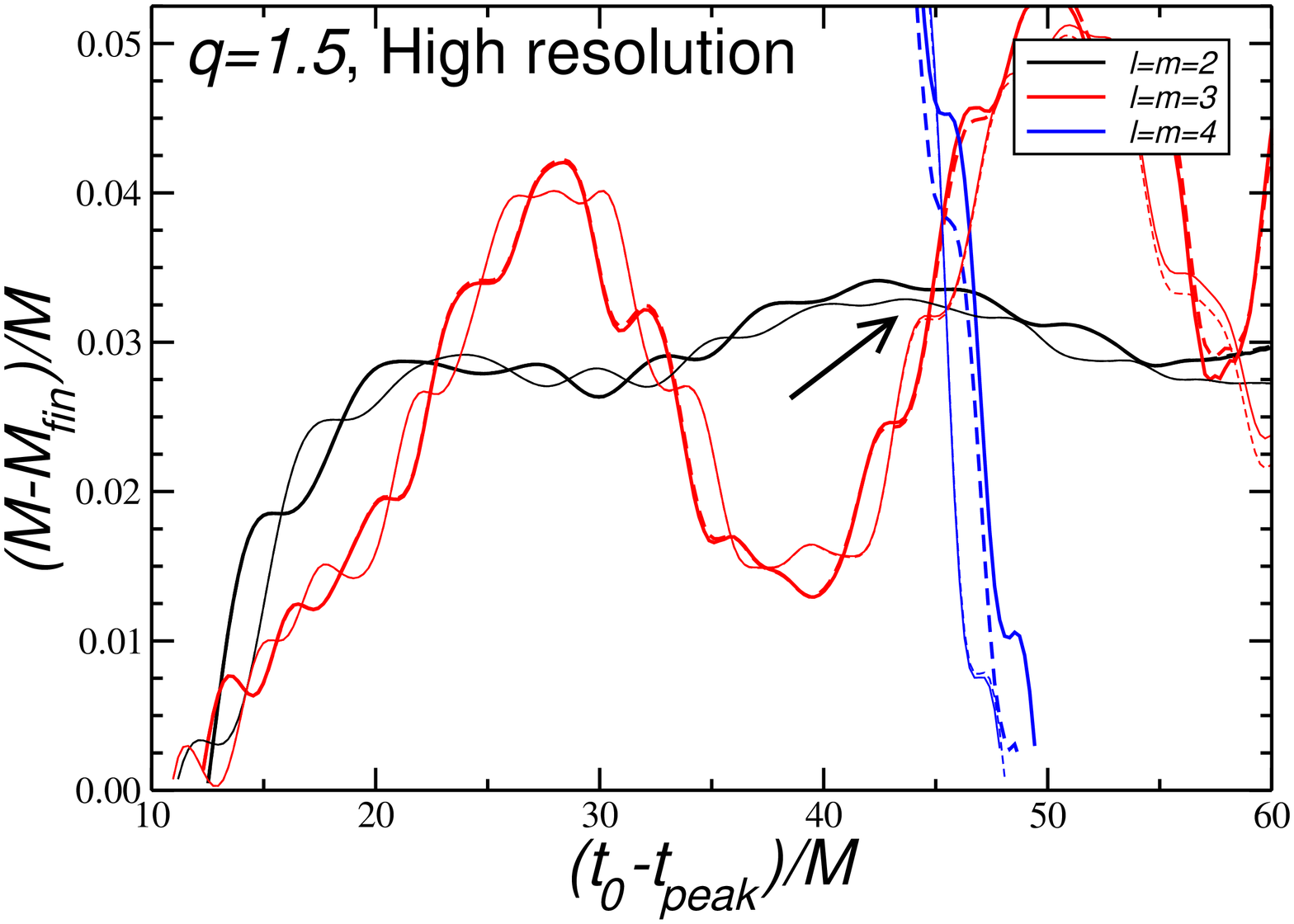,width=7cm,angle=-90}\\
\end{tabular}
\caption{Consistency in the energy radiated, as predicted by different
  multipoles, for a $q=1.5$, D7 run. Linestyles are the same as in
  Fig.~\ref{jconsistency}.
 \label{eradconsistency}}
\end{center}
\end{figure*}

In Fig.~\ref{eradconsistency} we use QNM fits of different multipoles to
extract the final black hole mass $M_{\rm fin}$. From $M_{\rm fin}$ we can
estimate the radiated energy as a function of $t_0$ by computing $(M-M_{\rm
  fin})/M$. The plots provide a remarkable consistency check of the results in
Fig.~\ref{jconsistency}: whenever results from numerical relativity are in
agreement with linear black hole perturbation theory for the angular momentum,
{\it they are also in agreement for the radiated energy}. In other words: when
angular momenta from $l=m=2$ and $l=m=3$ agree, also the masses do. In our
opinion this result is non-trivial, and it lends support to choosing this
``perturbation theory time'' (marked by arrows in the plots) as our best guess
to estimate the final black hole's parameters.

\begin{figure*}[ht]
\begin{center}
\begin{tabular}{cc}
\epsfig{file=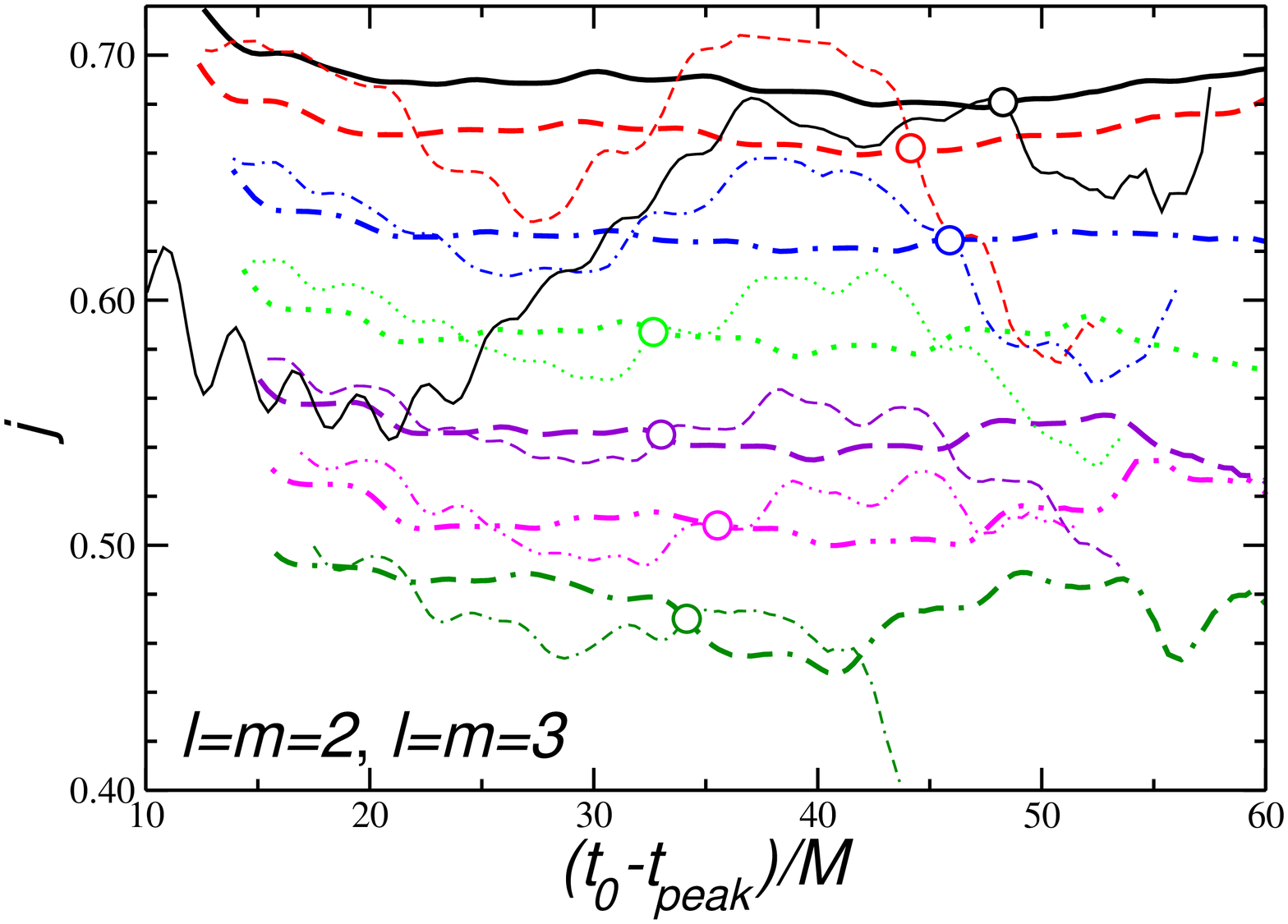,width=7cm,angle=-90} &
\epsfig{file=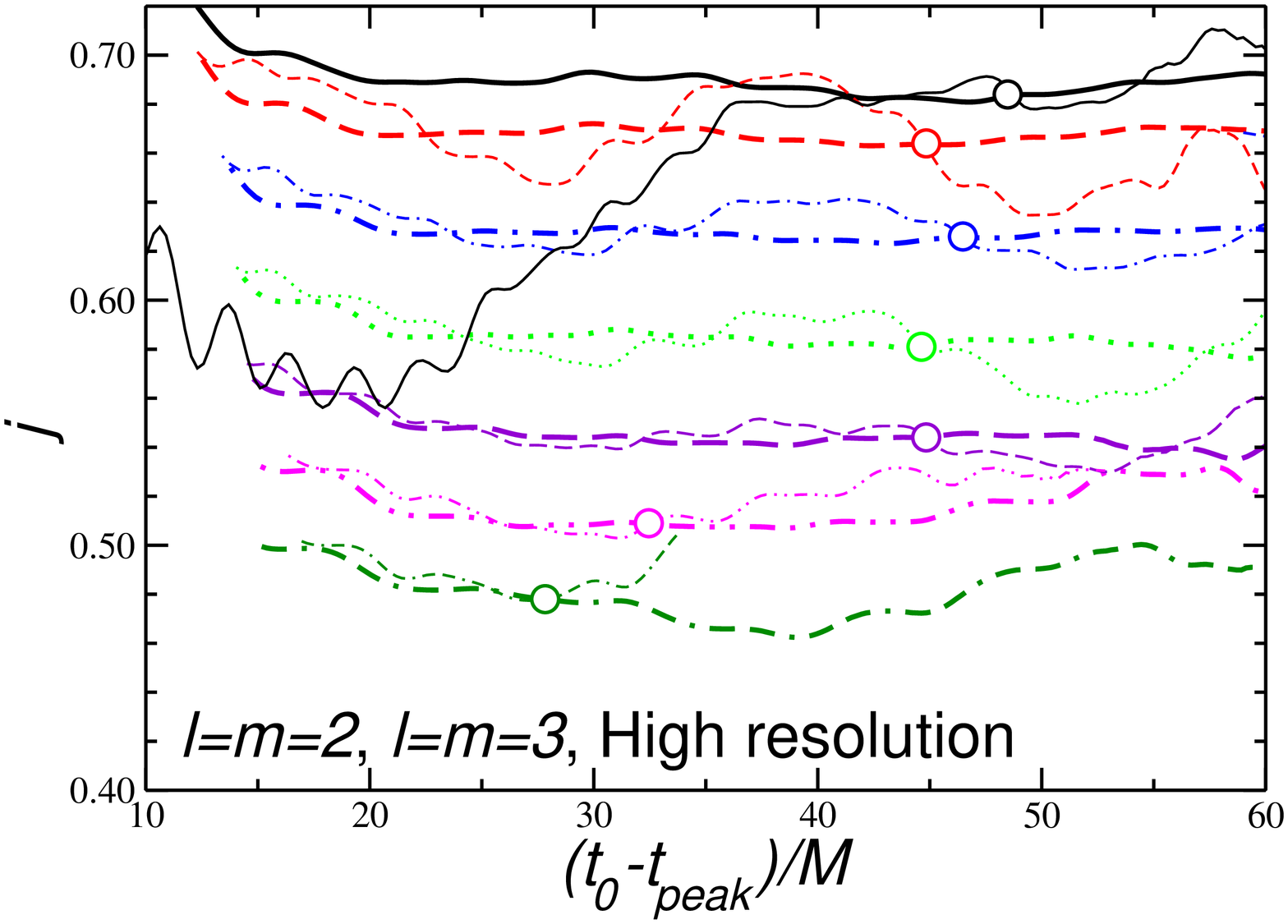,width=7cm,angle=-90} \\
\end{tabular}
\caption{Angular momentum estimated applying the  method to $l=m=2$ waveforms
 (thick lines) and to $l=m=3$ ($l=m=4$ in the case $q=1.0$) waveforms (thin
 lines). Hollow circles mark the ``perturbation theory time'' for each mass
 ratio (see text). From top to bottom, different linestyles refer to
 $q=1.0,~1.5,~2.0,~2.5,~3.0,~3.5$ and $4.0$, respectively.
 \label{jq-mp}}
\end{center}
\end{figure*}

In Fig.~\ref{jq-mp} we show the performance of the MP method in estimating
angular momenta for different mass ratios:
$q=1.0,~1.5,~2.0,~2.5,~3.0,~3.5,~4.0$. Increasing the resolution produces a
flattening of all curves, the effect being more pronounced for large mass
ratios. Remarkably, we find that the angular momenta and masses predicted from
fitting different multipolar components agree at some well-specified time {\it
  for all mass ratios}.
\begin{table}[ht]
  \centering \caption{\label{tab:FitResults-HR} Energy radiated $E_{\rm
      QNM}/M$ and final dimensionless angular momentum $j_{\rm QNM}$ as
    computed by a QNM fit of high-resolution D7 runs. We use the ``absolute''
    cutoff criterion for the fits. Results are presented using two fitting
    methods: MP and LM (the latter results are in parentheses). For ease of
    comparison, we also give the energy radiated as computed by wave
    extraction techniques. We denote by $t_{\rm PT}$ the ``perturbation theory
    time'' when the $l=m=2$ and $l=m=3$ predictions for mass and angular
    momentum are in agreement. For comparison we also show estimates of the
    radiated energy $E_{\rm tot}/M$ and of the final angular momentum $j_{\rm
      fin}$ obtained by subtracting the radiated energy and angular momentum
    from the initial ADM mass and angular momentum. Uncertainties are
    estimated using Richardson extrapolation as described in
    Sec.~\ref{setup}.}
\begin{tabular}{cccccccccccccccc}
\hline \hline
%
$q$ &
$E_{\rm tot}/M (\%)$ &
$E_{\rm QNM}/M (\%)$ &
$j_{\rm fin}$ &
$j_{\rm QNM}$ &
$\left (t_{\rm PT}-t_{\rm peak}\right )/M$ \\
\hline
1.0 &$3.718\pm0.069$ & $3.73\pm0.26$  & $0.688\pm0.002$ & $0.684\pm0.006$  & $48.5\pm2.3$ \\
    &                &($3.65\pm0.32$) &                 &($0.684\pm0.005$) &              \\
1.5 &$3.403\pm0.032$ & $3.35\pm0.02$  & $0.665\pm0.002$ & $0.664\pm0.003$  & $44.8\pm0.9$ \\
    &                &($3.23\pm0.41$) &                 &($0.665\pm0.008$) &              \\
2.0 &$2.858\pm0.055$ & $2.69\pm0.12$  & $0.626\pm0.003$ & $0.626\pm0.005$  & $46.5\pm4.0$ \\
    &                &($2.66\pm0.05$) &                 &($0.626\pm0.002$) &              \\
2.5 &$2.383\pm0.051$ & $2.43\pm0.55$  & $0.583\pm0.003$ & $0.581\pm0.009$  & $44.6\pm2.0$ \\
    &                &($2.36\pm0.37$) &                 &($0.581\pm0.008$) &              \\
3.0 &$2.000\pm0.035$ & $1.97\pm0.02$  & $0.543\pm0.002$ & $0.544\pm0.002$  & $44.8\pm1.8$ \\
    &                &($2.06\pm0.02$) &                 &($0.543\pm0.003$) &              \\
3.5 &$1.695\pm0.058$ & $1.63\pm0.06$  & $0.506\pm0.003$ & $0.509\pm0.002$  & $32.5\pm4.9$ \\
    &                &($1.47\pm0.02$) &                 &($0.512\pm0.002$) &              \\
4.0 &$1.451\pm0.034$ & $1.30\pm0.96$  & $0.473\pm0.013$ & $0.478\pm0.014$  & $27.8\pm26.6$\\
    &                &($1.50\pm0.55$) &                 &($0.473\pm0.008$) &              \\
\hline \hline
\end{tabular}
\end{table}
Now, in linearized theory different multipoles should be consistent with a
single linearly perturbed Kerr black hole. This means that the predictions for
$M$ or $j$ from one multipole should agree with the corresponding predictions
from any other multipole: lines in Figs.~\ref{eradconsistency} or \ref{jq-mp}
corresponding to different multipoles should lie on top of each other.
Accordingly, we will take the latest (in time) of these points as our ``best
guess'' to estimate the parameters of the final black hole, since by then the
background dynamical spacetime is as close as possible to a stationary Kerr
solution. In Table \ref{tab:FitResults-HR} we list the final angular momenta
and radiated energies extracted from a QNM fit at this ``optimal'' time,
comparing results against the corresponding estimates from wave extraction
techniques. The values thus obtained from the two independent methods agree
within the error estimates, indicating that QNM fits facilitate estimates of
the radiated energy and the final spin accurate to within a few percent or
better.

\subsection{\label{sec:rdstart} Criteria to determine the ringdown starting time}

There are important motivations to try and define the ringdown starting time
and to isolate, in a non-ambiguous way, the energy radiated in the ringdown
phase. For instance, from a detection-based point of view, the SNR of a
ringdown signal scales with the square root of the energy in the signal
\cite{fh,bcw}. To define the energy in ringdown waves we must somehow define
the ringdown starting time.  Being able to define the ringdown starting time
is also important when comparing numerical simulations with PN estimates of
the energy, angular and linear momentum. In fact, it has been suggested that
the discrepancy between PN estimates and numerical results for black hole
recoil is due to neglecting the ringdown in the former \cite{Gonzalez:2006md}.
To check the validity of this statement we must, again, define the starting
time of the ringdown phase.

Unfortunately, early studies in quasinormal ringing have established that
there is no such thing as ``the'' ringdown starting time (see eg.
\cite{Berti:2006wq} and references therein). In fact, the waveform can {\it
  never} be exactly described as a pure superposition of damped sinusoids: it
is always contaminated by noise or by other contributions (such as prompt
response or tails). This is essentially a consequence of the incompleteness of
QNMs.  However, from a practical viewpoint the signal {\it is} indeed
dominated by ringdown at some stage, and this is the reason why we can use
ringdown waves to estimate black hole parameters \cite{bcw}. The time span of
the ringdown phase can be defined in different ways, depending on context. In
the following we will discuss and implement three possible alternatives, two
of which have already been proposed in the past \cite{Nollert,Dorband:2006gg}.

\subsubsection{\label{sec:leastsquares} A least-squares approach}
A natural way to determine the QNM content of a given signal would be to
perform a non-linear fit of the data to an exponentially decaying sinusoid.
Here the unknown parameters are usually found in a least-squares sense, by
minimizing some functional of the form $\sum_{t=t_i} \left [h(t)-h^{\rm
    QNM}(t,\left\{\lambda\right\})\right ]^2$. In our specific case $h$ would
be the numerical data, sampled at instants $t=t_i$, and
$h^{\rm QNM}(t,\left\{\lambda\right\})$ is the model waveform (an exponentially
damped sinusoid).

The model depends on a set of unknown parameters $\left\{\lambda\right\}$ over
which the functional should be minimized.  It is of course very tempting to
treat the starting time as one of those parameters.  This is a possible way to
determine the ringdown starting time, and it served as the basis for the
proposal in \cite{Dorband:2006gg}.  There it was shown that the quality of a
QNM fit can be monitored by using some suitably defined norm.  In particular,
Ref.~\cite{Dorband:2006gg} proposed to use
\be\label{dorband-norm}
||N||(\tau_0)=
\f
{\int_{\tau_0}^{t_f} |\psi_{l\,,m}(t)-\psi_{\rm fit}^{l\,,m}|dt}
{\int_{\tau_0}^{t_f}|\psi_{l\,,m}(t)|dt}\,, \ee
where $\psi_{\rm fit}^{l\,,m}$ has been defined in Eq.~(\ref{fitfunc}).
Clearly $||N||\to 0$ when the fit is very close to the numerical waveform. The
idea is that the norm should have a local minimum when the ``trial'' starting
time $\tau_0$ tends to the ``true'' starting time, $\tau_0\to t_0$.

\begin{figure*}[ht]
\begin{center}
\epsfig{file=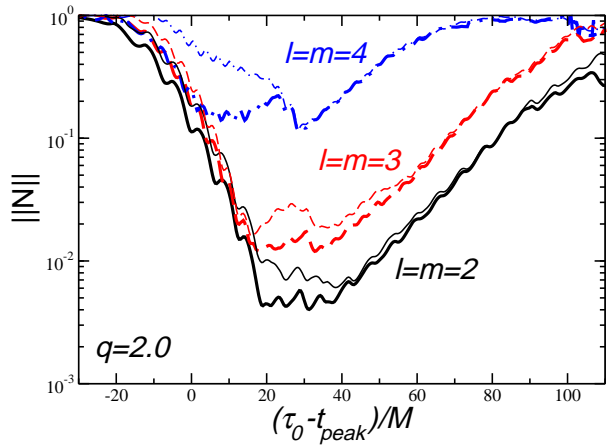,width=7cm,angle=-90} 
\caption{Norm (\ref{dorband-norm}) as a function of the trial starting time
 for the dominant components $(l,~m)=(2,~2),~(3,~3),~(4,~4)$ in a merger with
 $q=2.0$ (we consider a D7 run here). Thick lines are obtained by fitting the
 frequency each time we change $\tau_0$. Thin lines use the following  fixed
 values for the QNM frequency:
 $M\omega_R=0.51677$, $M\omega_I=0.08586$ for $l=m=2$,
 $M\omega_R=0.82210$, $M\omega_I=0.08571$ for $l=m=3$,
 $M\omega_R=1.12152$, $M\omega_I=0.08577$ for $l=m=4$.
 \label{dorband1_2}}
\end{center}
\end{figure*}

This idea works well for the classical perturbation theory problem of Gaussian
pulses scattered off a Kerr background \cite{Dorband:2006gg}, but
unfortunately it does not provide a very clear answer when tested on binary
black hole merger waveforms. The norm $||N||(\tau_0)$ for a binary with
$q=2.0$ is shown in Fig.~\ref{dorband1_2}, where it is computed in two
slightly different ways. The simplest way treats the QNM frequencies as known:
their values can be obtained once and for all by using Prony methods or
non-linear fits \cite{Berti:2007dg}, and kept fixed as we change $\tau_0$.
The second method achieves a marginal reduction of the norm by fitting for the
QNM frequency at each starting time $\tau_0$.

>From Fig.~\ref{dorband1_2} we see that the norm has some of the desired
properties. First of all, it grows as the quality of the QNM fit degrades: for
example, it is larger for the subdominant $(l,~m)$ components. In addition the
norm grows, as it should, when we try to extend the fit to encompass the
merger region, i.e. when $(\tau_0-t_{\rm peak})\lesssim 10$.

We find that the functional (\ref{dorband-norm}) has a minimum for most, but
not all of the waveforms. Even when it does have a minimum (as in the case of
Fig.~\ref{dorband1_2}) this minimum is very broad. In addition the norm
oscillates with a period which is basically the QNM period, and it has a
series of local minima and maxima.  The broad minimum and the oscillations
make it very hard to locate the starting time. Of course, the functional
(\ref{dorband-norm}) is by no means the only possibility. We experimented with
some alternative functional forms of the norm, but with no success. On a
positive note, the situation seems to improve when the waveform is computed at
large extraction radii \cite{tiglio}.

\subsubsection{Nollert's Energy Maximized Orthogonal Projection (EMOP)}
\label{EMOP}

A physically motivated notion of ringdown starting time time was introduced by
Nollert \cite{Nollert}. He realized that the problems with defining the
starting time arise immediately at the onset: QNMs are not complete and not
orthogonal with respect to any inner product, so a quantification of the
energy (and therefore of the starting time) going into each mode, using
standard ``basis expansion'' methods, is difficult (if not impossible). The
lack of orthogonality can be circumvented by formally defining an orthogonal
decomposition of the waveform into the contribution of one (or more) QNMs, and
some orthogonal remainder \cite{Nollert}:
\be h=h_{\parallel}+h_{\perp} \,.\ee
Here, $h_{\parallel}$ and $h_{\perp}$ are the part of $h$ parallel and
perpendicular, respectively, to a given QNM or a finite number $p$ of QNMs. We
therefore write
\be
h_{\parallel}=\sum_{i=1}^{p}a_{\parallel}^{(i)} h_{\rm QNM}^{(i)}\,,
\ee
where
\be
h_{\rm QNM}^{(i)}=\left\{ \begin{array}{ll}
            0   & \mbox{if $t<t_0$}\\
            e^{-\omega_i t}\sin{(\omega_r t+\phi)}    & \mbox{if $t>t_0$}\,.
\end{array}\right.
\label{qnmnollert}
\ee
is the QNM, assumed to start at some time $t_0$. The decomposition is achieved
using a standard orthogonal projection
\be \langle h_{\parallel},h_{\perp}\rangle=0\,, \ee
where the inner product, following arguments by Nollert \cite{Nollert}, is
defined in an energy-oriented way:
\be \langle\Psi,\Phi\rangle=\int \dot \Psi^* \dot \Phi dt\,. \ee
One can show that the energy ``parallel to the QNM component of the signal''
is given by
\be E_{\|}=\left|\int \dot h_{\rm QNM}^{(i)*} \dot h dt \right|^2 \left[\int \dot h_{\rm QNM}^{(i)*} \dot h_{\rm QNM}^{(i)}
dt\right]^{-1}\,. \ee
It is now meaningful to talk about (say) ``the fraction of energy going into
the first QNM''. This fraction obviously depends on the starting time $t_0$ in
Eq.~(\ref{qnmnollert}). Nollert observes that the ratio of the energy
``parallel to the QNM component'' to the total energy in the signal,
$E_{\|}/E_{\rm tot}$, has a maximum as a function of $t_0$. We can define the
ringdown starting time as the time $t_0$ corresponding to this {\it Energy
  Maximized Orthogonal Projection} (EMOP). In other words, according to
Nollert's criterion, the ringdown starting time $t_0=t_{\rm EMOP}$ is chosen
by looking for\footnote{Another conceivable definition would not use the total
  energy in the waveform $E_{\rm tot}$, but the energy in the waveform for
  $t>t_0$. It turns out that this quantity does not have a well-defined
  maximum. It is also possible to use a variable frequency in (\ref{emopexp}),
  in which case one could possibly obtain a larger maximum.  This method would
  be equivalent to matched filtering, which is discussed below.}
\be
\max_{t_0\,,\phi}
\frac{E_{\parallel}}{E_{\rm tot}}=
\max_{t_0\,,\phi}
\left(
\left|\int \dot h_{\rm QNM}^{(i)*} \dot h dt \right|^2
\left[\int \dot h_{\rm QNM}^{(i)*} \dot h_{\rm QNM}^{(i)} dt\right]^{-1}
\times \left[ \int \dot h^* \dot h dt\right]^{-1}
\right)
\label{emopexp}
\ee
The previous integral is evaluated separately for each polarization component.
To avoid memory effects, when we integrate $\Psi_4$ we fix the integration
constant so that $\dot h=0$ at the end of the simulation. We denote by $E_{\rm
  EMOP}$ the maximized energy parallel to the QNM component of the signal:
\be
E_{\rm EMOP}\equiv E_{\|}(t_0=t_{\rm EMOP})\,.
\ee

Using Prony methods or non-linear fits \cite{Berti:2007dg} we first determine
the QNM frequency and the damping time (for simplicity we consider a single
QNM). Then we compute $t_{\rm EMOP}$ and $E_{\rm EMOP}$ by maximizing
(\ref{emopexp}) over both $t_0$ and $\phi$.

\begin{table}[ht]
 \centering
 \caption{\label{tab:emop} EMOP data for $l=2$. Numbers separated by a
   comma correspond to the $+$ and $\times$ polarizations, respectively. The
   fraction of the total energy in the $l=2$ mode is about $42\%$ for all mass
   ratios. We find that, independently of mass ratio, the value of $t_{\rm
     EMOP}$ for a given polarization is generally at a fixed position relative
   to the maximum of the waveform's amplitude $t_{\rm peak}$. We measure this
   relative difference by $\Delta t_{\rm EMOP}\equiv t_{\rm peak}-t_{\rm
     EMOP}$, which turns out to be roughly independent of $q$.}
\begin{tabular}{cccccccc}
\hline  \hline
        $q$  & run  &$\frac{E_{\rm EMOP}}{E_{\rm tot}}$ &$\frac{\langle{t}_{\rm EMOP}\rangle}{M}$ &$\frac{\Delta t_{\rm EMOP}}{M}$ &$\frac{\langle{\Delta t_{\rm EMOP}}\rangle}{M}$   &$\frac{10^2 E_{\rm EMOP}}{M}$ &$\frac{\langle{10^2 E}_{\rm EMOP}\rangle}{M}$  \\
\hline
$1.0$   &   D7      & $0.41,0.42$     &  $225.5$      & $10.0,7.0$       &8.5   &$1.9,1.7$    &1.8\\
$1.5$ &   D7      & $0.41,0.43$      &  $227.2$     & $10.8,7.4$         &9.1   &$1.8,1.5$    &1.6\\
$2.0$   & D7      & $0.42,0.42$      &  $227.0$     & $9.9,6.9$          &8.4   &$1.4,1.2$    &1.3 \\
$2.5$ &   D7      & $0.41,0.43$      &  $229.2$     & $10.6,7.1$         &8.8   &$1.2,0.98$   &1.1 \\
$3.0$ &   D7      & $0.41,0.43$      &  $230.2$     & $11.2,7.8$         &9.5   &$0.95,0.82$  &0.88 \\
$3.5$ &   D7      & $0.40,0.43$      &  $232.0$     & $12.5,8.5$         &10.5  &$0.80,0.69$  &0.75  \\
$4.0$ &   D7      & $0.39,0.42$      &  $233.5$     & $13.3,9.3$         &11.3  &$0.68,0.59$  &0.64 \\
\hline
$2.0$ &   D8      & $0.40,0.41$      &  $453.0$     & $10.6,6.6$         &8.6   &$1.4,1.2$    &1.3 \\
$3.0$ &   D8      & $0.40,0.41$      &  $408.8$     & $11.0,7.6$         &9.3   &$0.95,0.81$  &0.88 \\
\hline \hline
\end{tabular}
\end{table}

\begin{figure*}[ht]
\begin{center}
\begin{tabular}{cc}
\epsfig{file=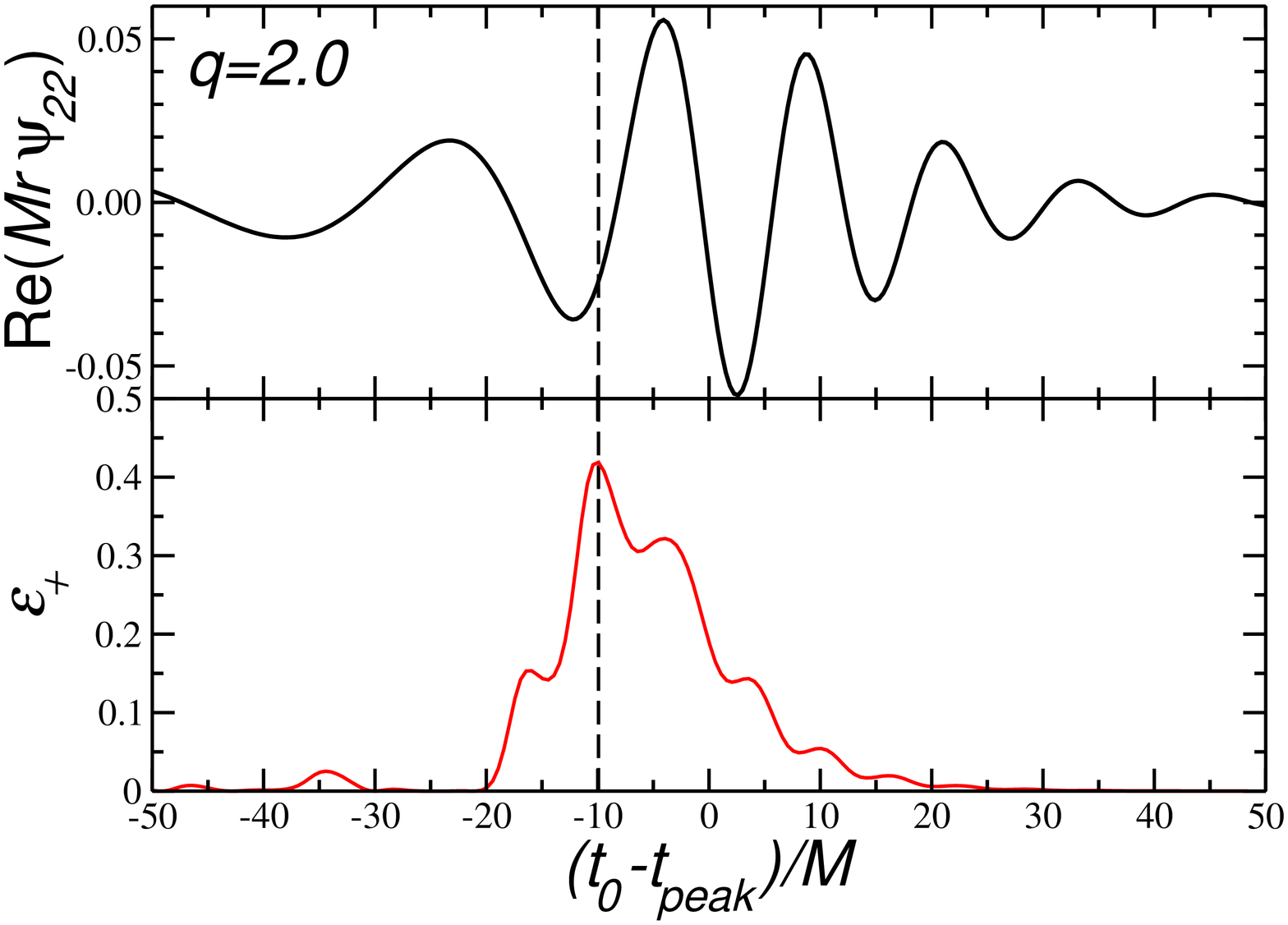,width=7cm,angle=-90}&
\epsfig{file=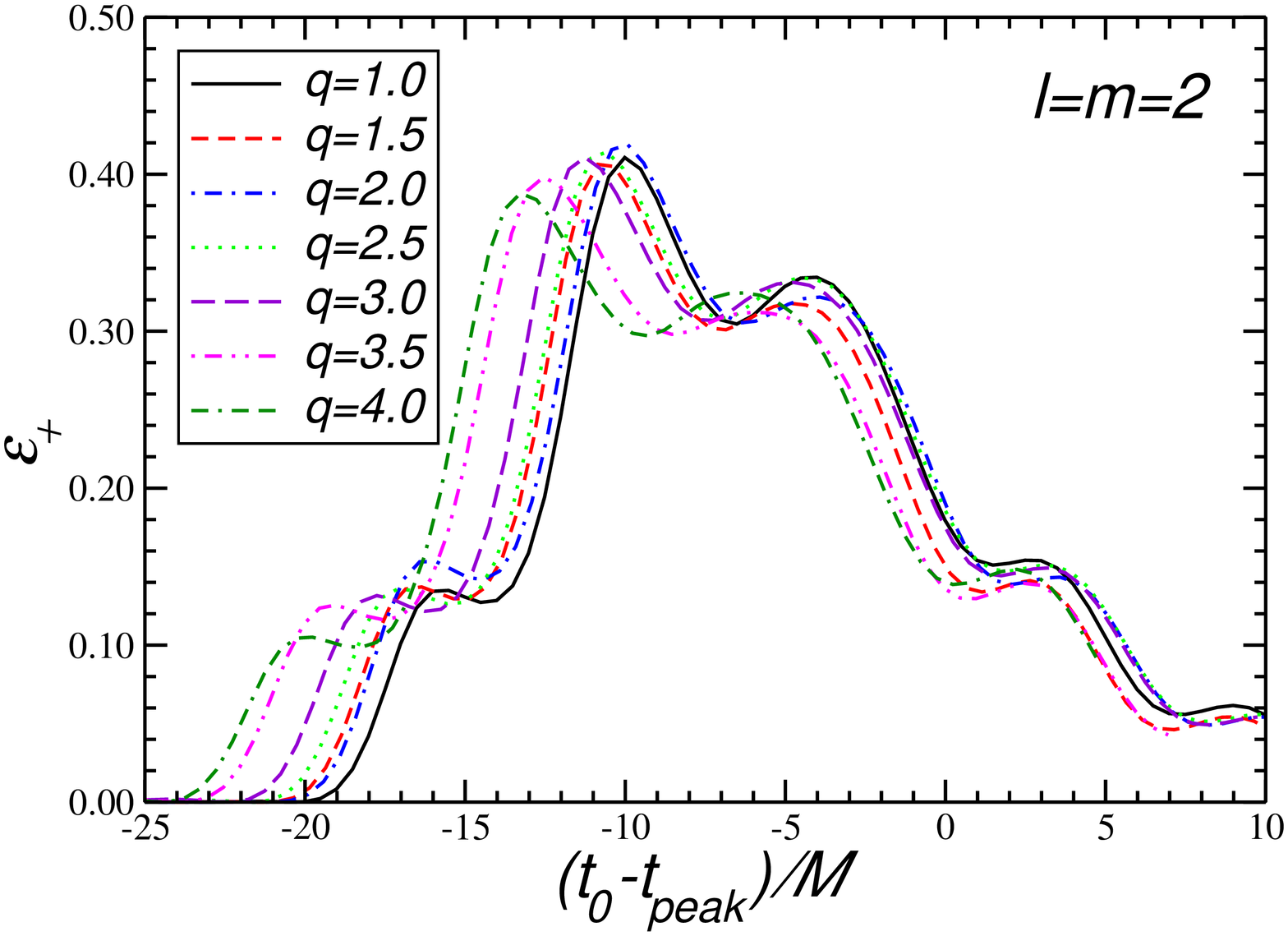,width=7cm,angle=-90}\\
\end{tabular}
\caption{EMOP for $l=2$ computed using run D7 (and high resolution). In the
  left panel we overplot $\epsilon_+$ and the actual waveform, marking $t_{\rm
    EMOP}$ by a vertical dashed line. In the right panel we show that results
  are quite insensitive to $q$: each line corresponds to a different mass
  ratio, and linestyles are the same as in Fig.~\ref{jq-mp}.
  \label{fig:EMOP}}
\end{center}
\end{figure*}

Our results for $l=m=2$ and run D7 are presented in Table \ref{tab:emop} and
Fig.~\ref{fig:EMOP}.  In the plots, $\epsilon_{+,\times}$ is the fraction of
energy radiated at $t>t_0$ in each of the two polarization components,
normalized to the total energy radiated in the simulation, and computed for
the value of the phase maximizing the EMOP.  The first thing to notice is that
there is a sharp maximum of the fractional energy going into ringdown. As seen
from Fig.~\ref{fig:EMOP}, $\sim 42\%$ of the total energy in the $l=2$ merger
waveform goes into ringdown. The results differ (very) slightly depending on
the chosen polarization state.

In Table \ref{tab:emop} we measure the ringdown starting time $t_{\rm EMOP}$
relative to the peak in $|Mr\,\psi_{22}|$, i.e., we compute $\Delta t_{\rm
  EMOP}\equiv t_{\rm peak}-t_{\rm EMOP}$. We see that $\Delta t_{\rm EMOP}$ is
basically constant for all mass ratios {\it and for all runs}, corresponding
to different initial separation of the binary.  This is an important
consistency test on the results. Notice also that $t_{\rm EMOP}$ is located
{\it before} the peak location.

EMOP times for the two polarizations are displaced by about $3M$ for run D7.
To define a unique ringdown starting time we take the average of both
polarizations (in Table \ref{tab:emop}, an average over the two polarizations
is denoted by angular brackets). Using this average starting time we can
define an energy radiated in ringdown, also shown in Table \ref{tab:emop}. We
find the following formula to be a good fit for the energy in the $l=2$ mode:
\be \frac{E_{\rm EMOP}}{M}=0.271 \frac{q^2}{(1+q)^4}\,,\quad l=2 \,.\ee

\begin{table}[ht]
 \centering
 \caption{\label{tab:emopl3} EMOP data for $l=3$. In this Table, by
   ``peak'' we mean the peak in the amplitude of the $l=3$ mode. Numbers
   separated by a comma correspond to the $+$ and $\times$ polarizations,
   respectively.  The fraction of the total energy in the $l=3$ mode is about
   $44\%$ for all mass ratios.}
\begin{tabular}{cccccccc}
\hline \hline
        $q$  & run  &$\frac{E_{\rm EMOP}}{E_{\rm tot}}$ &$\frac{\langle{t}_{\rm EMOP}\rangle}{M}$ &$\frac{\Delta t_{\rm EMOP}}{M}$ &$\frac{\langle{\Delta t_{\rm EMOP}}\rangle}{M}$   &$\frac{10^4E_{\rm EMOP}}{M}$ &$\frac{\langle{10^4E}_{\rm EMOP}\rangle}{M}$  \\
\hline
$1.5$ &   D7      & $0.44,0.45$      &  $233.0$     & $4.4,6.4$         &5.4  &$3.5,3.9$  &3.7\\
$2.0$ &   D7      & $0.45,0.45$      &  $230.5$     & $7.4,5.4$         &6.4  &$7.7,6.9$  &7.3 \\
$2.5$ &   D7      & $0.44,0.45$      &  $232.5$     & $8.2,6.2$         &7.2  &$9.6,8.6$  &9.1 \\
$3.0$ &   D7      & $0.44,0.45$      &  $234.0$     & $8.4,6.4$         &7.4  &$10,9.1$ &9.6 \\
$3.5$ &   D7      & $0.43,0.45$      &  $238.2$     & $4.8,7.3$         &6.0  &$7.8,9.2$  &8.5  \\
$4.0$ &   D7      & $0.43,0.45$      &  $240.2$     & $4.8,7.3$         &6.0  &$7.5,8.8$  &8.1 \\
\hline
$2.0$ &   D8      & $0.44,0.45$      &  $456.5$     & $7.1,5.1$         &6.1  &$7.6,6.8$  &7.2  \\
$3.0$ &   D8      & $0.45,0.44$      &  $412.7$     & $4.8,7.3$         &6.1  &$8.8,10$ &9.4 \\
\hline \hline
\end{tabular}
\end{table}
Results for $l=3$ follow the same pattern (see Table \ref{tab:emopl3}). The
average $\langle{t}_{\rm EMOP}\rangle$ for $l=3$ is located about $6M-7M$
after the average $\langle{t}_{\rm EMOP}\rangle$ for $l=2$. The following
formula provides a good fit for the energy in the $l=3$ mode:
\be
\frac{E_{\rm EMOP}}{M}=0.104 \frac{q^2(q-1)^2}{(1+q)^6}\,, \quad l=3\,.
\ee

If we take $t_{\rm EMOP}$ for $l=2$ as the fiducial ringdown starting time, we
can compute the energy, angular and linear momentum radiated during the
ringdown phase (as described by the EMOP). The results of this calculation are
listed in Table \ref{tab:summary}, and fitting formulas are provided in
Appendix \ref{app:postplunge} (see in particular Table \ref{tab:ISCO-EMOP}).

\subsubsection{A detection-based approach: energy deposited in matched filters}
\label{MFstart}

As we already stated QNMs do not form a complete set, so the signal will
always comprise quasinormal ringing plus some other component (such as prompt
response or tails). However, in most practical applications we are only
interested in some ``fairly good approximation'' to the ringdown waveform. The
notion of ``fairly good'' must be defined according to the specific context.

A possible definition, based on theoretical considerations, was introduced in
the previous Section.  Here we propose an alternative, practical definition of
the ringdown phase from a detection perspective.  Detection of ringdown waves
is likely to be achieved through matched filtering
\cite{fh,bcw}. The technique works by cross-correlating the
detector's output against a set of theoretical templates. It can be shown that
the maximum SNR is achieved when the template is equal in form to the
detector's output (hence the name matched filtering).  Matched filtering is the
method of choice to search for ringdown waves: it is quasi-optimal and
inexpensive, in the sense that it achieves the maximum SNR with a relatively
small number of templates or filters.

Now, for the purpose of a matched filtering detection, the ringdown definition
{\it must} be related to the use of ringdown templates. The relevant question
is therefore: what is the maximum SNR attainable through the use of a filter
which is a pure damped sinusoid?
By definition, given the numerical waveform $h(t)$, the SNR $\rho$ is
\be \rho=\max_{\left\{\lambda\right\},t_0}
\frac{(T(\left\{\lambda\right\},t_0)|h)}
{\sqrt{(T(\left\{\lambda\right\},t_0)|T(\left\{\lambda\right\},t_0)}}\,,\quad
(h_1|h_2)\equiv2\int_0^{\infty}\frac{h_1^*(f)h_2(f)+h_1(f)h_2^*(f)}{S_h(f)}\,,
\label{SNR}
\ee
where the template $T(\left\{\lambda\right\},t_0)$ is
\be
T(\left\{\lambda\right\},t_0)=
\left\{
\begin{array}{ll}
e^{-\omega_i^T (t-t_0)}\sin(\omega_r^T t+\phi^T)\,,
& \mbox{if $t\geq t_0$}\,,\\
0
& \mbox{if $t<t_0$}\,.
\end{array}\right.
\ee
$S_h(f)$ is the noise spectral density of the detector and
$\left\{\lambda\right\}$ is a set of parameters characterizing the templates.
The procedure is now simple: we ``slide'' this template backwards (starting at
large $t_0$ and decreasing it progressively) across the numerical waveforms,
and determine the maximum of the convolution (\ref{SNR}). A good initial guess
for the template parameters
$\left\{\lambda\right\}=(\omega_i^T\,,\omega_r^T\,,\phi^T)$ can be obtained
with Prony methods \cite{Berti:2007dg}.

As expected $t_0$ will depend on the observer, i.e., on the detector being
used, through the noise spectral density $S_h(f)$. In practice, however, the
dependence on the detector is usually very weak, since in general the largest
contribution to the convolution integral is near the resonant frequency
$\omega_r$. Thus, for all practical purposes, the detectors behave as if the
noise were white: the spectral density $S_h(f)$ can be approximated as
constant and moved out of the integral. This assumption also allows one to
sidestep the computation of the Fourier transform of the waveforms: by
Parseval's theorem, the frequency integral can be turned into a time integral.
A more complete analysis, taking into account the full structure of the
detector's noise, is in preparation.

A possible notion of effective ringdown starting time $t_{\rm MF}$ according
to a matched filter, which is useful to make contact with previous SNR
calculations \cite{bcw}, can be given simply as follows\footnote{Nollert's
  ``theoretical'' definition, explained in the previous Section, is not too
  dissimilar from a ``detection-oriented'' definition. Indeed, expression
  (139) in \cite{Nollert} can be interpreted as the fitting factor between
  actual waveforms and ringdown templates (for white noise).  If we take the
  ringdown frequencies as unknown parameters and choose them to maximize the
  EMOP (\ref{emopexp}), the results we get are very close to the present
  matched-filtering criterion.}.  Define the effective starting time $t_{\rm
  MF}$ as the instant for which
\be
\rho=\sqrt{(h_{t_{\rm MF}}|h_{t_{\rm MF}})}\,,\quad \quad
h_{t_{\rm MF}}\equiv\left\{
\begin{array}{ll}
            T(\{\lambda\},t_{\rm MF})   & \mbox{if $t\geq t_{\rm MF}$}\,,\\
            0    & \mbox{if $t<t_{\rm MF}$}\,,
\end{array}\right. \label{defstart}\ee
where $\rho$ is computed from Eq.~(\ref{SNR}). Notice that, in general,
$t_{\rm MF}$ does {\it not} coincide with the instant at which the convolution
between the signal and the template has a maximum.  By using
Eq.~(\ref{defstart}) the SNR can be expressed in terms of energy in the actual
signal. This is a common approach in engineering, introduced in the context of
gravitational wave detection by Flanagan and Hughes \cite{fh} (see also
\cite{bcw}).

\begin{table}[ht]
  \centering \caption{\label{tab:mergerrd} Estimated, polarization-averaged
    effective starting times $t_{MF}$ and energy radiated in ringdown from a
    matched-filter detection perspective, as functions of mass ratio. The
    listed energies should be taken as rough estimates,
    depending on the number of filters one is willing (and able)
    to use. We list also $\Delta t_{\rm MF}$, which is the ``effective''
    starting time as measure from the peak of the $l=m=2$ waveform: $\Delta
    t_{\rm MF}\equiv t_{\rm peak}-t_{\rm  MF}$.}
\begin{tabular}{cccc}
\hline \hline
$q$           &$\langle t_{\rm MF}/M\rangle$
&$\langle \Delta t_{\rm MF}/M \rangle$
& $\langle E_{\rm MF}/M\rangle$      \\
\hline
1.0    & 207             &27.0                    & 0.028\\
1.5    & 208             &28.4                    & 0.026\\
2.0    & 209             &26.4                    & 0.021\\
2.5    & 209             &29.1                    & 0.018\\
3.0    & 210             &29.8                    & 0.015\\
3.5    & 211             &31.5                    & 0.012\\
4.0    & 212             &32.8                    & 0.011\\
\hline \hline
\end{tabular}
\end{table}

The detection-based criterion, when applied to the merger waveforms considered
in this paper, yields the results shown in Table \ref{tab:mergerrd}.  From the
above discussion, it is clear that the values we list for the energy radiated
during ringdown are effective energies measured by the detector.  These
correspond to the values used in data analysis (see for instance \cite{bcw}).
From the Table we see that the effective energy radiated in ringdown for an
equal mass merger is $\sim 3\%$, in very good agreement with the
``guesstimate'' by Flanagan and Hughes \cite{fh}, which has often been used in
the literature to compute SNRs and measurement errors. We also note that this
value is much larger than the energy estimated by the EMOP, typically twice as
large.  This happens because the filter is looking for the maximum
correlation, usually implying that the best-match parameters ($\omega_r$ and
$\omega_i$) will differ significantly from the true signal parameters.

Also notice that different polarizations yield slightly different energies and
starting times.  For instance, for equal mass mergers, we get $t_{\rm MF}\sim
205$ and $t_{\rm MF}\sim 208$ for the plus and cross polarizations,
respectively. If we average over polarization states, this yields an effective
radiated energy of $\sim 2.8\%$.

We also point out that the amount of energy depends on the parameter space to
be searched. In principle, the correlation (\ref{SNR}) is to be maximized over
all possible values of $\omega_r\,,\omega_i$. In practice this would lead to a
very large number of filters, so we must choose reasonable cutoffs on the
parameters.  For instance, in black hole ringdown searches one looks for modes
with a quality factor typically smaller than $\sim 20$. It may be possible to
increase the SNR and the amount of effective energy in ringdown by enlarging
the parameter search (this would also allow us to search for ringdown modes of
other objects, such as neutron stars or boson stars). A discussion of these
issues will be presented elsewhere.

To conclude this Section, we point out that a fit of the total effective
energy radiated in ringdown, according to a matched filtering criterion, is:
\be
\frac{E_{\rm MF}}{M}\approx 0.44\frac{q^2}{(1+q)^4} \,.
\ee


\section{Conclusions and outlook}

The present study of binary black hole waveforms is, in many ways, only
preliminary. The following is a partial list of important open problems.

\begin{center}
{\bf Using ``hybrid'' waveforms in data analysis}
\end{center}

The present study explored the physical properties of numerical waveforms and
their relation with analytical methods. Our focus has been on providing
analytical insight into the structure of the waveforms. For this reason, we
deliberately avoided problems at the interface between numerical relativity
and data analysis (see Ref.~\cite{Baumgarte:2006en} for some steps in this
direction). We strongly believe that an analytical understanding of the
numerical simulations will be useful, or even {\it necessary}, to bridge the
gap between the (daunting) numerical task of generating waveforms, and the
injection of these waveforms into a data analysis pipeline.

The PNQC approximation studied in this paper provides a concrete example. We
showed that the physical content of any given simulation can be reproduced
quite accurately by substituting the orbital frequency $\Omega$ in the
dominant waveform amplitudes, Eqs.~(\ref{dominant-psis}). These ``hybrid''
PNQC waveforms can be used to create simple but accurate templates, and to
interpolate between numerical waveforms with different physical parameters.

Despite the recent progress in numerical relativity, simulations are still
computationally expensive. Hybrid template families could be injected in LIGO,
or used in connection with LISA simulators in future rounds of the Mock LISA
Data Challenges \cite{Arnaud:2007vr}.  Semianalytical waveforms may
significantly reduce the number of simulations needed for detection and
parameter estimation, and they should be particularly useful when spins are
included in the model.

\begin{center}
{\bf Removing spurious eccentricity and including additional physical parameters}
\end{center}

Our study clearly shows that the simulations have some small, but
non-negligible, eccentricity.  The eccentricity shows up as a typical
modulation of all physical quantities of interest: the punctures' orbital
velocity (Fig.~\ref{velratio}), the binary's orbital frequency
(Fig.~\ref{omegas}), the energy and angular momentum fluxes
(Fig.~\ref{fig:flux} and Fig.~\ref{fig:fluxj}), and so on. Measuring this
spurious eccentricity, and possibly removing it by fine-tuning initial data,
is an important open problem
\cite{Berti:2006bj,Miller:2003pd,Pfeiffer:2007yz}.  Incidentally, the study of
{\it truly} eccentric binaries could be relevant for massive black hole
binaries to be observed by LISA \cite{Berti:2006ew}.

In the present study we completely neglect spins. There is mounting evidence,
based for example on recent studies of binary black hole recoil, that spins
will have a dramatic effect on the inspiral-merger-ringdown transition. An
extension of our study to spinning, precessing black hole binaries is urgently
needed.

\begin{center}
{\bf Stitching numerical and analytical waveforms}
\end{center}

For reasons of space, we decided not to address the important problem of
comparing the PN phase evolution with the numerical phase evolution. This
problem is central to connect the early inspiral phase with the merger phase,
and it is a topic of active investigation. Since numerical evolutions show
signs of eccentricity, comparisons of the phase evolution may benefit from the
inclusion of eccentricity in the PN models as well.

Another active research field concerns the problem of ``stitching'' PN and
numerical waveforms. For the purpose of this stitching, do we need the full PN
waveforms, or does the restricted PN approximation (including PN corrections
in the phase, but not in the amplitude) work well enough? Does the number of
cycles to be simulated numerically depend on the mass ratio and other physical
parameters (eg. the spins)? We plan to return to these problems in the future.

\begin{center}
{\bf Bridging the gap with black hole perturbation theory}
\end{center}

Computational resources and resolution limitations reduce the accuracy of
numerical simulations for large mass ratios. Unfortunately, many astrophysical
black hole binaries could have mass ratio $q=10$ or larger (see eg.
\cite{Berti:2006ew} and references therein). It is important to determine the
maximum value of $q$ that {\it should} be simulated in numerical relativity,
or equivalently, the smallest value of $q$ for which black hole perturbation
theory can be considered adequate for detection and/or parameter estimation.
In Appendix \ref{app:pointparticles} we collect some results that may be
useful in this context.

\begin{center}
{\bf Astrophysics and gravitational wave detection}
\end{center}

The most interesting applications of our results should be in astrophysics and
gravitational wave detection. For example, the multipolar analysis of the
radiation performed in this paper can be used to determine the cosmological
distance at which we can test the no-hair theorem with LISA, LIGO or Virgo.
Future extensions of this analysis to spinning binaries could also predict the
parameter range (mass ratio, spin magnitudes and directions) in which the
recoil velocity is astrophysically relevant, and the probability for these
regions of the parameter space to be populated in astrophysical scenarios.

\section*{Acknowledgments}
We are particularly grateful to El-Hadi Djermoune for providing us with MATLAB
codes implementing the matrix pencil and Kumaresan-Tufts algorithms. We thank
Clifford Will for constant encouragement, and for computing the PN estimates
of energy, angular momentum and linear momentum radiated after plunge shown in
Figs.~\ref{plunge} and \ref{pplunge}.  We also thank John Baker, Luc Blanchet,
Alessandra Buonanno, Jaime Cardoso, Greg Cook, Pablo Laguna, Deirdre Shoemaker
and Manuel Tiglio for discussions, and the anonymous referee for a very useful
and detailed report. V.~C.~acknowledges financial support from CNPq - Conselho
Nacional de Desenvolvimento Cient\'{\i}fico e Tecnol\'ogico through grant
number 453463/2006-1.  J.~G.~and U.~S. acknowledge support from the ILIAS
network.  This work was supported in part by DFG grant SFB/Transregio 7
"Gravitational Wave Astronomy", by Funda\c c\~ao para a Ci\^encia e Tecnologia
(FCT) -- Portugal through project PTDC/FIS/64175/2006, by the National Science
Foundation under grant number PHY 03-53180, and by NASA under grant number
NNG06GI60 to Washington University. Numerical computations were performed at
HLRS Stuttgart and LRZ Munich.


\appendix

\section{Multipolar decomposition of the Post-Newtonian waveforms}
\label{app:multipoles}

Here we list the spin-weighted spherical harmonic components of a
PN expansion of the Weyl scalar $\Psi_4$. For $l=2$:
\begin{subequations}
\beq Mr\, \psi_{2\,, 2}e^{i\tilde{\phi}}&=& 
32\sqrt{\frac{\pi}{5}}\eta (M\Omega)^{8/3}
\left[1+\f{55\eta-107}{42}(M\Omega)^{2/3}+2\pi
(M\Omega)-\frac{2173+7483\eta-2047\eta^2}{1512}(M\Omega)^{4/3}\right.\nn\\
&+&\left. \left(\f{-107+34\eta}{21}\pi+\left(\varpi+\f{112}{5}\right)
i\eta\right)(M\Omega)^{5/3} \right]\,,
\label{c22e}\\
Mr\, \psi_{2\,, 1}e^{i\tilde{\phi}}&=&\f{8}{3}\sqrt{\frac{\pi}{5}}\eta \frac{\delta M}{M} (M\Omega)^{3} \left[
1+\f{20\eta-17}{28}(M\Omega)^{2/3}+\frac{2\pi-i(1+\ln 16)}{2}M\Omega\right.
\nn\\
&+&\left. \left(-\frac{43}{126}-\frac{509}{126}\eta+\frac{79}{168}\eta^2\right )(M\Omega)^{4/3} \right]\,.
\label{c21e} \eeq
\end{subequations}

For $l=3$:
\begin{subequations}
\beq Mr\, \psi_{3\,, 3}e^{i\tilde{\phi}}&=&27\sqrt{\frac{6\pi}{7}}\eta \frac{\delta M}{M}(M\Omega)^{3} \left[
1-(4-2\eta)(M\Omega)^{2/3}+\left [3\pi-i\left(\f{21}{5}-6\ln{(3/2)}\right)\right ]M\Omega\right.\nn\\
&+&\left.\left(\frac{123}{110}-\frac{1838}{165}\eta+\frac{887}{330}\eta^2\right)(M\Omega)^{4/3} \right] \,,
\label{c33e}\\
Mr\, \psi_{3\,, 2}e^{i\tilde{\phi}}&=&\f{32}{3}\sqrt{\frac{\pi}{7}}\eta
(M\Omega)^{10/3}\left[(1-3\eta)-\frac{193-725\eta+365\eta^2}{90}(M\Omega)^{2/3}
\right.\nn\\
&+&\left.\left(-3i+2\pi-\frac{6\eta}{5}(5\pi-11i)\right )M\Omega\right]\,,
\label{c32e}\\
Mr\, \psi_{3\,, 1}e^{i\tilde{\phi}}&=&\f{2}{3}\sqrt{\frac{\pi}{70}}\eta \frac{\delta M}{M}(M\Omega)^{3} \left[
1-\f{2(\eta+4)}{3}(M\Omega)^{2/3}+\frac{5\pi-i(7+10\ln
2)}{5}M\Omega\right.\nn\\
&+&\left.\left(\frac{607}{198}-\frac{136}{99}\eta-\frac{247}{198}\eta^2\right)(M\Omega)^{4/3} \right]\,,
\label{c31e} \eeq
\end{subequations}

For $l=4$:
\begin{subequations}
\beq Mr\, \psi_{4\,, 4}e^{i\tilde{\phi}}&=& \frac{1024}{9}\sqrt{\frac{\pi}{7}}\eta
(M\Omega)^{10/3}\left\{(1-3\eta)
-\frac{1779-6365\eta+2625\eta^2}{330}(M\Omega)^{2/3}\right.\nn\\
&+&\left. \left[4\pi-i\left(\f{42}{5}-8\ln(2)\right)
-\eta\left(12\pi-i\left(\f{1193}{40}-24\ln(2)\right)\right) \right](M\Omega) \right\}\,,
\label{c44e}\\
Mr\, \psi_{4\,, 3}e^{i\tilde{\phi}}&=&\f{162}{5}\sqrt{\f{\pi}{14}}\eta \frac{\delta M}{M}(M\Omega)^{11/3}
\left[(1-2\eta)+\left(-\frac{39}{11}+\frac{1267}{132}\eta-\frac{131}{33}\eta^2\right)(M\Omega)^{2/3}\right]\,,
\label{c43e}\\
Mr\, \psi_{4\,, 2}e^{i\tilde{\phi}}&=& \frac{32}{63}\sqrt{\pi}\eta
(M\Omega)^{10/3}\left[(1-3\eta)-\frac{1311-4025\eta+285\eta^2}{330}
(M\Omega)^{2/3}\right.\nn\\
&+&\left.\left(-\frac{21}{5}(1-4\eta)i+2\pi(1-3\eta)\right)M\Omega\right]\,,
\label{c42e}\\
Mr\, \psi_{4\,, 1}e^{i\tilde{\phi}}&=&\f{\sqrt{2\pi}}{105}\eta\frac{\delta M}{M} (M\Omega)^{11/3}
\left[(1-2\eta)+\left(-\frac{101}{33}+\frac{337}{44}\eta-\frac{88}{33}\eta^2\right)(M\Omega)^{2/3}\right]\,,
\label{c41e} \eeq
\end{subequations}

For $l=5$:
\begin{subequations}
\beq Mr\, \psi_{5\,, 5}e^{i\tilde{\phi}}&=&\f{3125}{12}\sqrt{\f{5\pi}{66}}\eta \frac{\delta
M}{M}(M\Omega)^{11/3}\left[(1-2\eta)+\left(-\frac{263}{39}+\frac{688}{39}\eta-\frac{256}{39}\eta^2\right)(M\Omega)^{2/3}\right]\,,
\label{c55e}\\
Mr\, \psi_{5\,, 4}e^{i\tilde{\phi}}&=& \f{4096}{45}\sqrt{\f{\pi}{33}}\eta \left(1-5\eta+5\eta^2\right
)(M\Omega)^{4}\,,
\label{c54e}\\
Mr\, \psi_{5\,, 3}e^{i\tilde{\phi}}&=&\f{81}{20}\sqrt{\f{3\pi}{22}}\eta \frac{\delta
M}{M}(M\Omega)^{11/3}\left[(1-2\eta)+\left(-\frac{69}{13}-\frac{8\eta}{39}(11\eta-58)\right)(M\Omega)^{2/3}\right]\,,
\label{c53e}\\
Mr\, \psi_{5\,, 2}e^{i\tilde{\phi}}&=& \f{64}{135}\sqrt{\f{\pi}{11}}\eta \left(1-5\eta+5\eta^2\right
)(M\Omega)^{4}\,,
\label{c52e}\\
Mr\, \psi_{5\,, 1}e^{i\tilde{\phi}}&=&\f{1}{180}\sqrt{\f{\pi}{77}}\eta\frac{\delta M}{M}
(M\Omega)^{11/3}\left[(1-2\eta)+\left(-\frac{179}{39}-\frac{4}{39}\eta(\eta-88)\right)(M\Omega)^{2/3}\right]\,,\label{c51e}
\eeq
\end{subequations}

For $l=6$:
\begin{subequations}
\beq Mr\, \psi_{6\,, 6}e^{i\tilde{\phi}}&=& \f{15552}{5}\sqrt{\f{\pi}{715}}\eta
\left(1-5\eta+5\eta^2\right)(M\Omega)^{4}\,,
\label{c66e}\\
Mr\, \psi_{6\,, 5}e^{i\tilde{\phi}}&=&\f{15625}{63}\sqrt{\f{5\pi}{429}}\eta(1-\eta)(1-3\eta)\frac{\delta M}{M}(M\Omega)^{13/3}\,, \label{c65e}\\
Mr\, \psi_{6\,, 4}e^{i\tilde{\phi}}&=& \f{16384}{495}\sqrt{\f{2\pi}{195}}\eta \left(1-5\eta+5\eta^2\right
)(M\Omega)^{4}\,,
\label{c64e}\\
Mr\, \psi_{6\,, 3}e^{i\tilde{\phi}}&=&\f{729}{385}\sqrt{\f{\pi}{13}}\eta(1-\eta)(1-3\eta)\frac{\delta M}{M}(M\Omega)^{13/3}\,, \label{c63e}\\
Mr\, \psi_{6\,, 2}e^{i\tilde{\phi}}&=& \f{64}{1485}\sqrt{\f{\pi}{13}}\eta \left(1-5\eta+5\eta^2\right)(M\Omega)^{4}\,, \label{c62e}\\
Mr\, \psi_{6\,, 1}e^{i\tilde{\phi}}&=&\f{1}{2079}\sqrt{\f{2\pi}{65}}\eta(1-\eta)(1-3\eta)\frac{\delta
M}{M}(M\Omega)^{13/3}\,, \label{c61e} \eeq
\end{subequations}

For $l=7$:
\begin{subequations}
\beq
Mr\, \psi_{7\,, 7}e^{i\tilde{\phi}}&=&\f{823543}{180}\sqrt{\f{7\pi}{4290}}\eta(1-\eta)(1-3\eta)\frac{\delta M}{M}(M\Omega)^{13/3}\,, \label{c77e}\\
Mr\, \psi_{7\,, 5}e^{i\tilde{\phi}}&=&\f{78125}{3276}\sqrt{\f{5\pi}{66}}\eta(1-\eta)(1-3\eta)\frac{\delta M}{M}(M\Omega)^{13/3}\,, \label{c75e}\\
Mr\, \psi_{7\,, 3}e^{i\tilde{\phi}}&=&\f{2187}{20020}\sqrt{\f{3\pi}{10}}\eta(1-\eta)(1-3\eta)\frac{\delta M}{M}(M\Omega)^{13/3}\,, \label{c73e}\\
Mr\, \psi_{7\,, 1}e^{i\tilde{\phi}}&=&\f{1}{108108}\sqrt{\f{\pi}{10}}\eta(1-\eta)(1-3\eta)\frac{\delta
M}{M}(M\Omega)^{13/3}\,, \label{c71e}
\eeq
\end{subequations}
Our $\theta$ is the same as the inclination angle $\iota$ in Blanchet {\it et
  al.} \cite{Blanchet:1996pi}. We recall that
\be e^{i\tilde{\phi}}\equiv e^{im( \int \Omega dt-2M\Omega\ln{\Omega/\Omega_0})}\,,\ee
and therefore the phase $\tilde{\phi}$ is defined up to an additive term $mc$,
with $c$ a constant factor. By fixing the constant to be $c=\pi/2$ we recover,
in the limit $\eta\to 0$, Poisson's results from perturbation theory.
If we include radiation reaction terms in the waveform by using Eq.~(27) of
\cite{Kidder:2007gz}, we find $\varpi=-24$. If we neglect those terms by
redefining the phase and using their Eq.~(32), then $\varpi=-8/7$.

The spin-weighted spherical harmonic components of $h_+\,,h_{\times}$ can be
obtained from the corresponding components of $\Psi_4$,
Eqs.~(\ref{c22e})-(\ref{c71e}), as
\be
\left(h_+-ih_{\times}\right)_{l\,,m}=
-\frac{1}{m^2\Omega^2}\psi_{l\,,m}\,.\ee
The resulting expressions do include the logarithmic corrections to the phase.
They are valid up to 2.5PN, with the only exception of the $l=m=2$ component,
which is given in Eq.~(\ref{h22m}).

\section{Estimates of the post-plunge energy, angular momentum and linear momentum}
\label{app:postplunge}

In this Section we give estimates of the energy, angular momentum and linear
momentum radiated in the last phases of a binary black hole inspiral.

\begin{table}[ht]
  \centering \caption{\label{tab:CAH} Energy, angular momentum and linear
    momentum emitted after the estimated time of CAH formation, as listed in
    Table \ref{tab:pars}. The CAH formation time is measured relative to the
    peak of the $l=m=2$ waveform: $\Delta t_{\rm CAH}\equiv \left (t_{\rm
        peak}-t_{\rm CAH}\right )/M$.}
\begin{tabular}{ccccccccccc}
\hline \hline
$q$  &run&  $\Delta t_{\rm CAH}/M$
&$E_{\rm CAH}/M$   &$J_{\rm CAH}/M^2$ & $E_{\rm CAH}M/J_{\rm CAH}$
&$10^4P_{x\,,{\rm CAH}}/M$& $10^4P_{y\,,{\rm CAH}}/M$ &$10^4P_{\rm CAH}/M$  \\
\hline
$1.0$ & D7 &19.0 &0.0259 &0.1261 &4.869 &0     &  0   &    0\\
$1.5$ & D7 &18.2 &0.0232 &0.1193 &5.142 &-2.56 & -0.28&   2.58\\
$2.0$ & D7 &17.9 &0.0193 &0.1004 &5.202 &-3.98 & -0.15&   3.98\\
$2.5$ & D7 &17.1 &0.0156 &0.0852 &5.462 &-5.01 & -0.20&   5.01\\
$3.0$ & D7 &16.6 &0.0128 &0.0716 &5.594 &-5.38 & -0.53&   5.41\\
$3.5$ & D7 &15.7 &0.0105 &0.0580 &5.524 &-5.62 & -1.34&   5.78 \\
$4.0$ & D7 &14.3 &0.0086 &0.0442 &5.139 &-5.78 & -2.30&   6.22\\
\hline
$2.0$ & D8 &17.7 &0.0192 &0.1055 &5.495 &4.15  & -0.23&   4.16\\
$3.0$ & D8 &15.2 &0.0125 &0.0493 &3.944 &0.28  & -5.90&   5.91\\
\hline \hline
\end{tabular}
\end{table}

Table \ref{tab:CAH} lists the energy, angular momentum and linear momentum
radiated after the time of CAH formation, estimated as the point where the
ratio of the radial and tangential puncture velocities $v_r/v_t>0.3$ (see
Section \ref{sec:transition}).

Given the uncertainties in our estimate of the CAH formation we also consider
another useful (if somewhat conventional) indicator of the regime of validity
of PN expansions: the Innermost Stable Circular Orbit (ISCO). The ISCO is
defined by the condition that the energy of the two-body system ${\cal E}$,
which is a function of the orbital frequency $\Omega$, has a minimum: $d{\cal
  E}/d\Omega=0$. Since ${\cal E}$ is only known as a PN series in $\Omega$,
the location of the ISCO depends on the PN order \cite{Buonanno:2002ft}.
\begin{table}[ht]
\centering \caption{\label{tab:ISCO2pn} ISCO data using the 2PN Taylor
  expansion of the energy. The ISCO time is measured relative to the
    peak of the $l=m=2$ waveform: $\Delta t_{\rm ISCO}\equiv \left (t_{\rm
        peak}-t_{\rm ISCO}\right )/M$.}
\begin{tabular}{cccccccccc}
\hline  \hline
$q$  &run &$M\Omega_{\rm ISCO}$ &$\Delta t_{\rm ISCO}/M$
&$E_{\rm ISCO}/M$   &$J_{\rm ISCO}/M^2$&$E_{\rm ISCO}M/J_{\rm ISCO}$
&$10^4P_{x\,,{\rm ISCO}}/M$& $10^4P_{y\,,{\rm ISCO}}/M$ &$10^4P_{\rm ISCO}/M$  \\
\hline
$1.0$   & D7   &0.137  &19.6 &0.0256  &0.1243  &0.206  & 0      &0        &0\\
$1.5$   & D7   &0.136  &19.2 &0.0232  &0.1181  &0.196  &-2.29   & -0.40   &2.32\\
$2.0$   & D7   &0.136  &18.6 &0.0192  &0.0994  &0.193  &-3.79   & -0.07   &3.79\\
$2.5$   & D7   &0.134  &18.2 &0.0156  &0.0853  &0.183  &-4.53   & -0.33   &4.54\\
$3.0$   & D7   &0.134  &17.9 &0.0129  &0.0727  &0.177  &-4.83   & -0.60   &4.86\\
$3.5$   & D7   &0.133  &17.9 &0.0108  &0.0619  &0.174  &-4.73   &-1.24    &4.89\\
$4.0$   & D7   &0.132  &17.6 &0.0092  &0.0524  &0.176  &-4.48   &-1.75    &4.81\\
\hline
$2.0$   & D8   &0.136  &18.8 &0.0197  &0.1068  &0.184 &-3.65   &-0.00     &3.65\\
$3.0$   & D8   &0.134  &17.9 &0.0130  &0.0551  &0.236  &-4.63   &-0.40    &4.65\\
\hline \hline
\end{tabular}
\end{table}
\begin{table}[ht]
\centering \caption{\label{tab:ISCO3pn} ISCO data using the 3PN Taylor
  expansion of the energy. The ISCO time is measured relative to the
    peak of the $l=m=2$ waveform: $\Delta t_{\rm ISCO}\equiv \left (t_{\rm
        peak}-t_{\rm ISCO}\right )/M$.}
\begin{tabular}{cccccccccc}
\hline \hline
$q$  &run &$M\Omega_{\rm ISCO}$ &$\Delta t_{\rm ISCO}/M$
&$E_{\rm ISCO}/M$   &$J_{\rm ISCO}/M^2$&$E_{\rm ISCO}M/J_{\rm ISCO}$
&$10^4P_{x\,,{\rm ISCO}}/M$& $10^4P_{y\,,{\rm ISCO}}/M$ &$10^4P_{\rm ISCO}/M$\\
\hline
$1.0$ &   D7      &0.129    &22.1 &0.0266  &  0.1231  &0.216&0      &  0       &    0\\
$1.5$ &   D7      &0.126    &22.8 &0.0245   & 0.1165 &0.210 &-1.74   &-1.08     &2.05\\
$2.0$ &   D7      &0.120    &23.8 &0.0208   & 0.0985  & 0.211 &-2.86   &-1.37     &3.17\\
$2.5$ &   D7      &0.116    &24.9 &0.0172   & 0.0850 &0.202 &-3.11   &-1.97     &3.68\\
$3.0$ &   D7      &0.112    &26.5 &0.0146   & 0.0737 &0.198 &-3.05   &-2.46     &3.92 \\
$3.5$ &   D7      &0.109    &27.5 &0.0123   & 0.0645 &0.191 &-2.70   &-2.78     &3.88 \\
$4.0$ &   D7      &0.107    &28.1 &0.0106   & 0.0568 &0.187 &-2.28   &-3.07     &3.82\\
\hline
$2.0$ &   D8      &0.120    &23.6 &0.0210   & 0.0940 &0.223 &+2.75   &+1.26    &3.02\\
$3.0$ &   D8      &0.112    &25.8 &0.0146   & 0.0872 &0.167 &-3.10   &-1.86     &3.62 \\
\hline \hline
\end{tabular}
\end{table}

In Tables \ref{tab:ISCO2pn} and \ref{tab:ISCO3pn} we list the orbital
frequency at the ISCO $M\Omega_{\rm ISCO}$ computed by including terms in the
energy function up to 2PN and 3PN, respectively.  Notice that 3PN corrections
{\it lower} the ISCO frequency for all mass ratios, the reduction being larger
for larger mass ratios\footnote{For comparison, the ISCO for point particles
  is at $r_0=6M$, or equivalently at an orbital frequency of $M\Omega
  =6^{-3/2}\simeq 0.068$.  Corrections to this point particle limit were
  worked out by Clark and Eardley \cite{ClarkEardley}, yielding a simple
  analytical estimate for large but finite mass ratios:
  $r_0/M=6q/(1+q)\,,M\Omega =r_0^{-3/2}$. For $q=4$ this yields $M\Omega
  \simeq 0.095$, not too far from the 3PN estimate of $0.107$.}.  We also list
the time location of the ISCO (relative to the peak in the amplitude of the
$l=m=2$ mode). We identified this time location as the instant when the ISCO
frequency equals the orbital frequency from our simulations, as estimated from
the gravitational wave emission of the dominant multipolar component $l=m=2$
(see Section \ref{freq-ests}): $M\Omega_{\rm ISCO}=M\omega_{D2}$. The 3PN ISCO
``absolute'' location in terms of the total simulation time is also shown in
Table \ref{tab:pars}, and it should be compared with the CAH formation
estimates in the same Table. As $q\to 1$ the CAH formation time is very close
to the 3PN ISCO time.  For large mass ratios, when one of the holes is very
small, the difference is larger, as expected on physical grounds.

Tables \ref{tab:ISCO2pn} and \ref{tab:ISCO3pn} also list the energy, angular
momentum and linear momentum emitted after the ISCO, with the ISCO location
estimated by PN methods. While the energy emitted is a robust quantity, with
very weak dependence on the PN order, angular and linear momenta are very
sensitive to a variation of the PN order from 2PN to 3PN (i.e., they are very
sensitive to small variations in the starting time of the integration). The
reason for this behavior is apparent from an inspection of
Figs.~\ref{fig:flux} and \ref{fig:fluxj}. While the energy flux is a smooth
function, even in the strong-field region, the angular momentum flux is a
strongly oscillating function of time.

The functional dependence of energy and angular momentum on mass ratio $q$ can
be inferred by combining the multipolar decomposition of the PN expansion
(Appendix \ref{app:multipoles}) with Eq.~(\ref{Flm}). We find that good fits
to the total angular momentum, energy, and multipolar energy distribution in
the dominant modes for times $t>t_0=t_{\rm ISCO}^{\rm 3PN}$ are:
\beq \left.J_{\rm ISCO}/M^2\right|_{t>t_0}
=j_{\rm tot} \frac{q^2}{(1+q)^4}\,,\label{JISCOfit}
\eeq
\beq
\label{enISCOfit}
\left.E_{\rm ISCO}/M\right|_{t>t_0}=
\epsilon_{\rm tot} \frac{q^2}{(1+q)^4}\,,\qquad
\left.E_{\rm ISCO\,,2}/M\right|_{t>t_0}=
\epsilon_2 \frac{q^2}{(1+q)^4}\,,\qquad
\left.E_{\rm ISCO\,,3}/M\right|_{t>t_0}= \epsilon_3
\frac{q^2(q-1)^2}{(1+q)^6}\,.
\eeq
with the fitting coefficients listed in Table \ref{tab:ISCO-EMOP}.

\begin{table}[ht]
  \centering
  \caption{\label{tab:ISCO-EMOP} Fitting coefficients for the energy and
    angular momentum emitted after the ISCO and EMOP times.}
\begin{tabular}{ccccc}
\hline \hline
$t_0$ & $j_{\rm tot}$ & $\epsilon_{\rm tot}$ & $\epsilon_2$ & $\epsilon_3$ \\
\hline
3PN ISCO & 2.029 & 0.421 & 0.397 & 0.168\\
EMOP     & 1.173 & 0.295 & 0.271 & 0.104\\
\hline \hline
\end{tabular}
\end{table}

In the same Table, for comparison, we also list the corresponding coefficients
for $t>t_0=t_{\rm EMOP}$. The ringdown and plunge phase are strongly related
with each other, and the numbers are roughly proportional.  According to the
EMOP criterion ringdown always starts after the ISCO. Therefore the post-EMOP
radiation of energy, angular and linear momentum is always smaller than the
corresponding radiation after the ISCO.

The PN expansion breaks down after the ISCO. An estimate of the linear
momentum emitted {\it after} the ISCO, within PN theory, was obtained in
\cite{Blanchet:2005rj} by integrating the PN linear momentum flux along a
plunge geodesic of the Schwarzschild metric. In \cite{Blanchet:2005rj}, the
integration is performed all the way from the ISCO ($r\simeq 6M$) to the
Schwarzschild horizon ($r\simeq 2M$). The energy and angular momentum radiated
after the ISCO can be computed using the same method, and they were kindly
provided to us by Clifford Will \cite{Cliff}.

\begin{figure*}[ht]
\begin{center}
\begin{tabular}{cc}
\epsfig{file=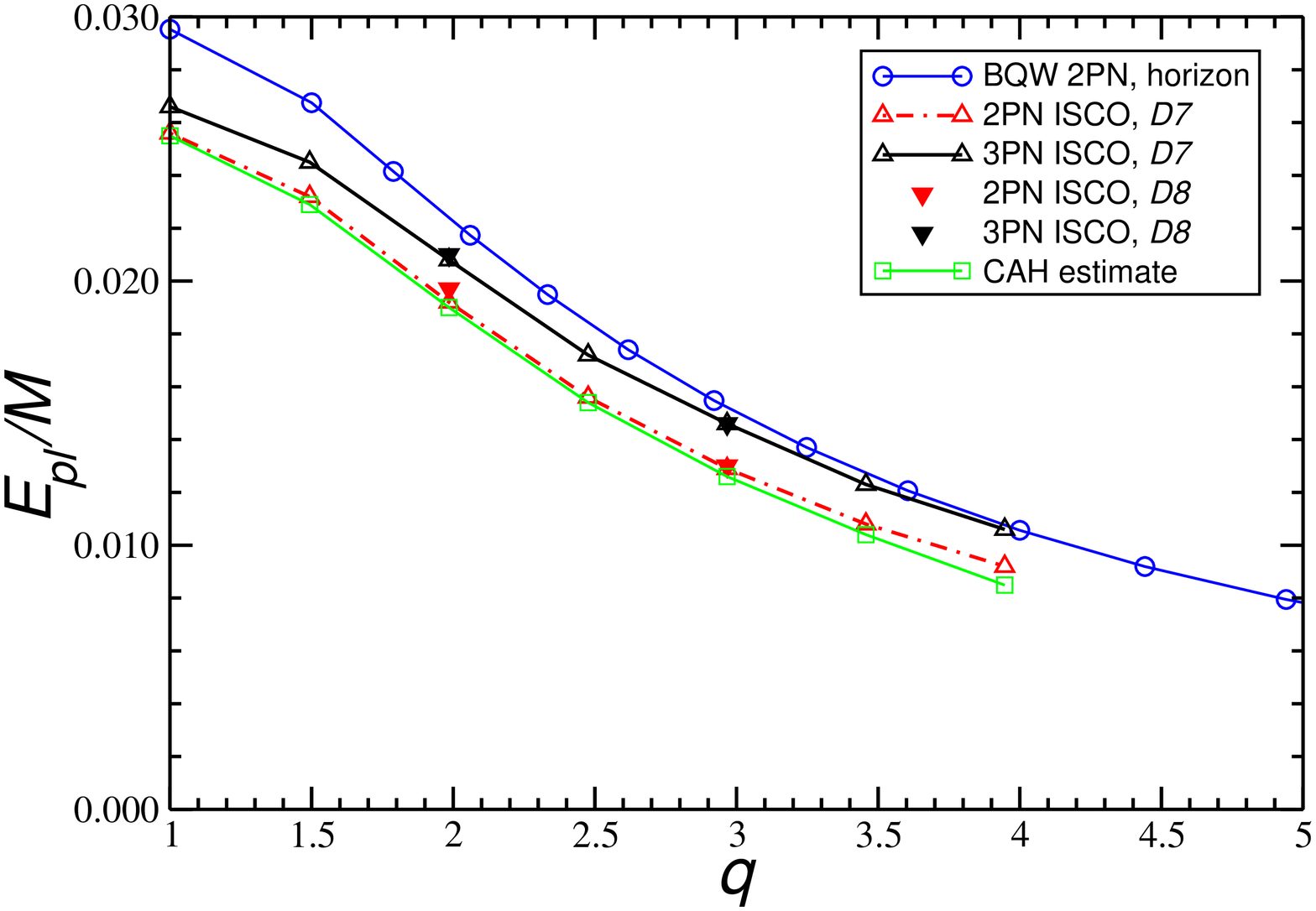,width=7cm,angle=-90} &
\epsfig{file=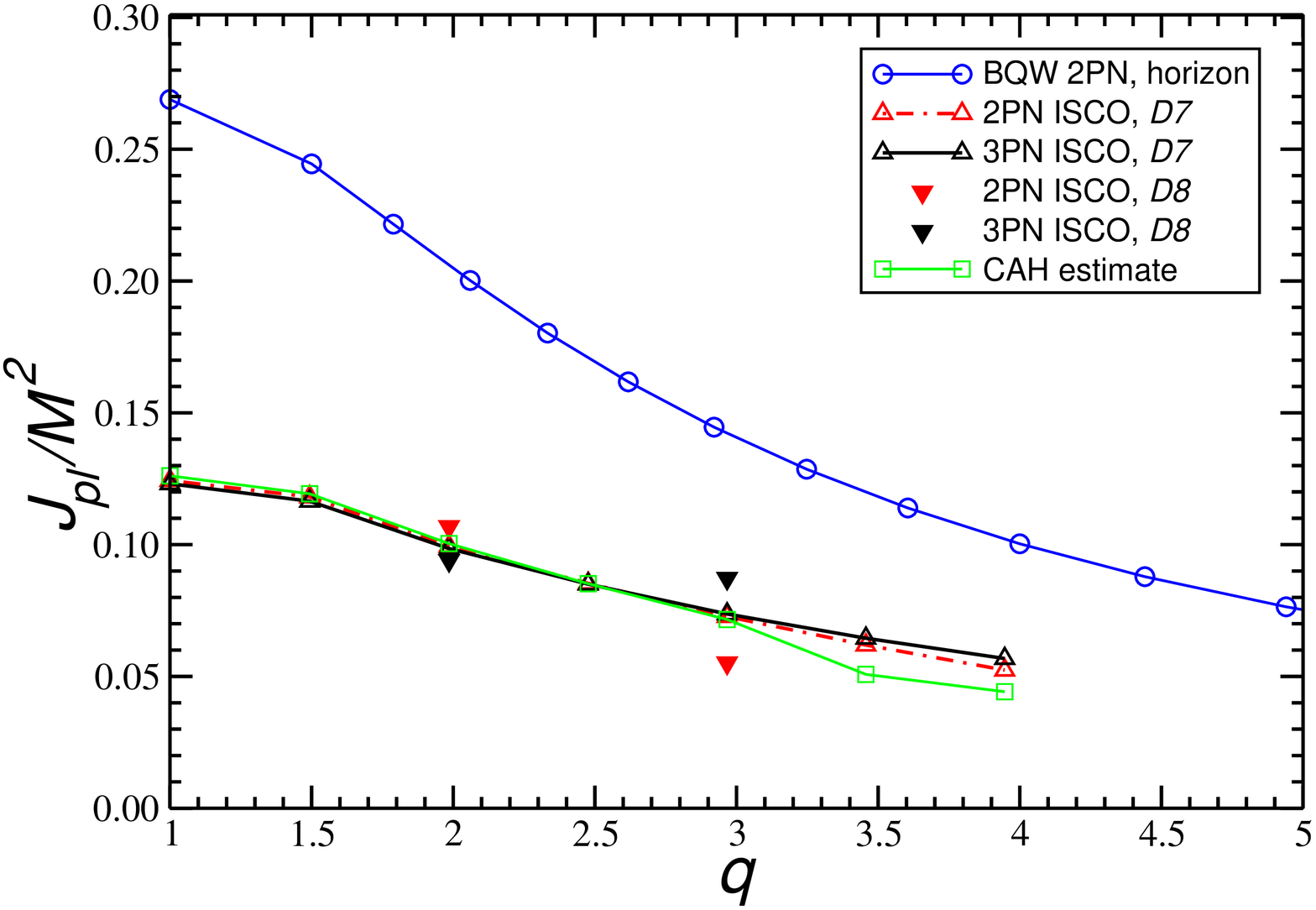,width=7cm,angle=-90} \\
\end{tabular}
\caption{Energy and angular momentum radiated in the plunge using the 2PN and
  3PN definitions of the ISCO are compared against the simple estimate by
  Blanchet {\it et al.} \cite{Blanchet:2005rj} (BQW in the legend). All
  estimates were computed using the D7 runs (except for the inverse
  triangles, which refer to D8 runs).
  \label{plunge}}
\end{center}
\end{figure*}

Results of a 2PN estimate of the energy and angular momentum radiated after
the ISCO are shown in Fig.~\ref{plunge} for different mass ratios, along with
different estimates of the corresponding quantities from our numerical
simulations.  In particular, we show numerical estimates of the energy and
angular momentum radiated after the 2PN and 3PN ISCO, and after the CAH
formation, as functions of the mass ratio.  To check the robustness of our
results against initial conditions we also considered two runs starting at
larger initial separations (namely, D8 simulations with $q=2.0$ and $q=3.0$).

Some comments are in order. The radiated {\it energy} from the simple PN
estimate is in surprisingly good agreement with numerical results, the
agreement getting better as we increase the PN order used to estimate the ISCO
location. The agreement is particularly good when we consider radiation
emitted after the 3PN estimate of the ISCO and relatively large mass ratios.
The agreement in the radiated angular momenta is much worse. This seems to be
a general feature when comparing PN estimates against numerical simulations.
For example, Fig.~2 and 3 in \cite{Berti:2006bj} show that the eccentricity
required to match PN predictions for a binary's angular momentum against
numerical calculations in quasiequilibrium is significantly larger than the
eccentricity required to match the corresponding energies. In the present
case, the disagreement is partially affected by the strongly oscillating
functional dependence of the angular momentum flux (see eg.
Fig.~\ref{fig:fluxj}).  This is confirmed by the fact that a relatively small
change in the initial separation (using D8 runs instead of D7 runs) produces a
significant change in the numerical estimate of the angular momentum radiated
after plunge. Given the large uncertainties associated with both numerical and
analytical estimates, we cannot draw reliable conclusions from the observed
disagreement.

It may be tempting to attribute the observed differences in the angular
momentum to the fact that the PN estimates of \cite{Blanchet:2005rj} neglect
the ringdown phase. We can naively try to correct for this effect by
integrating the energy and angular momentum fluxes from the ISCO {\em up to
  the CAH only}. However, this will result in serious disagreement for both
the radiated energy and angular momentum: they turn out to be extremely small,
especially for mass ratio $q\to 1$ (in this limit the CAH formation time and
the ISCO are very close, see Table \ref{tab:pars}).  Thus, ringdown alone
cannot explain the disagreement. A more detailed analysis, possibly combining
the PN approach and the close-limit approximation, is necessary.

\begin{figure*}[ht]
\begin{center}
\epsfig{file=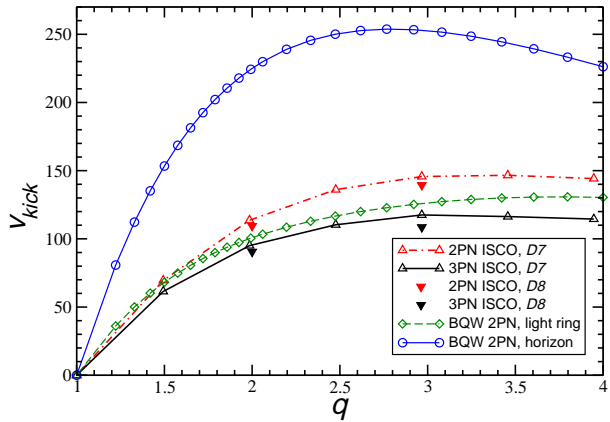,width=7cm,angle=-90}
\caption{The kick velocity accumulated after plunge using the 2PN and 3PN
  definitions of the ISCO is compared against the corresponding estimates by
  Blanchet {\it et al.} \cite{Blanchet:2005rj} (BQW in the legend). In the two
  BQW estimates the integration is truncated at the horizon or at the light
  ring, respectively. \label{pplunge}}
\end{center}
\end{figure*}
An interesting possibility is that the agreement between PN estimates and
numerical results could improve if the PN integration is truncated at the {\it
  light ring}, instead of integrating all the way to the horizon (as
originally done in \cite{Blanchet:2005rj}). The physical argument for
truncating at the light ring is that most of the radiation emitted after
$r\simeq 3M$ would be filtered by the potential barrier surrounding the black
hole, and that this potential barrier (for Schwarzschild black holes) has a
peak at the light ring \cite{Buonanno:2000ef}. Fig.~\ref{pplunge} shows that
truncating at the light ring sensibly improves the estimate of the {\it linear
  momentum} radiated after plunge, correspondingly improving the estimate of
the total kick velocity. Given the uncertainties involved in the
extrapolation, this may be little more than a coincidence. In any case this
problem is worth investigation, given the potential astrophysical relevance of
recoil velocities.

\section{Multipolar distribution of radiation for extreme mass ratios}
\label{app:pointparticles}

In this Appendix we collect the main results for the energy, angular momentum
and linear momentum radiated by particles falling into (rotating or
non-rotating) black holes. Our purpose is to provide a quick reference for the
extreme-mass ratio limit of numerical relativity simulations, to be compared
with present and future numerical relativity calculations of binaries with
large mass ratio (and possibly spin). Research in this direction is already
under way: for example, evolutions of large mass ratio binaries using finite
element methods can be found in \cite{Sopuerta:2005gz}.

\subsection{The energy radiated by plunging particles: non-rotating black holes}

The first investigations of particles plunging into black holes began with
Zerilli \cite{zerilli}, who laid down the perturbation formalism to analyze
gravitational radiation from a point-like particle with mass $m_p$ around a
Schwarzschild black hole with mass $M\gg m_p$. His analysis was completed by
Davis, Ruffini, Press and Price \cite{drpp} (hereafter referred to as DRPP),
who numerically computed the gravitational radiation generated when a small
particle at rest falls from infinity into a Schwarzschild black hole. DRPP
found that the total energy emitted in the process (in geometrized units) is
given by
\be
E_{\rm tot}=0.0104 (m_p^2/M) \,.
\ee

Detweiler and Szedenits \cite{detweilerszedenits} and Oohara and Nakamura
\cite{OoharaNakamura} generalized DRPP's results to particles plunging into a
Schwarzschild black hole with non-zero orbital angular momentum. In the
perturbation framework under consideration, the particle's trajectory as it
plunges down the hole, with zero velocity at infinity, is described by
\beq \theta &=&\pi/2\,,\quad \frac{dt}{d\tau}=\frac{1}{1-2M/r}\,,\quad \frac{d\phi}{d\tau}=\frac{L_z}{r^2}\,,\\
\frac{dr}{d\tau}&=&\pm  {\Bigl [} 1-\left (1-2M/r\right )\left (1+\frac{L_z^2}{r^2}\right ){\Bigr ]}^{1/2}\,.
\eeq
The particle has an orbital angular momentum $J_p=m_p L_z$. When $L_z=0$ the
particle falls straight into the black hole. For $L_z$ between zero and $4M$,
the particle spirals a finite number of times around the hole before crossing
the event horizon. For $L_z=4M$, the particle spirals an infinite number of
times around the marginally bound circular orbit at $r=4M$. For $L_z>4M$ the
particle never falls into the black hole, so we discard this case.

\begin{table}[ht]
  \centering \caption{\label{tab:Energyplungenonrotating} Energy radiated in
    each of the three lowest multipoles for a particle with mass $m_p$ and
    angular momentum $m_p L_z$ falling from infinity into a Schwarzschild
    black hole (from \cite{OoharaNakamura}). We show the percentage
    radiated in each mode relative to the total energy radiated (as
    extrapolated from the data, which typically yields an error of less than
    5$\%$). We also show the number $N=|\Delta \phi|/(2\pi)$ of ``laps'' the
    particle performs before plunging. The coefficients $a$ and $b$ are
    defined in the text. 
  }
\begin{tabular}{ccccccccccc}
%
\hline \hline
 $L_z/M$      & $a$  & $b$&   $N$   &$E_{\rm tot}$& $E_2$   &$\%$   &$E_3 $&$\%$   &$E_4 $&$\%$ \\
\hline
$0$  & $0.44$ & $2$ &  0    &   0.010            & $9.1\times 10^{-3}$&88  &$ 1.1\times 10^{-3}$ &10& $ 1.5\times 10^{-4}$&1.4  \\
$1$  & $0.21$ &$1.5$ &  0.15   &  0.013            & $1\times 10^{-2}$& 78&$ 2.3\times 10^{-3}$&18 &$ 5\times 10^{-4}$&4 \\
$2$  & $0.22$ &$1.1$  & 0.32   &  0.036            &$2.4\times 10^{-2}$ &67  &$ 8.1\times 10^{-3}$&22 &$ 2.7\times 10^{-3}$&7.5  \\
$3$  & $0.36$ &$0.86$  & 0.55  &  0.112            & $6.4\times 10^{-2}$&57 &$ 2.7\times 10^{-2}$&24 &$ 1.2\times 10^{-2}$&10.7 \\
$3.5$&$0.53$ &$0.76$   & 0.75  &  0.218            & $1.2\times 10^{-1}$&55 &$ 5.4\times 10^{-2}$&25 &$ 2.5\times 10^{-2}$&11.5  \\
$3.9$ &$1.0$ & $0.71$   & 1.15  &   0.485            & $2.4\times 10^{-1}$&49 & $1.2\times 10^{-1}$&25 &$ 5.8\times 10^{-2}$ &11.9\\
\hline \hline
\end{tabular}
\end{table}

In Table \ref{tab:Energyplungenonrotating} we show results from
\cite{OoharaNakamura} for the total energy radiated in the first three
multipoles ($l=2,~3,~4$) as a function of $L_z$. From the Table it is apparent
that the total energy output grows with $L_z$, and so does the energy in each
multipole $l$. The {\it relative} energy output in each mode behaves somewhat
differently: as $L_z$ grows, the percentile energy going into $l=2$ decreases.
The opposite happens for all other modes.

For $L_z=4M$ the total energy is obviously infinite: the particle spirals an
infinite number of times around $r=4M$ and therefore radiates for an infinite
time. The energy radiated as a function of $l$ is well approximated by a
simple function: $E_l\sim ae^{-b l} m_p^2/M$, where the coefficients
$a,~b$ (which are functions of $L_z$) are listed in Table
\ref{tab:Energyplungenonrotating}. The total energy radiated is well
approximated by $E_{\rm tot} \sim ae^{-2b}(1-e^{-b})^{-1}m_p^2/M$
\cite{OoharaNakamura}. Remarkably, the relative contribution of the $l=3$ mode
is always larger than $10\%$, and that of the $l=4$ mode is always larger than
$1\%$. As usual the $l=2$ mode dominates, with a relative contribution always
larger than $\sim 50\%$.

Also shown in Table \ref{tab:Energyplungenonrotating} is the number of spirals
the particle completes before entering the horizon. This number is useful for
two reasons. The first reason is that, if the particle falls with angular
momentum very close to the marginal value $4M$, it will complete many
revolutions and radiate a huge amount of radiation. In this case the
perturbation expansion would no longer be valid, and therefore we must make
sure that $N$ is not much larger than one. The second reason is that $N$ gives
us an estimate of how much of the output energy is due to the actual, almost
radial plunge motion (see eg. Fig.~\ref{velratio}), and how much of it comes
from the particle circling around the black hole.

\subsection{The energy radiated by plunging particles: rotating black holes}
The standard formalism for small perturbations of Kerr black holes was
formulated by Teukolsky \cite{teukolsky}. The equations decouple and separate,
reducing to two coupled ordinary differential equations with a source term. In
the case of gravitational waves emitted by particles plunging into the hole
the source term diverges at the boundaries, so this is not the most convenient
formalism (but see \cite{detweilerszedenits} for a way to get around these
difficulties). Using the alternative formalism developed by Sasaki and
Nakamura \cite{sn}, a series of papers by Nakamura and co-workers (see eg.
\cite{KojimaNakamura,KojimaNakamura2,OoharaNakamura,nakamurahaugan} and
references therein) examined the gravitational radiation emitted by point
particles moving in the vicinities of a Kerr black hole.

\begin{table}[ht]
  \centering \caption{\label{tab:Energyplungekerr} Energy radiated in each of
    the three lowest multipoles for a particle with zero angular momentum
    falling from infinity into a Kerr black hole along the equator, as a
    function of $j\equiv J/M^2$. Taken from Fig. 3 in \cite{KojimaNakamura2}.}
\begin{tabular}{cccccccccc}
\hline \hline
 $j$  & $E_{\rm tot}$& $E_2$   &$\%$   &$E_3 $&$\%$   &$E_4 $&$\%$ \\
\hline
$0.0$& $1.0\times 10^{-2}$&$9.1\times 10^{-3}$&88  &$ 1.1\times 10^{-3}$ &10& $ 1.5\times 10^{-4}$&1.4  \\
$0.7$& $1.8\times 10^{-2}$&$1.5\times 10^{-2}$& 83&$ 2.2\times 10^{-3}$&12 &$ 3.9\times 10^{-4}$&2.2 \\
$0.85$&$2.3\times 10^{-2}$&$1.9\times 10^{-2}$ &83&$ 3.4\times 10^{-3}$&15&$ 7.3\times 10^{-4}$&3.2 \\
$0.99$&$4.7\times 10^{-2}$&$3.3\times 10^{-2}$&70 &$ 9.6\times 10^{-3}$&20 &$ 2.7\times 10^{-3}$&5.7 \\
\hline \hline
\end{tabular}
\end{table}
\begin{table}[ht]
\centering \caption{\label{tab:Energyplungekerr2} Energy radiated in each of
  the three lowest multipoles for a particle with angular momentum $m_p L_z$
  falling from infinity into a Kerr black hole with $j=0.85$, along the
  equator. Taken from Fig. 5 in \cite{KojimaNakamura}.}
\begin{tabular}{ccccccccccc}
\hline \hline
 $L_z/M$   &$E_{\rm tot}$ &$N$    & $E_2$   &$\%$   &$E_3 $&$\%$   &$E_4 $&$\%$ \\
\hline
$2.6$ &   1.2       &0.97 & $5.0\times 10^{-1}$&42  &$ 3.0\times 10^{-1}$ &25& $ 1.8\times 10^{-1}$&15  \\
$1.3$&$1.0\times 10^{-1}$&0.2  & $6.5\times 10^{-2}$& 61&$ 2.5\times 10^{-2}$&23 &$ 9.0\times 10^{-3}$&8.4 \\
$0.65$&$4.8\times 10^{-2}$&0.06 &$3.5\times 10^{-2}$ &73  &$ 8.7\times 10^{-3}$&18 &$ 2.5\times 10^{-3}$&5.2  \\
$0.0$ &$2.3\times 10^{-2}$ &0.05 &  $1.9\times 10^{-2}$&81 &$ 3.4\times 10^{-3}$&15 &$ 7.3\times 10^{-4}$&3.2 \\
$-0.8$&$9.1\times 10^{-3}$&0.17 & $8.0\times 10^{-3}$&88 &$ 9.0\times 10^{-4}$&10 &$ 1.3\times 10^{-4}$&1.4  \\
$-2.25$& $1.4\times 10^{-2}$&0.38 & $1.2\times 10^{-2}$&81 & $1.8\times 10^{-3}$&12 &$ 3.5\times 10^{-4}$ &2.3\\
$-4.5$& $7.7\times 10^{-2}$ &1.02   & $5.5\times 10^{-2}$&71 & $1.3\times 10^{-2}$&17 &$ 5.0\times 10^{-3}$ &6.4\\
\hline \hline
\end{tabular}
\end{table}

The results for the total energy, as well as the energy radiated in each mode,
are summarized in Tables \ref{tab:Energyplungekerr} and
\ref{tab:Energyplungekerr2}. These Tables refer to particles falling along the
equator of the black hole. Starting with particles falling from infinity with
zero orbital angular momentum, Table \ref{tab:Energyplungekerr} shows a
familiar pattern. The total energy increases with increasing $j$. For
near-extremal black holes ($j=0.99$) the total energy radiated is almost five
times the Schwarzschild value.  Again, the total energy going into each
multipole $l$ increases. As previously pointed out, the relative contribution
of each mode increases with $j$ for $l>2$, but it decreases for $l=2$. The
$l=3$ mode always contributes more than $10\%$ of the total energy, and the
$l=4$ mode always contributes more than $1\%$.

Table \ref{tab:Energyplungekerr2} shows results for a $j=0.85$ black hole, for
several values of orbital angular momentum $L_z$. The total energy varies by
two orders of magnitude between $L_z=2.6M$ and $L_z=-0.8M$.  The energy
emitted is larger than $10\%$ for $l=3$, and larger than $1\%$ for $l=4$. We
also list the number of spirals before plunge. This number is always smaller
than unity, so the results can in principle be interpreted as a plunging
motion and applied to the merger phase.

Not shown here is the contribution of different $m$'s to the total energy. For
large black hole rotation and large positive values of $L_z$ most of the
energy goes into $l=m$ modes.  Negative-$m$ modes emit a negligible amount of
radiation in this regime. For negative $L_z$ the situation is different: all
modes seem to be excited to comparable amplitudes. See \cite{KojimaNakamura}
for more details.

\subsection{Linear and angular momentum radiated by plunging particles}

Computing the linear momentum carried by gravitational waves is of great
astrophysical importance. Coalescing binary black hole systems may abound in
galactic disks and in the centers of galactic nuclei. Due to the emission of
gravitational radiation the final black hole receives a ``kick,'' i.e., it
acquires a non-zero recoil velocity because of momentum conservation.
Depending on the momentum emitted, the recoil velocity may be large enough to
release black holes from the host galaxy. If so, gravitational radiation
effects will have considerable observable consequences for astrophysics and
cosmology, such as the depletion of black holes from host galaxies, the
disruption of active galactic core energetics, and the ejection of black holes
and stellar material into the intergalactic medium. In the following we
briefly review some perturbative calculations of the recoil velocity.

\subsubsection{Linear momentum}

In a Schwarzschild background, the linear momentum carried by gravitational
waves when a zero angular momentum point particle falls into the black hole is
$|\Delta P|=8.73 \times 10^{-4} m_p^2/M$ \cite{nakamurahaugan}.  This leads to
a recoil velocity $v \sim 2.63 (10m_p/M)^2{\rm ~km/s}$. For the general case
of a particle plunging with non-zero angular momentum, the linear momentum
carried by gravitational waves is well approximated by
\beq |\Delta P|&=&9 \times 10^{-6}\left [4\left (\frac{L_z}{M}\right )^2+5\frac{L_z}{M}+10\right ]^2
m_p^2/M\,,\quad 0<\frac{L_z}{M}<3.4\,, \\
&=&4.5\times 10^{-2}m_p^2/M\,,\quad 3.4<\frac{L_z}{M}<4\,. \eeq
This leads to a recoil velocity of
\beq v &\sim& 2.7 \left [\frac{2}{5}\left (\frac{L_z}{M}\right )^2+\frac{1}{2}\frac{L_z}{M}+1\right ]^2
(10m_p/M)^2{\rm ~km/s}\,,\quad 0<\frac{L_z}{M}<3.4\,,\\
&\sim&130~ (10m_p/M)^2{\rm ~km/s}\,,\quad 3.4<\frac{L_z}{M}<4\,. \eeq
These results are not very sensitive to the rotation of the black hole (see
eg. Fig.~6 in \cite{KojimaNakamura}). A hint to extrapolate recoil velocities
to mass ratios close to unity comes from the Newtonian result: replace $\mu/M$
by $f(q)$ \cite{detweilerfitchett}, where
\be f(q)=q^2\frac{q-1} {\left (q+1\right )^{5}}\,. \ee
The function $f(q)$ has two extrema at
\beq q&=& \frac{3-\sqrt{5}}{2}\sim 0.38\,, \quad f\left (\frac{3-\sqrt{5}}{2}\right)=-
\frac{1}{25\sqrt{5}}\equiv f_{\rm min}(q)\,,\label{max}\\
q&=&\left (\frac{3-\sqrt{5}}{2}\right )^{-1}\sim 2.62\,,\quad f \left (\frac{3+\sqrt{5}}{2}\right)=
\frac{1}{25\sqrt{5}}\equiv f_{\rm max}(q) \,. \label{min} \eeq
We then get
\beq v &\sim& 4.8 \left [\frac{2}{5}\left (\frac{L_z}{M}\right )^2+\frac{1}{2}\frac{L_z}{M}+1\right ]^2
\frac{f(q)}{f_{\rm max}(q)}{\rm ~km/s}\,,\quad 0<\frac{L_z}{M}<3.4\,,\\
&\sim&232\frac{f(q)}{f_{\rm max}(q)}{\rm ~km/s}\,, \quad 3.4<\frac{L_z}{M}<4\,. \eeq
%

\subsubsection{Angular momentum}

Radiation of angular momentum demands either a non-zero orbital angular
momentum of the particle, or a non-zero angular momentum of the black hole.
For Schwarzschild black holes, Oohara and Nakamura \cite{OoharaNakamura} found
that, even though both the radiated energy $E_{\rm tot}$ and angular
momentum $\Delta J$ diverge in the limit $L_z/M \rightarrow 4$, their ratio is
to a good approximation independent of $L_z$:
\be 
\left|\frac{M\, E_{\rm tot}}{\Delta J}\right|=0.15\,, 
\quad {\rm for}\,\quad |L_z/M| \gtrsim 1\,. \ee
For rotating black holes, the angular momentum radiated depends on the
relative sign between the rotation of the black hole and $L_z$. We denote by
$x$ the ratio of radiated energy to radiated angular momentum:
\be 
x(j,L_z)=\frac{M\, E_{\rm tot}}{\Delta J}\,. 
\ee
This quantity is listed in Table \ref{tab:xfactor} (after Table I in
\cite{KojimaNakamura}) for some values of $L_z$ and $j$.
\begin{table}[ht]
  \centering \caption{\label{tab:xfactor} The factor $x(j,L_z) \equiv M\,
    E_{\rm tot}/\Delta J$ for some values of $L_z$ and $j$ (from Table I in
    \cite{KojimaNakamura}).}
\begin{tabular}{cccccccc}
\hline
\hline \multicolumn{2}{c}{j=0} & \multicolumn{2}{c}{j=0.7}& \multicolumn{2}{c}{j=0.85}& \multicolumn{2}{c}{j=0.99}\\
\hline
         $L_z/M$  & $x$  &$L_z/M$& $x$   &$L_z/M$       &$x$  &$L_z/M$     &$x$  \\
\hline  $-4$ & $0.15$ & $-4.4$& 0.03  & $-4.5$   &0.04  &$-4.7$    &0.07\\
$-3$ &$0.15$  & $-3.3$& 0.04  &$-3.375$  &0.05  &$-3.5$    &0.13\\
$-1$ &$0.15$  & $-2.2$& 0.04  &$-2.25$   &0.06  &$-2.35$   &0.15\\
$1$ &$0.15$   & 1.5   & 0.19  &1.3       &0.21  &1         &0.30\\
$2$ &$0.15$   & 2.25  & 0.20  &1.95      &0.22  &1.5       &0.30\\
$3$ &$0.15$   & 3.0   & 0.22  &2.6       &0.24  &2         & 0.34 \\
\hline \hline
\end{tabular}
\end{table}

\subsection{\label{poinparticleapproach} Perturbation theory as a guide to numerical results for comparable mass ratios}

One of the most prominent features borne out of binary black hole simulations
seems to be the absence of strong non-linearities: the potential barrier close
to the black hole horizon acts as a very effective cloak, filtering out many
non-linear features of the dynamics \cite{bhcollisionpullinPRD}. For this
reason results from perturbation theory (i.e., $q\gg1$) can usually be
extrapolated to the equal mass ratio case, yielding very good agreement with
full blown numerical simulations. To quote Smarr, ``the agreement is so
remarkable that something deep must be at work'' \cite{smarr}. This was also
found to be the case for the head-on collision of two black holes. The
perturbation theory result $E_{\rm tot}=0.0104 (M_1^2/M_2)$, with $M_1\ll M_2$
\cite{drpp} can be compared to the full numerical result for $q=1$: $E_{\rm
  tot}\approx 0.0013M =0.001\times 16\eta^2M=0.0104\times (2M ) \eta^2$ (see
\cite{bhcollisionpullinPRD}). Simple scaling arguments applied to perturbative
results work surprisingly well.

On this basis, a natural conjecture is that particles plunging with large but
sub-critical orbital angular momentum should describe reasonably well the
final stages of a binary black hole inspiral \cite{detweilerszedenits}.
Indeed, some of the results discussed in the main text suggest that the merger
of two equal mass black holes can be described by extrapolating results from
perturbation theory. For instance, the final spin of the black hole can be
predicted by using the small mass ratio approximation: see
Eq.~(\ref{iscofinalj}) and the related discussion.

For inspiralling binaries evolving through quasi-circular orbits, the plunge
(``merger'') happens when the smaller body crosses the ISCO, even though this
notion is not well defined for comparable-mass bodies. We argue that the
merger phase should be reasonably well described by a particle plunging with
an orbital angular momentum $L_z$ only slightly smaller than the marginal
value $L_z=4M$ (corresponding to a radius equal to $4M$). In this case the
trajectory resembles that of a particle going through the merger phase: a
quasi-circular orbit followed by a plunge. It is important to specify how
close $L_z$ should be to the marginal value.  We look for orbits that complete
barely less than one lap before plunging. This guarantees that the energy
output is due only to a plunge trajectory; it also guarantees that the orbit
was quasi-circular before the plunge. We therefore argue that, as seen from
the Table, $L_z\simeq 3.9M$ is a near-optimal value. An obvious objection is
that the ISCO for point particles is not at $r=4M$ (the location of the
marginally bound orbit), but rather at $r=6M$.  Fortunately, for more massive
bodies an approximate ISCO can be defined, and it is usually closer to $r=4M$.
We can thus try to extend these perturbative results to understand the merger
phase.

Extrapolating the point particle results presented in the previous subsections
we get, for non-rotating holes:
\be
\frac{E}{M}=0.485\frac{q^2}{(1+q)^4}\,,\qquad
\frac{M\, E}{\Delta J}=0.15\,,\qquad
v_{\rm recoil}=232\frac{f(q)}{f(q)_{\rm max}}\,{\rm km\cdot s^{-1}}\,.
\ee

For instance, the value $0.485 q^2/(1+q)^4$ for the energy is found
extrapolating the last row in Table \ref{tab:Energyplungenonrotating} by the
substitution $m_p^2/M^2 \rightarrow q^2/(1+q)^4$. From fits of the
corresponding numerical quantities at the 3PN ISCO (which should work as a
good reference point, for lack of a better guess) we get:

\be
\frac{E_{\rm ISCO}}{M}=0.421\frac{q^2}{(1+q)^4}\,,\qquad
\frac{M\, E_{\rm ISCO}}{\Delta J_{\rm ISCO}}=0.19\,,\qquad
v_{\rm recoil}^{\rm ISCO}=120\frac{f(q)}{f(q)_{\rm max}}\,{\rm km\cdot s^{-1}}\,,
\ee

The two sets of values are reasonably consistent. The largest disagreement
refers to the linear momentum radiated, and therefore to the recoil velocity
of the final hole.

\section{Polarization of the waveforms}
\label{app:polarization}

Getting information about the polarization content of a waveform is simple in
the presence of a monochromatic wave.  In more general settings, making
statements about the polarization state is easier in Fourier space. Methods to
compute the polarization content of a waveform solely from a time-domain
analysis were presented in \cite{polarization}. Here we shall adapt these
techniques for the case at hand.

From the (real) polarization components $h_+(t)\,,h_{\times}(t)$\footnote{In
  practice we do not use the gravitational wave amplitudes, but the real and
  imaginary components of $\Psi_4$, in order to avoid problems with the
  integration constants and to sidestep the memory effect discussed in Section
  \ref{memory}. This should have no influence on the final results.} we can
define the so-called {\it analytic signal} $H_+\,,H_{\times}$ in the following
way:
\beq H_+&\equiv& h_+(t)+i{\cal H}_{+}(t)\,,\\
H_{\times}&\equiv& h_{\times}(t)+i{\cal H}_{\times}(t)\,.\eeq
The imaginary part of the analytic signal is the Hilbert transform ${\cal
  H}(t)$ of the signal $h(t)$, defined as
\be {\cal H}(t)=\frac{1}{\pi}\int_{-\infty}^{+\infty} \frac{h(\tau)}{t-\tau}d\tau \,,\ee
where the integral is taken as the Cauchy principal value. For reference we
note that the Hilbert transform of $\sin t$ is $-\cos t$ and the transform of
$\cos t$ is $\sin t$.

From the analytic signal we define a covariance matrix ${\bf C}$ as
\begin{eqnarray}
{\bf C}=\left(
  \begin{array}{ccccc}
 H_+ H_+^* & H_+H_{\times}^* \\
 H_{\times}H_+^* & H_{\times}H_{\times}^*
  \end{array}
\right)\,,
\end{eqnarray}
where an asterisk stands for complex conjugation. We note that the covariance
matrix is Hermitian, and thus its eigenvalues $\lambda_0\,,\lambda_1$ are real
and positive. Without loss of generality, we assume $\lambda_0>\lambda_1$. It
can be shown that the normalized eigenvector ${\bf v}=(x_0\,,y_0)$ ($|{\bf
  v}|=1$) associated with $\lambda_0$ points in the direction of the largest
amount of polarization \cite{polarization}. One can define an elliptical
component of polarization $P_E$ as
\be P_E=\frac{\sqrt{1-X^2}}{X}\,, \ee
where
\be
X=\max_{\alpha}\sqrt{{\rm Re}[e^{i\alpha}x_0]^2+{\rm Re}[e^{i\alpha}y_0]^2}\,.
\ee
The quantity $P_E=1$ for circular polarization, and $P_E=0$ for linear
polarization. For illustration, consider the waveform $h_+=\sin
t\,,h_{\times}=\cos t$. This is a good approximation to a typical inspiral
waveform at large orbital separation, as viewed from the normal to the orbital
plane, and it is obviously circularly polarized. For this waveform we have
\begin{eqnarray}
{\bf C}=\left(
\begin{array}{ccccc}
 1 & -i \\
 i & 1
  \end{array}
\right)\,,
\end{eqnarray}
and ${\bf v}=(-i/\sqrt{2},1/\sqrt{2})$. This implies $X=1/\sqrt{2}$ and
$P_E=1$, as expected. For $h_{\times}=0$ (linear polarization) we would have
$X=1$ and $P_E=0$. Thus $P_E$ is a good indicator of the degree of circular or
linear polarization. We can also define a polarization strength as follows:
\be P_S=1-\f{\lambda_1}{\lambda_0}\,. \ee
$P_S=1$ means that the waveform is entirely linearly polarized or circularly
polarized (there is only one polarization component), and $P_S=0$ means that
the two polarization states have comparable magnitude.

\begin{figure*}[ht]
\begin{center}
\epsfig{file=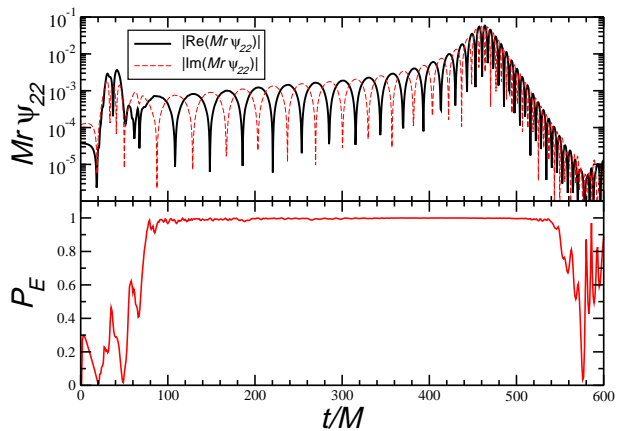,width=7cm,angle=-90}
\caption{Top: $|{\rm Re}(Mr\,\psi_{22})|$ and $|{\rm Im}(Mr\,\psi_{22})|$
  for $q=2.0$, D8 (top). Bottom: the degree of elliptic polarization $P_E$ for
  the $l=2$ waveform as viewed from the normal to the orbital plane.
  \label{wf1-pol}}
\end{center}
\end{figure*}

In Fig.~\ref{wf1-pol} we show the result of computing $P_E$ using the dominant
($l=|m|=2$) component of a binary black hole merger waveform with $q=2.0$.
This plot clearly shows that the polarization is circular for both inspiral
and ringdown, with the exception of the unphysical portions of the wave: the
initial data burst and the final, noise-dominated part of the ringdown
waveform.



\end{document}